%% file: Fission_fragment_1.tex
\documentclass[10pt,aps,prc,floatfix,twocolumn,nofootinbib]{revtex4-1}

\usepackage[subpreambles=true]{standalone}

\usepackage[labelfont={small},subrefformat=parens,caption=false]{subfig}
\usepackage{graphicx,amsmath,amssymb,bm}
\usepackage{amsfonts}
\usepackage[utf8]{inputenc}

\usepackage{verbatim}
\usepackage{float}
\usepackage{cancel}
\usepackage{multirow}
\usepackage{array}
\usepackage{xparse}
\usepackage{makecell}

\usepackage{physics}
\usepackage{color}
\usepackage{dsfont}
\usepackage{tabularx}

\usepackage[pdfencoding=auto, pdfpagelabels]{hyperref}

\usepackage[overload]{textcase}

\definecolor{linkcolor}{rgb}{0,0,0.40} 
\hypersetup{%
    pdfsubject=Paper,
    pdfkeywords={nuclear physics} {Bayesian} {chiral EFT},
    unicode = true,
    breaklinks = true,
    colorlinks = true,
    linkcolor = linkcolor,
    citecolor = linkcolor,
    menucolor = linkcolor,
    urlcolor = linkcolor
}

\usepackage[linesnumbered,algoruled,lined,commentsnumbered]{algorithm2e}

\usepackage{cellspace}
\graphicspath{{./figures/}}

\usepackage{silence}
\usepackage[T1]{fontenc}

\usepackage[]{standalone}

\newcommand{\phantomsublabel}[3]{%
\unitlength=1in%
\put(#1,#2){%
    \subfloat[]{%
        \label{#3}%
    }}%
}

\input{buqeye_macros}

\newcommand{\Lambdab}{\Lambda_b}
\newcommand{\mpi}{m_{\pi}}
\newcommand{\Qpoly}{Q_{\mathrm{smax}}}

\newcommand{\Qsmax}{Q_{\mathrm{smax}}}

\newcommand{\Qsum}{Q_{\mathrm{sum}}}
\newcommand{\qcm}{q_{\mathrm{CM}}}
\newcommand{\mpieff}{m_{\mathrm{eff}}} 
\newcommand{\psmax}{p_{\mathrm{smax}}}
\newcommand{\negcos}{-\cos(\theta)}

\newcommand{\dcip}{P}

\begin{document}


\title{Assessing Correlated Truncation Errors in Modern Nucleon-Nucleon Potentials}

\author{P.~J. Millican}
\email{millican.7@osu.edu}
\affiliation{Department of Physics, The Ohio State University, Columbus, OH 43210, USA}

\author{R.~J. Furnstahl}
\email{furnstahl.1@osu.edu}
\affiliation{Department of Physics, The Ohio State University, Columbus, OH 43210, USA}

\author{J.~A. Melendez}
\email{melendez.27@osu.edu}
\affiliation{Department of Physics, The Ohio State University, Columbus, OH 43210, USA}

\author{D.~R.~Phillips}
\email{phillid1@ohio.edu}
\affiliation{Department of Physics \& Astronomy and Institute of Nuclear \& Particle Physics, Ohio University, Athens, OH 45701, USA}

\affiliation{Department of Physics, Chalmers University of Technology, SE-41296 G\"oteborg, Sweden}

\author{M.~T. Pratola}
\email{mpratola@stat.osu.edu}
\affiliation{Department of Statistics, The Ohio State University, Columbus, OH 43210, USA}

\date{\today}

\begin{abstract}
We test the BUQEYE model of correlated effective field theory (EFT) truncation errors on Reinert, Krebs, and Epelbaum's semi-local momentum-space implementation of the chiral EFT (\chiEFT) expansion of the nucleon-nucleon (NN) potential.
This Bayesian model hypothesizes that dimensionless coefficient functions extracted from the order-by-order corrections to NN observables can be treated as draws from a Gaussian process (GP).
We combine a 
variety of graphical and statistical diagnostics
to assess when predicted observables have a \chiEFT\ convergence pattern consistent with the hypothesized GP statistical model.
Our conclusions are: First, the BUQEYE model is generally applicable to the potential investigated here, 
which
enables statistically principled estimates of the impact of higher EFT orders on observables.
Second, parameters defining the extracted coefficients such as the expansion parameter $Q$ must be well chosen for the coefficients to exhibit a regular convergence pattern---a property we exploit to obtain posterior distributions for such quantities.
Third, the assumption of GP stationarity across lab energy and scattering angle is not generally met; this necessitates adjustments in future work.
We provide a workflow and interpretive guide for our analysis framework, and show what can be inferred about 
probability distributions for $Q$, 
the EFT breakdown scale $\Lambdab$, the scale associated with soft physics in the \chiEFT\ potential $\mpieff$, and the GP hyperparameters.
All our results can be reproduced using a publicly available Jupyter notebook, which can be straightforwardly modified to analyze other \chiEFT\ NN potentials. 
\end{abstract}

\maketitle


\section{Introduction}
\label{sec:introduction}

Nucleon-nucleon (NN) potentials based on a chiral effective field theory (EFT) expansion, 
suitably augmented by three-nucleon forces, have found fruitful application in computations of finite nuclei and of nuclear and neutron matter~\cite{Epelbaum:2008ga,Machleidt:2011zz,Hammer:2019poc,Furnstahl:2021rfk,Tews:2022yfb}.
However, given that these nuclear forces are derived from an EFT at finite order, we need to know about the error induced by the truncation of that expansion. 
This is essential for robust uncertainty quantification (UQ) because such errors are expected to be at least comparable to
uncertainties induced by fitting the rather accurate data in the NN sector and precisely known binding energies of particular nuclides.
Failure to account for these truncation errors---including their correlations in energy and angle---could lead to 
bias and prevent us from properly propagating uncertainties to predictions.
In this paper, we test the BUQEYE model of correlated EFT truncation errors.

Various chiral NN potentials have been developed recently~\cite{Epelbaum:2014efa,Gezerlis:2014zia,Piarulli:2014bda,Ekstrom:2015rta,Carlsson:2015vda,Reinert:2017usi,Ekstrom:2017koy,Entem:2017gor}. 
These potentials differ in the way they are regulated (local, non-local, or a mix of the two called semi-local), the value of the associated regulator parameter, the order to which the EFT expansion is carried out (labeled \NNLO, \NNNLO, \NNNNLO, and even beyond; see below), and the data to which the parameters in the potentials are fit. 
In some cases the potentials include explicit $\Delta(1232)$ degrees of freedom. 
However, all claim to be an implementation of the idea---originally proposed by Weinberg over thirty years ago---that the NN potential can be organized as an expansion in powers of the parameter $Q \equiv f(\prel,m_\pi)/\Lambdab$, where $\prel$ is the relative momentum of the two nucleons, $\mpi$ is the rest mass of the pion, and $\Lambdab$ is the breakdown scale of the theory)~\cite{Weinberg:1990rz,Weinberg:1991um}. The benefit of such an organization is that systematically smaller effects occur at higher order in the expansion, and so calculations of observables should become progressively more accurate when potentials of higher order are employed.

In Refs.~\cite{Furnstahl:2015rha,Melendez:2017phj}  we proposed a pointwise Bayesian statistical model (building on Refs.~\cite{Cacciari:2011ze, Epelbaum:2014efa}) to estimate EFT truncation errors for predicted observables $\genobs$.
``Pointwise'' in this context means that the error model is applied independently at each energy or angle in the input space (generically denoted $x$).
The model incorporates expert knowledge~\cite{RePEc:taf:amstat:v:73:y:2019:i:s1:p:69-81} on the convergence pattern (inherited from the EFT power counting) into prior distributions, and subsequently updates these beliefs given order-by-order predictions of observables $\{\genobs_n\}$, thereby formalizing the notion of EFT convergence for $\genobs$.

Subsequently, we extended this pointwise model 
to a curvewise treatment of dimensionless coefficient \emph{functions} $c_{n}(x)$ extracted from the order-by-order predictions $\{y_{n}(x)\}$~\cite{Melendez:2019izc}.
The BUQEYE model postulates that the $c_{n}(x)$ can be modeled as random draws from a Gaussian process (GP).
GPs are powerful tools for both regression and classification, and have become popular in statistics, physics, applied mathematics, machine learning, and geostatistics~\cite{sacks1989design,cressie1992statistics,rasmussen_gaussian_2006}.
The BUQEYE model uses them in an atypical way by learning the GP parameters characterizing the distribution of the known coefficient functions and, by induction, using this same GP to predict error bands for unknown higher-order contributions.
This correlated truncation error model has  been applied to infinite nuclear matter~\cite{Drischler:2020hwi,Drischler:2020yad}, nucleon-nucleus elastic scattering~\cite{Baker:2021iqy},  $np \leftrightarrow d \gamma$ reactions~\cite{Acharya:2021lrv}, muon capture by deuterium~\cite{Gnech:2023mvb}, experimental design for proton Compton scattering~\cite{Melendez:2020ikd}, and parameter estimation of NN low-energy constants (LECs)~\cite{Svensson:2023twt} (see also ~\cite{Wesolowski:2018lzj,Wesolowski:2021cni,Alnamlah:2022eae,Poudel:2021mii}).

In this paper, we exemplify a workflow for using the BUQEYE correlated error model.
We apply it to examine in detail whether the model
can accurately describe order-by-order convergence of the observables predicted by modern NN potentials.
However, in order to achieve this the input space and hyperparameters of the GP that encodes our error model must be carefully chosen. 
We limit ourselves to the neutron-proton (\emph{np}) scattering observables predicted using the semi-local momentum-space (SMS) potential of Reinert, Krebs, and Epelbaum with cutoff 500 MeV~\cite{Reinert:2017usi}. 
(Other potentials and proton-proton scattering will be examined in future work.)
We emphasize that 
we do not refit the LECs of this potential to data here;
instead, the LECs for the N$^{k}$LO potential are taken from the N$^{k}$LO fit as given by the original publication~\cite{Reinert:2017usi}. 
However, because our truncation-error model can be embedded within a Bayesian parameter estimation framework (see Ref.~\cite{Wesolowski:2018lzj}), we are setting the stage for a full Bayesian treatment of such potentials with correlated errors.

In Sec.~\ref{sec:the_model} we review the GP model of EFT truncation errors~\cite{Melendez:2019izc}, 
starting with the extraction of the coefficient functions from order-by-order predictions of the total cross section based on a chosen \chiEFT\ potential.
With appropriate choices of input parameters, visual inspection of these functions suggests they are consistent with statistical draws from a common GP. 
The tools for a more complete analysis that uses complementary statistical diagnostics and outputs of our GP analysis are summarized in
Sec.~\ref{sec:diagnosticsandoutputs}, which includes a schematic workflow for model-checking.
We then extract posteriors for the expansion parameter $Q$ in Sec.~\ref{sec:gp_stationarity}
and thereby expose the inadequacy of 
a GP model that is stationary in an input space defined by the scattering angle and energy. 
This flaw is mitigated by the requirement of statistical consistency for the EFT's error bands, which implies Bayesian credibility intervals for both the EFT breakdown momentum $\Lambda_b$ and the effective soft scale $\mpieff$ (typically identified with the pion mass for \chiEFT).
We show that consistent choices for these EFT scales produce approximate stationarity provided we employ a suitable input space and consider only observables corresponding to relative momenta above the pion mass.
In Sec.~\ref{sec:results_model} we demonstrate how this informed choice of input parameters leads to coefficient functions consistent with the BUQEYE model
across observables.
Two representative applications are given in Sec.~\ref{sec:applications} before 
Sec.~\ref{sec:outlook} provides a summary 
and highlights outstanding problems and future work. 


\section{A model of EFT correlated truncation errors}
\label{sec:the_model}

We recapitulate briefly in Sec.~\ref{subsec:buqeye_recap} the details of the BUQEYE model for correlated truncation errors~\cite{Melendez:2019izc}.
For our analysis we need to convert  a potential's order-by-order predictions to dimensionless coefficients, which requires choosing a set of parametrizations.
These are introduced and their rationales and functional forms explained in Sec.~\ref{subsec:parametrizations}.
Then, in Sec.~\ref{subsec:bayesian_vars_intro} we give a simple example of the impact that these parametrization choices can have and discuss how the convergence pattern explained in Sec.~\ref{subsec:buqeye_recap} can be enhanced for a given set of parametrization choices.

\subsection{Recap of BUQEYE model}
\label{subsec:buqeye_recap}

We follow Refs.~\cite{Furnstahl:2015rha,Melendez:2019izc,Melendez:2019izc} and formalize the power counting of the EFT by writing an observable $\genobs$ as
\begin{align} \label{eq:obs_k_expansion}
    \genobs_k(\kinparvec) = \genobsref(\kinparvec) \sum_{n=0}^k c_n(\kinparvec) Q^n(\kinparvec),
\end{align}
where $k$ is the highest computed order of the EFT, $Q$ is the dimensionless expansion parameter, 
and $x$ is the input space(s).
In the case of angular NN scattering observables $\kinparvec=(\Elab,\theta)$, the lab-frame energy and center-of-mass scattering angle, respectively, whereas $\kinparvec=\Elab$ for the total cross section $\sigmatot$.
The dimensionful reference scale $\genobsref$ is taken to be $\genobsref(\kinparvec) \equiv \genobs_k(\kinparvec)$ for the total and differential cross sections and $\genobsref(\kinparvec) \equiv 1$ for spin observables, except in Sec.~\ref{subsec:yref} where an example of a mischosen $\genobsref$ is provided.
The theoretical uncertainty, $\delta\genobs_k$, is then written as:
\begin{equation} \label{eq:obs_k_truncation}
    \delta \genobs_k(\kinparvec)= \genobsref(\kinparvec) \sum_{n=k+1}^\infty c_n(\kinparvec) Q^n(\kinparvec).
\end{equation}
Given choices of $\genobsref$, the expansion parameter $Q$ and the input space $x$, the observable coefficients $c_n$ for $n=0, \ldots, k$ are completely determined by the order-by-order predictions $\genobs_n$.

The formulation of Eqs.~\eqref{eq:obs_k_expansion} and~\eqref{eq:obs_k_truncation} allows us to estimate the size and correlations of the truncation error $\delta\genobs_k$ if we can relate the properties of these low-order $c_n$s to the properties of the higher-order $c_n$s.
An underlying assumption of the BUQEYE model is that the $c_n$, if properly extracted, should share common properties due to the regularity with which a well-constructed EFT should converge.
For most NN observables the \chiEFT\ expansion does not converge monotonically from above or below. 
The coefficients thus show no preference for positive or negative values and their mean is about zero.
The smoothness of the EFT corrections as functions of $\kinparvec$ is also inherited by the coefficients.
But the particular choices made for $Q$, $x$, and $\genobsref$ can affect whether the $c_n$ have similar properties across the domain; this is the property of \emph{stationarity}. 

If we can show that the known $c_n$ are effectively random and embody similar properties, we can learn these properties to inform ourselves of what higher-order $c_n$ should look like.
To this end, we hypothesize that the $c_n$ are independent and identically distributed (iid) draws from a Gaussian process\footnote{The $z \sim \cdots$ notation is a common shorthand in statistical literature for ``$z$ is distributed as.'' We use ``$\pr(z) = \cdots$'' as well. Also, ``$z\,|\,I$'' is read as ``$z$ given $I$.''}
\begin{align}
    c_n(\kinparvec) \given \param & \overset{\text{\tiny iid}}{\sim} \GP[\mu=0, \sdth^2 r(\kinparvec,\kinparvec';L)] , \label{eq:cn_iid} 
\end{align}
which is a collection of random variables, any finite number of which have a joint Gaussian distribution~\cite{rasmussen2006gaussian}.
We take the mean function $\mu$ to be identically zero.
The \emph{a~priori} unknown GP parameters are $\param = \{\sdth^2, L\}$,%
\footnote{We follow statistical convention and use $\param$ to denote the vector of these parameters but also use $\theta$ for the center-of-mass scattering angle.}
where $\sdth$, the marginal standard deviation, controls the average magnitude of the curves, and $L$, the correlation length scale matrix, controls the approximate frequencies with which the curves oscillate.
The correlation function $r$ governs the smoothness properties of the curves.

Our approach to modeling truncation errors is then based on the idea that from the knowledge of a few coefficients, we can create a statistically meaningful distribution for the truncation error $ \delta\genobs_k(\kinparvec)$. 
In particular, 
Ref.~\cite{Melendez:2019izc} showed that, given \eqref{eq:cn_iid}, 
\begin{align} \label{eq:discr_k_prior}
    \delta\genobs_k(\kinparvec) \given \param, Q \sim \GP[0, \sdth^2\discrcorr{k}(\kinparvec, \kinparvec'; L)],
\end{align}
where
\begin{align}
    \discrcorr{k}(\kinparvec, \kinparvec';L) & \equiv \genobsref(\kinparvec)\genobsref(\kinparvec')\frac{[Q(\kinparvec)Q(\kinparvec')]^{k+1}}{1 - Q(\kinparvec)Q(\kinparvec')} r(\kinparvec, \kinparvec';L). \label{eq:truncation_corrfunc}
\end{align}
Given a prior $\pr(\param)$, the $\param$ needed to define Eq.~\eqref{eq:discr_k_prior} can be estimated---and subsequently integrated over, if desired---by using the low-order $c_n$ (see Ref.~\cite{Melendez:2019izc}).

To match the observed smoothness of calculated coefficient functions, we choose the squared exponential (a.k.a.\ radial basis function or Gaussian) for $r$ (see Table~\ref{tab:observable_assumptions}).
But this choice also assumes stationarity, which will prove to be questionable for the energy/momentum dependence of many of the coefficients. 
The length scale prior is taken to be uninformative: 
a uniform distribution over all positive values (an improper prior, but one that nonetheless expresses the wide range of values that the length scale can take).
In addition, we use a weakly informative conjugate prior for the $c_n$ variance that accurately reflects our prior beliefs: If the $c_n$ are naturally sized, we expect that $\sdth \approx 1$, but allow it to vary if the data support other values.
We have tested prior dependence in previous works~\cite{Furnstahl:2015rha,Melendez:2017phj} and found it to be slight.%
\footnote{The insensitivity to the precise form of the prior motivates the use of conjugate priors, for which updating from prior to posterior as we accumulate data on low-order coefficients is analytic. The posterior for $\mu, \sdth^2$ has the same functional form as the prior (see Ref.~\cite{Melendez:2019izc} for additional discussion). 
There are no conjugate priors for $L$ or $Q$.}

Our statistical model can also incorporate symmetry constraints on the values of scattering observables~\cite{Melendez:2019izc}.
In particular, some spin observables are constrained to take the value of zero at particular angles when time-reversal invariance is imposed as a necessary symmetry.
Specifically, time-reversal symmetry fixes $A(\Elab, \theta = 0^{\circ}) = 0$ and $A_{y}(\Elab, \theta = \left\{0^{\circ}, 180^{\circ}\right\}) = 0$ for all $\Elab$\,\cite{lafrance:jpa-00208966}, where we follow the nomenclature of Ref.~\cite{Melendez:2017phj}.
These symmetry constraints 
shrink the uncertainty band in the immediate vicinity of the constraint point(s);
for how this can impact statistical diagnostics, see Sec.~\ref{subsec:traintestsplit}.

\subsection{Options for parametrization}
\label{subsec:parametrizations}

Inverting Eq.~\eqref{eq:obs_k_expansion} enables the extraction from NN observable data of coefficient functions given some choice of $\genobsref$ and $Q$ parametrization, and these functions can be plotted against an input space $x$.
Then, we assess those coefficients visually and statistically to see if our choices of $Q$, $\genobsref$, and $x$ yield coefficients that are stationary across the domain of interest and manifest 
expected 
power counting by exhibiting common properties across orders.
Here we 
focus on  choices for $Q$ and $x$.

\begin{figure}[tbh]
    \centering
    \includegraphics[width=\columnwidth]{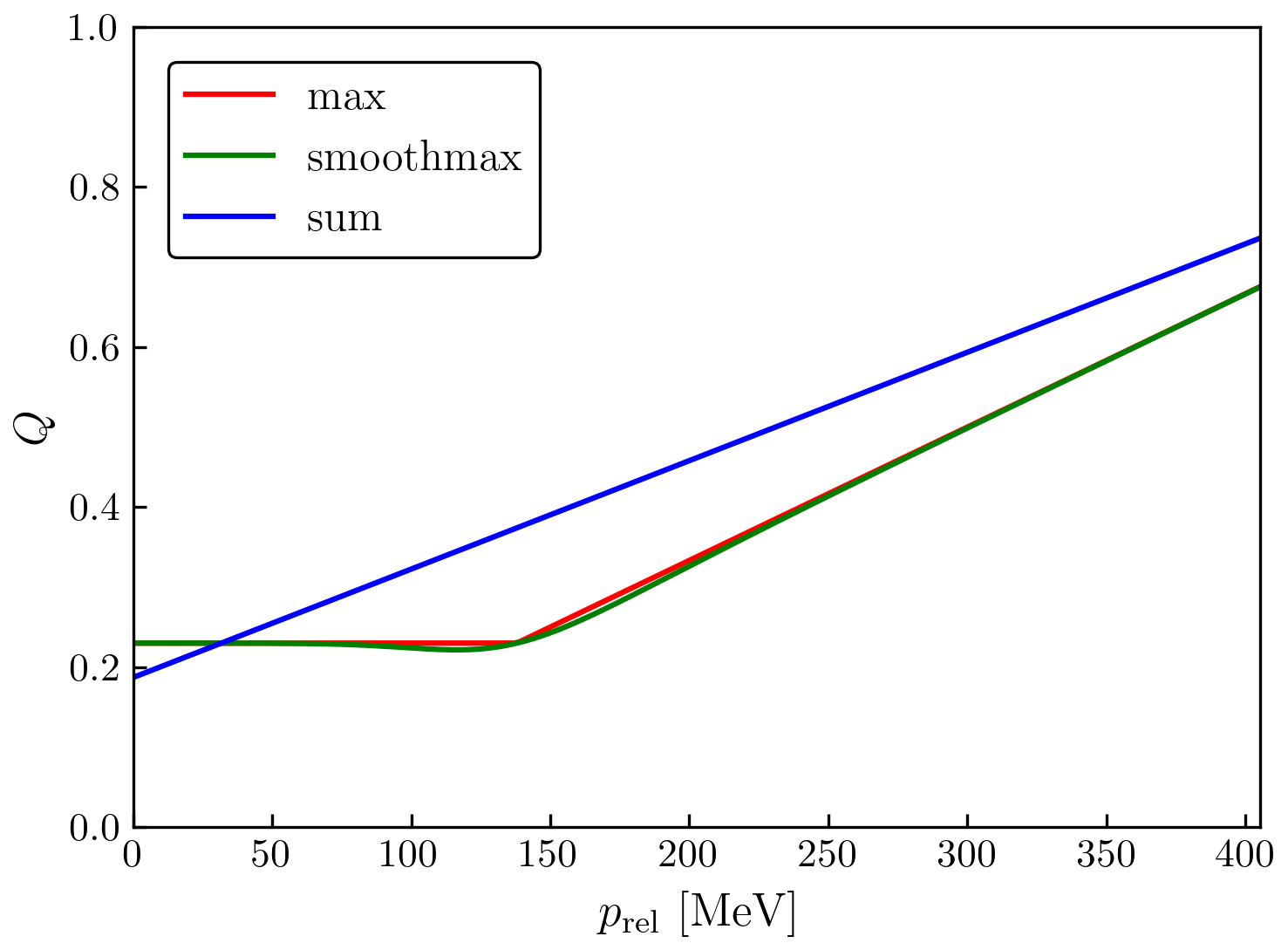}
    \caption{Plot of the three parametrizations of the dimensionless expansion parameter $Q$ that are tested in this paper --- $\Qmax$, $\Qpoly$, and $\Qsum$ --- versus relative momentum $\prel$.
    In plotting all three prescriptions here, it is assumed that $\Lambdab = 600\,\mathrm{MeV}$ and $\mpieff = \mpi = 138\,\mathrm{MeV}$, but that is not the case elsewhere in this work.
    }
    \label{fig:Q_params_vs_p}
\end{figure}

The conventional form of $Q(p, \mpieff)$~\cite{Epelbaum:2014efa} is
\begin{align}
\label{eq:Q_max}
    \Qmax(p, \mpieff) = \frac{\max(p, \mpieff)}{\Lambdab} \;,
\end{align} 
with $p = \prel$ and $\mpieff = m_{\pi} = 138 \, \mathrm{MeV}$.
This choice of $Q$ prescription was made in previous work (e.g., in~\cite{Melendez:2017phj,Wesolowski:2018lzj}).
$\Qmax$ can pose issues for GP fitting due to the cusp at $\prel = \mpieff$, so we introduce
a differentiable smooth-maximum (``smoothmax'' or ``smax'') version given by~\cite{Melendez:2017phj}
\begin{align}
\label{eq:Q_poly}
    \Qpoly(p, \mpieff) = \frac{1}{\Lambdab} \frac{p^{i} + \mpieff^{i}}{p^{i - 1} + \mpieff^{i - 1}} \;,
\end{align}
where we choose $i=8$ (see Fig.~\ref{fig:Q_params_vs_p} for a comparison).

In Figs.~\subref*{fig:sgt_coeff_Qmax_Elab_mpi138} and \subref*{fig:sgt_coeff_Qmax_prel_mpi138}, we compare coefficient functions for the \emph{np} total cross section ($\sigmatot$) for $\Qpoly$ and two choices of the input space,  $x = \Elab$ and $x = \prel$.
Here,
\begin{align} \label{eq:preltoElab}
    \Elab = \frac{\prel^{2} - m_{1} m_{2} + \sqrt{(\prel^{2} + m_{1}^{2}) (\prel^{2} + m_{2}^{2})}}{m_{2}},\;
\end{align}
where $m_{1}$ is the mass of the beam particle and $m_{2}$ the mass of the target particle in MeV, 
which simplifies to
\begin{align}
    \Elab = \frac{2 \prel^{2}}{M}
    \label{eq:prel_to_Elab}
\end{align}
in the case where $m_{1} = m_{2} = M$.
This choice of independent variable $\kinparvec$ can cause the $c_n$ to have a shorter local wavelength at low $\Elab$ compared to high $\Elab$.

As seen in 
Figs.~\subref*{fig:sgt_coeff_Qmax_Elab_mpi138} and \subref*{fig:sgt_coeff_Qmax_prel_mpi138},
choosing to parametrize $Q$ with  $\Qpoly$ (or $\Qmax$) can lead to the $c_n$ growing systematically with $n$ for $\prel \lesssim m_\pi$.
This behavior would violate the BUQEYE model's assumption that the coefficients share common properties across orders.
A third prescription,
\begin{align}
\label{eq:Q_sum}
    \Qsum(p, \mpieff) = \frac{p + \mpieff}{\Lambdab + \mpieff} \;,
\end{align} 
was originally devised to ameliorate this issue
as seen in Figs.~\subref*{fig:sgt_coeff_Qsum_Elab_mpi138} and~\subref*{fig:sgt_coeff_Qsum_prel_mpi138}; a more direct motivation for Eq.~\eqref{eq:Q_sum} will be given in Sec.~\ref{sec:gp_stationarity}.

Both $\Qsmax$ and $\Qsum$ are defined so that the breakdown scale $\Lambdab$ is defined by $Q(p = \Lambdab, \mpieff) = 1$.
However, their differing functional forms should 
be borne in mind when comparing values of $\Lambdab$ and $\mpieff$ between the two prescriptions; more details are given in Sec.~\ref{sec:gp_stationarity}.
We note that none of these three forms for $Q$ are based on analytical arguments regarding the combinations of $p$ and $m_\pi$ that appear in \chiEFT\ Feynman diagrams.

\begin{figure}[tbh]
    \centering
    \includegraphics{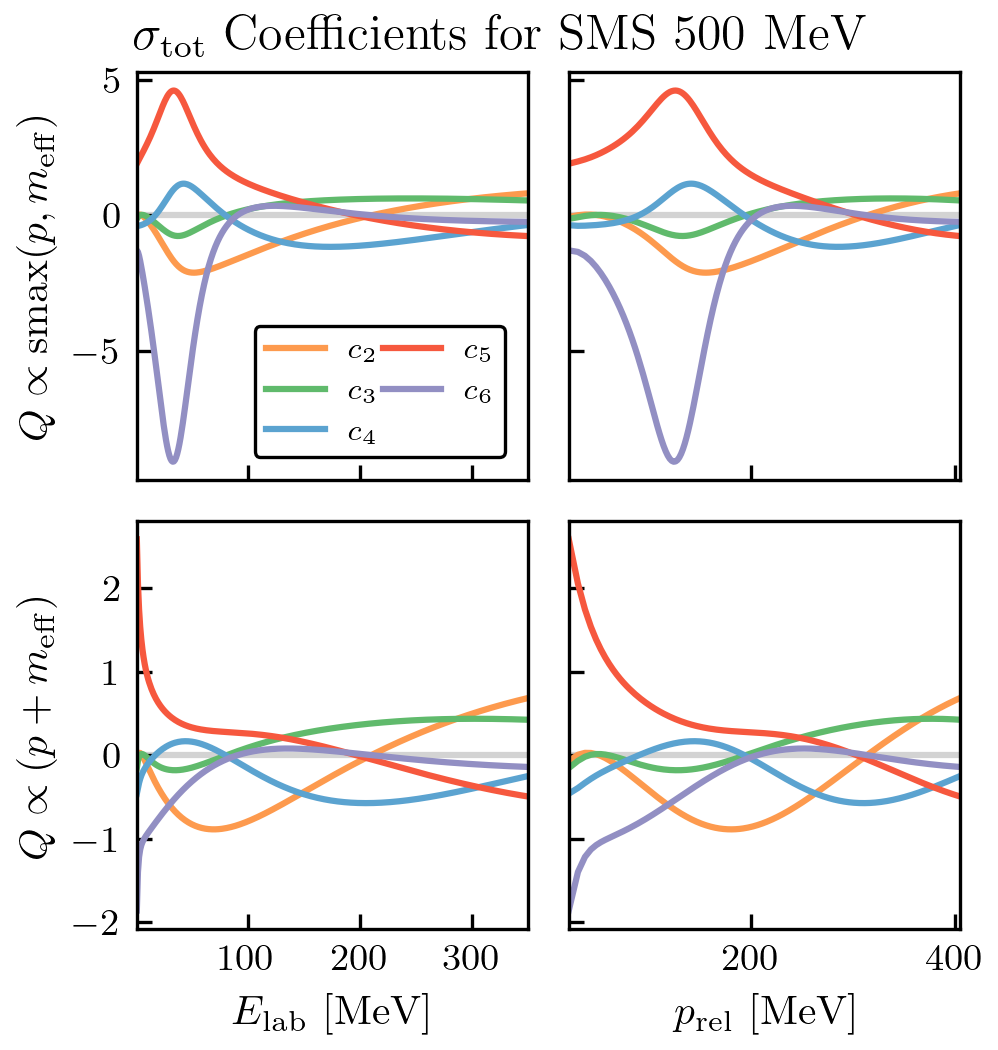}
    \phantomsublabel{-1.85}{3.00}{fig:sgt_coeff_Qmax_Elab_mpi138}
    \phantomsublabel{-0.34}{3.00}{fig:sgt_coeff_Qmax_prel_mpi138}
    \phantomsublabel{-1.85}{1.43}{fig:sgt_coeff_Qsum_Elab_mpi138}
    \phantomsublabel{-0.34}{1.43}{fig:sgt_coeff_Qsum_prel_mpi138}
    \caption{Observable coefficients for the total neutron-proton cross section ($\sigmatot$), under various assumptions for $Q(p)$ and the input space $\kinparvec$.
    Predictions are generated with the 500 MeV SMS potential from Ref.~\cite{Reinert:2017usi} and assume a fixed value of $\mpieff = 138\, \mathrm{MeV}$ (i.e., the approximate pion rest mass) and $\Lambda_b= 600$\,MeV.
    The top row uses a smoothed maximum $\Qpoly$ [Eq.~\eqref{eq:Q_poly})] while the bottom row uses $\Qsum$ [Eq.~\eqref{eq:Q_sum}].
    The left column uses $\Elab$ as the $x$-variable while the right column uses $\prel$. (The relationship between the two is given by Eq.~\eqref{eq:prel_to_Elab}.)
    Choosing $x = \prel$ and $Q = \Qsum$ results in coefficients that look the most stationary, i.e., similarly sized and with similar length scale across the domain and across coefficients.
    }    \label{fig:sgt_coeff_assumptions_mpi138}
\end{figure}

\begin{table}[tb]
    \centering
    \renewcommand{\arraystretch}{1.7}
    \setlength{\tabcolsep}{1pt}
    \begin{ruledtabular}
    \begin{tabularx}{0.2\textwidth}{c c c c}
        \multicolumn{4}{c}{NN Observable-Specific Quantities} \\
        Quantity & $\sigmatot$ & $\sigma(\theta)$ & Spin \\
        \hline
        $\genobsref$ & $\genobs_k$ & $\genobs_k$ & 1 \\
        ($\kinparvec_{E}$, $\kinparvec_{\theta}$) & ($\prel$) & $(\prel, \negcos)$ & $(\prel, \negcos)$ \\
        $L$ & $\ell_E$ & $\text{diag}(\ell_E, \ell_\theta)$ & $\text{diag}(\ell_E, \ell_\theta)$ \\
        \hline\hline
        \multicolumn{4}{c}{Generic Gaussian Process Quantities for NN} \\
        \hline
         \multicolumn{2}{c}{Characteristic Momentum} &
         \multicolumn{2}{c}{$p = \prel$} \\
         \multicolumn{2}{c}{Expansion Parameter} &
         \multicolumn{2}{c}{$Q = (p + m)/(\Lambda_b + m)$} \\
         \multicolumn{2}{c}{Correlation Function} &
         \multicolumn{2}{c}{\makecell{$r(\kinparvec, \kinparvec';L) = $ \\ $e^{-\frac{1}{2}(\kinparvec-\kinparvec')^\trans L^{-1}(\kinparvec-\kinparvec')}$}} \\
         \multicolumn{2}{c}{Variance Prior} &
         \multicolumn{2}{c}{$\sdth^2 \sim \chi^{-2}(\nu_0=1, \tau_0^2=1^2)$} \\
         \multicolumn{2}{c}{Length Scale Prior} &
        \multicolumn{2}{c}{$\pr(\ell_i) \propto 1$, $\ell_i > 0$}
    \end{tabularx}
    \end{ruledtabular}
    \caption{Preferred assumptions about the model parametrization choices for each NN observable, under the constraint of stationary GPs.
    (See Sec.~III of Ref.~\cite{Melendez:2019izc} for further details.)
    The correlation function is the squared exponential, which assumes that the $c_n$ are very smooth and stationary.
    The GP variance prior is a weakly informative inverse chi-squared distribution, while the length scales priors are (positive) uninformative and uniform.
    The angular observables live in a two-dimensional (2D) space, so the matrix of correlation lengths, $L$,  is a 2D diagonal matrix with $\ell_E$ and $\ell_\theta$ on the diagonal.
    }
    \label{tab:observable_assumptions}
\end{table}

Additionally, we must choose a functional form for the characteristic momentum $p$.
We consider
three options for the momentum scale that appears in the expansion parameter: $p = \prel$;
$p = \qcm$,
where 
\begin{align}
    \qcm^{2} = (\vec{p} - \vec{p}^{\, \prime})^{2} 
    \, \Rightarrow \, \qcm = \prel \sqrt{2 (1 - \cos\theta)} \;
    \label{eq:qcm}
\end{align}
when $p = p^{\prime} = \prel$; and a combination of the two, $p = \psmax(\prel, \qcm)$,
where
\begin{equation}
    \psmax(x, y) = \frac{1}{N} \log_{1.01}(1.01^{N x} + 1.01^{N y}) 
    \label{eq:psmax}
\end{equation}
with $N = 5$, which is a smooth maximum interpolation function borrowed from deep-learning applications~\cite{blanchard2021}.

Besides parametrizations of $Q$ and $p$, we have also tested parametrizations of the input space $x = (x_{E}, x_{\theta})$, where $x_{E}$ is the energy-dependent input space and $x_{\theta}$ the angle-dependent one.
Which options for $x$ are available depends upon which physical quantities the observables depend upon.
The observables considered in this paper are the total cross section $\sigmatot$; the differential cross section; and the spin observables $D$, $A_{xx}$, $A_{yy}$, $A$, and $A_{y}$, the latter of which is elsewhere referred to as $P$ or $PB$~\cite{Bystricky:1976jr,lafrance:jpa-00208966,Carlsson:2015vda,Melendez:2017phj}.
All are functions of scattering angle and lab energy except for the total cross section, which is a function of lab energy alone.
When the observable depends on both quantities, it is of course also possible to plot 
that observable (or the dimensionless coefficients derived therefrom) at some fixed lab energy or scattering angle against the remaining other quantity, which is allowed to vary.
In the fixed-angle case, the two options for $x_{E}$ are those discussed already, namely $\Elab$ and $\prel$.
In the fixed-energy case, there are four options for parametrizing $x_{\theta}$: $\theta$, $\negcos$, $\qcm$, and $\qcm^{2}$.

After exploring many parametrizations for use with the assumed stationary GPs (case studies for $Q(p)$, and $(x_{E}, x_{\theta})$ can be found in Secs.~\ref{subsec:Q_param} and~\ref{subsec:input_space}), we have collected our 
preferred choices for NN observable-specific and GP quantities in Table~\ref{tab:observable_assumptions}.
These are the same physically grounded choices we use to
generate many of the figures in Secs.~\ref{sec:gp_stationarity} and~\ref{sec:results_model}, where we show that---with additional restriction on the input space---they lead in many (but not all) cases to coefficient functions manifesting 
the model assumptions adopted here: naturalness and stationarity.
A more complete implementation of the correlated error model will require
future generalizations to nonstationary GPs that can provide the additional flexibility needed to accommodate the structure observed in NN data.

\subsection{Varying the breakdown scale and effective mass}
\label{subsec:bayesian_vars_intro}

As we have discussed, Fig.~\ref{fig:sgt_coeff_assumptions_mpi138} shows $c_n(\kinparvec)$ extracted with two choices of $Q$ (namely $\Qpoly$ and $\Qsum$), plotted against two choices of independent variables $\kinparvec$ (namely $\Elab$ and $\prel$) for the \emph{np} $\sigmatot$ calculated with the 500\,MeV cutoff potential from Ref.~\cite{Reinert:2017usi}. 
In order to comport with our model, the coefficients should exhibit \emph{naturalness}, the tendency to have 
variances
not too  
much different
than unity across orders.
Additionally, they should exhibit \emph{stationarity}, the tendency to have roughly the same length scale and variance (or magnitude, in this case) across the input space.
When these qualities of the coefficients are not in evidence, an informed change in the parametrization of $Q(p)$, $\genobsref$, and/or $x$ can sometimes turn a failure of our model into a success.

Here, the behavior of the coefficients in Figs.~\subref*{fig:sgt_coeff_Qmax_Elab_mpi138}~and~\subref*{fig:sgt_coeff_Qmax_prel_mpi138} at low energies/momenta suggests that the coefficients behave unusually there.
Specifically, compared to elsewhere in the input space, the length scale appears shorter and the variance larger.
But with the choice of $Q = \Qsum(p = \prel, \mpieff = \mpi, \Lambdab = 600\,\mathrm{MeV})$ and $\kinparvec = \prel$ (see Fig.~\subref*{fig:sgt_coeff_Qsum_prel_mpi138}), then all $c_n$ appear more regular, except possibly at very small $\prel$ and for rather high-order coefficients.
(The putative failure there is perhaps because specifics of the fitting procedure used in Ref.~\cite{Reinert:2017usi} get amplified by the smallness of $O(Q^5)$ and $O(Q^6)$ contributions, or because the operative power-counting scheme for that low-momentum regime is that of pionless EFT.)
Even in Fig.~\subref*{fig:sgt_coeff_Qsum_prel_mpi138}, however, we notice discrepancies between the exhibited and ideal order-by-order convergence in the coefficients.
The question is, then: Can we do better?

\begin{figure}[htb]
    \centering
    \includegraphics{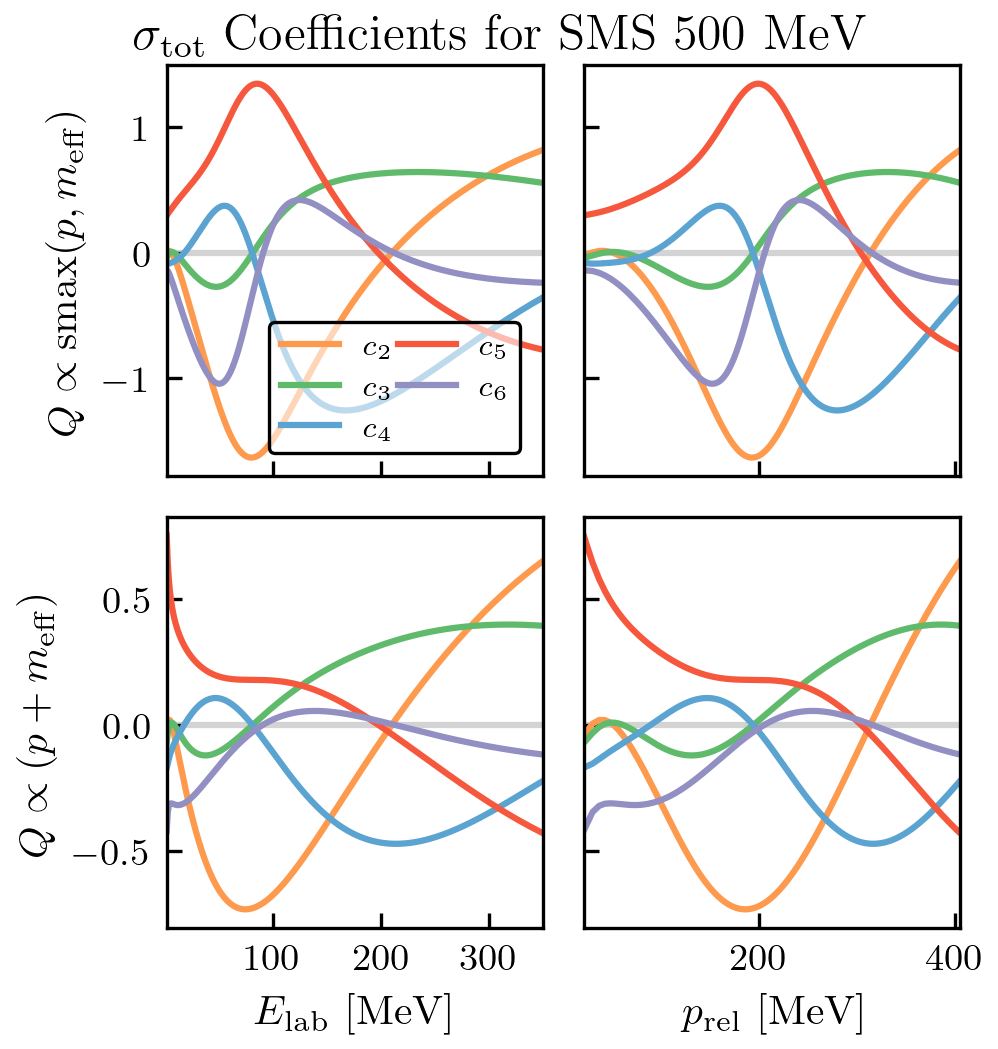}
    \phantomsublabel{-1.85}{3.00}{fig:sgt_coeff_Qmax_Elab_mpi200}
    \phantomsublabel{-0.34}{3.00}{fig:sgt_coeff_Qmax_prel_mpi200}
    \phantomsublabel{-1.85}{1.43}{fig:sgt_coeff_Qsum_Elab_mpi200}
    \phantomsublabel{-0.34}{1.43}{fig:sgt_coeff_Qsum_prel_mpi200}
    \caption{Observable coefficients for the total neutron-proton cross section ($\sigmatot$), under various assumptions for $Q(p)$ and the input space $\kinparvec$.
    The coefficients presented here are extracted the same way as those in Fig.~\ref{fig:sgt_coeff_assumptions_mpi138}, with $\Lambdab = 600\,\mathrm{MeV}$ and $\mpieff = 200\,\mathrm{MeV}$.
}\label{fig:sgt_coeff_assumptions_mpi200}
\end{figure}

In Ref.~\cite{Epelbaum:2019zqc}, an effective value of 200--225\,MeV for the soft scale $\mpieff$, which differs from the heretofore assumed value of 138 MeV,  was extracted from credible-interval (``weather'') plots (see Sec.~\ref{subsec:credible_intervals}) based upon $\sigmatot$.
When we make this change, we observe (see Fig.~\ref{fig:sgt_coeff_assumptions_mpi200}) in the coefficients much stronger evidence of the features we hope to see --- namely, naturalness and stationarity in the length scale.
Nevertheless, the region of $\prel < \mpi$ remains problematic.

The visible improvements between Fig.~\ref{fig:sgt_coeff_assumptions_mpi138} and Fig.~\ref{fig:sgt_coeff_assumptions_mpi200} suggest the possibility that the values of $\mpieff$ and $\Lambdab$ can be chosen to optimize the convergence pattern of a potential under a given parametrization (such as the choices of $Q(p)$ and $x$).
By means of a Bayesian-statistical approach to parameter estimation (outlined in Sec.~\ref{subsec:corner_plots}), we can extract a maximum \emph{a posteriori} (MAP) value for these parameters and re-plot the coefficients \emph{mutatis mutandis}.
A further discussion of this possibility, with accompanying figures, is given in Sec.~\ref{sec:gp_stationarity}.

One final note: There may be a region in the input space that is resistant to our efforts to obtain stationarity and naturalness through reparametrization because of physics reasons.
In this case, it remains an option to simply exclude that region from training and testing the GP (or use a nonstationary GP; see Sec.~\ref{sec:outlook}).
For examples, see Figs.~\ref{fig:axx_90degrees_SMS500MeV_failure} and~\ref{fig:axx_90degrees_SMS500MeV_success} (when the EFT is ill-suited to a particular domain), and Figs.~\ref{fig:a_50MeV_SMS500MeV_cos}--\ref{fig:a_50MeV_SMS500MeV_deg} and~\ref{fig:ay_200MeV_SMS500MeV_cos}--\ref{fig:ay_200MeV_SMS500MeV_deg} (where constraints distort the GP fitting process in regions where data may not be measured).
The particular regime that bears mentioning is that of momenta below the pion rest mass of approximately 138 MeV, where pionless EFT is expected to describe the power counting more accurately than \chiEFT.
Imposing the assumption of stationarity across all momenta 
may lead to inconsistencies and ill-fitting due to an underlying nonstationarity.
Specifically, we note just such dependence of the length scale on the momentum---i.e., manifest nonstationarity in the GP---and its implications in Sec.~\ref{sec:gp_stationarity}.

\section{Diagnostics and outputs}
\label{sec:diagnosticsandoutputs}

\begin{turnpage}

\begin{table*}[p]
    \centering
    \renewcommand{\arraystretch}{1.7}
   \includegraphics[]{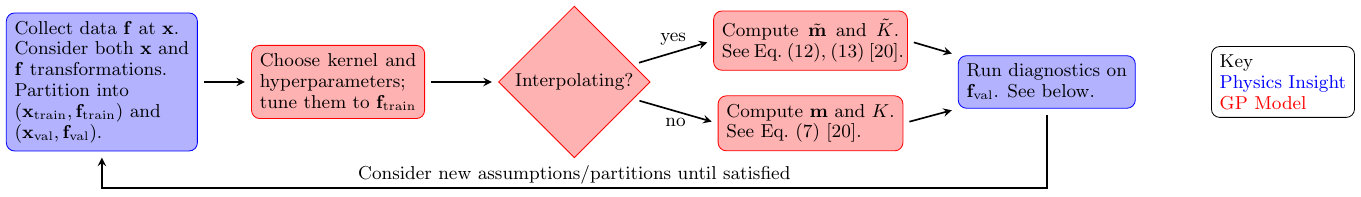}
    \vspace{0.5cm}
    \begin{ruledtabular}
    \begin{tabular}{>{\centering\arraybackslash}S{p{3.2cm}}ScS{p{4cm}}S{p{4cm}}S{p{4cm}}}
        Diagnostic & Formula & Motivation & Success & Failure \\
        \hline
        Visualize the function
        & ---
        & Does $\fvalid$ look like a draw from a GP? What kind of GP?
        & $\fvalid$ ``looks similar'' to draws from a GP
        & $\fvalid$ ``stands out'' compared to GP draws \\
        Mahalanobis Distance\newline
        $\DMD^2$
        & $(\fvalid - \mathbf{m})^\trans K^{-1} (\fvalid - \mathbf{m})$
        & Can we \emph{quantify} how much the $\fvalid$ looks like a GP?
        & $\DMD^2$ follows its theoretical distribution ($\chi_M^2$)
        & $\DMD^2$ lies too far away from the expected value of $M$ \\
        Pivoted Cholesky\newline
        $\DVAR{PC}$
        &
        $G^{-1}(\fvalid - \mathbf{m})$
        &
        Can we understand why $\DMD^2$ is failing?
        & At each index, points follow standard Gaussian
        & Many cases (see below)
        \\
        Credible Interval\newline
        $\DCI(\dcip)$ for $\dcip\in[0, 1]$
        & $\displaystyle\frac{1}{M} \sum_{i=1}^M \indicator\!\left[{\fvalid}_{,i} \in \CI_i(\dcip)\right]$
        & Do $100\dcip\%$ credible intervals capture data roughly $100\dcip\%$ of the time?
        & Plot $\DCI(\dcip)$ for $\dcip\in[0, 1]$; the curve should be within errors of  $\DCI(\dcip)=\dcip$, 
        & $\DCI(\dcip)$ is far from $100\dcip\%$, particularly for large $100\dcip\%$ (e.g., $68\%$ and $95\%$).
    \end{tabular}
    \end{ruledtabular}
    \vspace{0.1in}
    \begin{ruledtabular}
    \begin{tabular}{ScScSl}
        Variance & Length Scale & Observed Pattern in $\DVAR{PC}$ \\
        \hline
        $\sigma_{\textrm{est}} = \sigma_{\textrm{true}}$
        & $\ell_{\textrm{est}} = \ell_{\textrm{true}}$
        & Points are distributed as a standard Gaussian, with no pattern across index
        (e.g., only $\approx5\%$ of points outside $2\sigma$ lines).
        \\
        $\sigma_{\textrm{est}} = \sigma_{\textrm{true}}$
        & $\ell_{\textrm{est}} > \ell_{\textrm{true}}$
        & Points look well distributed at small index but expand to a too-large range at high index.
        \\
        $\sigma_{\textrm{est}} = \sigma_{\textrm{true}}$
        & $\ell_{\textrm{est}} < \ell_{\textrm{true}}$
        & Points look well distributed at small index but shrink to a too-small range at high index.
        \\
        $\sigma_{\textrm{est}} > \sigma_{\textrm{true}}$
        & $\ell_{\textrm{est}} = \ell_{\textrm{true}}$
        & Points are distributed in a too-small range at all indices.
        \\
        $\sigma_{\textrm{est}} < \sigma_{\textrm{true}}$
        & $\ell_{\textrm{est}} = \ell_{\textrm{true}}$
        & Points are distributed in a too-large range at all indices.
    \end{tabular}
    \end{ruledtabular}
    
    \caption{A cheatsheet for diagnostics. The interpretation of all variables and the workflow that we have found valuable is described in detail in the text.
    }
    \label{tab:workflow}
\end{table*}
\end{turnpage}

\begin{figure*}
    \centering
    \includegraphics{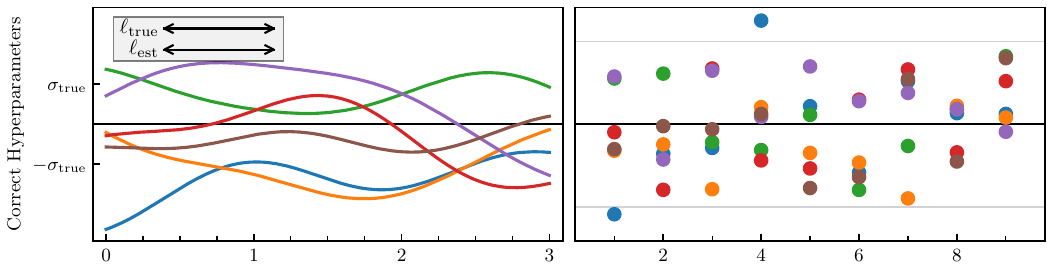}
    \includegraphics{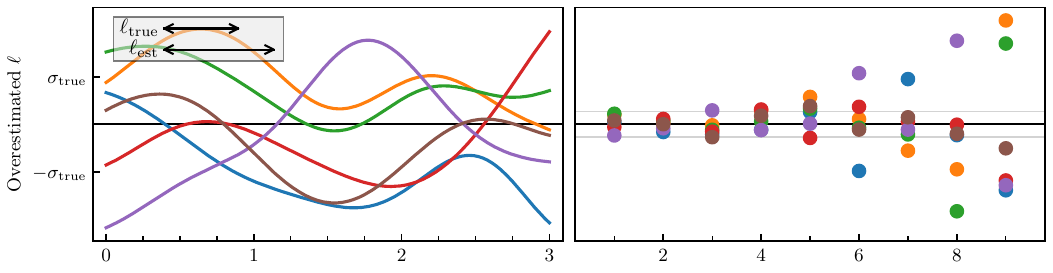}
    \includegraphics{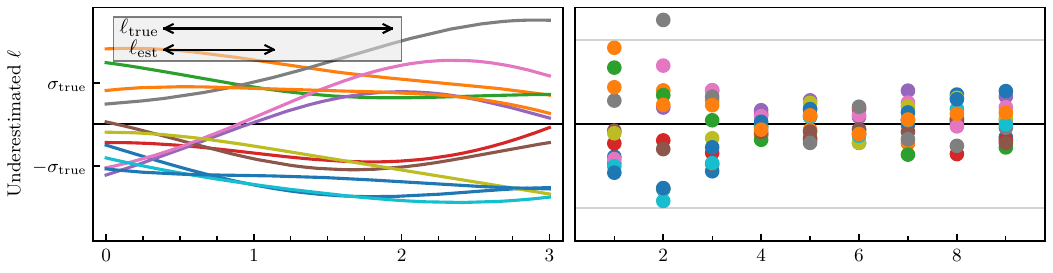}
    \includegraphics{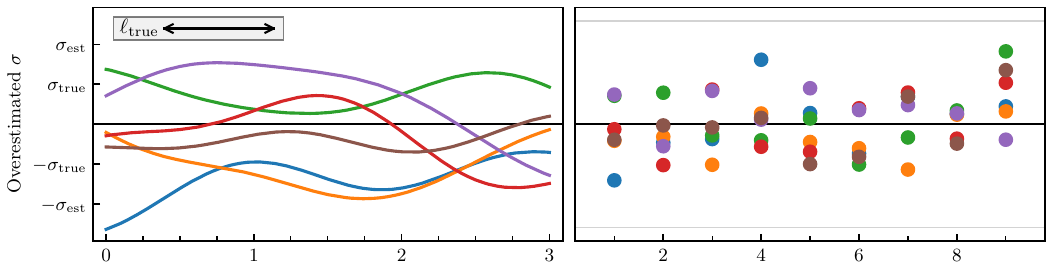}
    \includegraphics{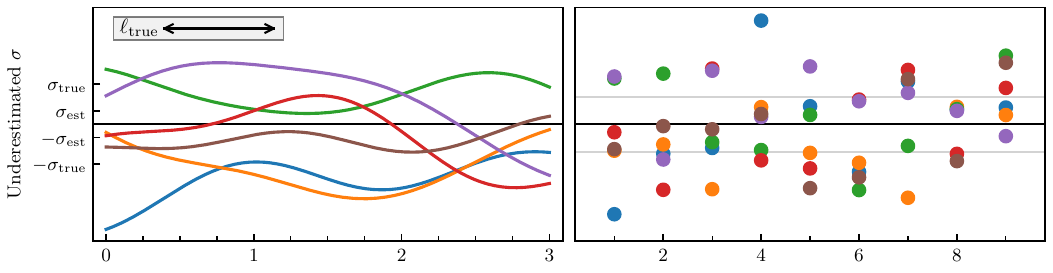}
    
    \vspace*{-13pt}
    \caption{Illustrations of Table~\ref{tab:workflow} using a toy model. See text for details.}
    \label{fig:diagnostic_cheatsheet}
\end{figure*}

An important part of a complete Bayesian analysis is model checking~\cite{GelmanBayesianDataAnalysis2013}.
Here this means assessing whether the assumptions 
of the BUQEYE model of \chiEFT\ convergence are validated or violated in practice.
The first step is a visual examination of the order-by-order EFT coefficient curves as we saw in Sec.~\ref{sec:the_model}: It often permits preliminary identification of flaws in the GP model for those coefficients.
If one (or more than one) curve fluctuates markedly farther away from zero and/or fluctuates at a different rate than the others, that may signal a problem.
Such behavior contradicts our hypothesis that all $n_c$ of the $c_n$s are drawn from a GP with a common length scale and variance.
And, if the fluctuations in an EFT coefficient change significantly in size or length scale over the kinematic domain being considered, this violates the assumption of a stationary GP; i.e., that $\cbar^2$ and $\ell$ do not depend on the NN kinematics at which $c_n$ is evaluated.
We note that GP stationarity is not a necessary assumption of the BUQEYE model (see Sec.~\ref{sec:gp_stationarity}), but it is taken for granted throughout the work presented in this paper because of our choice of GP kernel.

But detecting these patterns is non-trivial;
humans may see patterns where there are none. 
Thus we consider the general statistical diagnostics for GPs that were carefully explained and tested in Ref.~\cite{BastosDiagnosticsGaussianProcess2009}. 
In Ref.~\cite{Melendez:2019izc} we adapted Ref.~\cite{BastosDiagnosticsGaussianProcess2009} to the particular case of EFTs.
Out of the several diagnostics considered in Ref.~\cite{BastosDiagnosticsGaussianProcess2009} we emphasize three that we have 
found most illuminating in our analysis of
the EFT expansion coefficients for NN potentials.
These diagnostics are presented in this section, and their use is demonstrated in a publicly available Python package~\cite{modern_nn_potentials_package}. 
Our claims regarding what the different diagnostics show can be verified in that notebook.
A general summary of the diagnostic workflow is given in Table~\ref{tab:workflow}.

The default diagnostic for determining the ``quality of fit'' in the nuclear physics literature has usually been the reduced chi squared $\chi^2_{\text{red}}$ (a.k.a. the chi squared per degree of freedom)~\cite{Andrae:2010gh}.
It is often wielded as ``proof'' of a model's success or failure, but the theory behind its use and hence the interpretation of its value is too frequently omitted from discussion.
The structure of our discussion is designed to build upon the intuition of $\chi^2_{\text{red}}$
and extend it to the case of correlated errors.
Some important theoretical properties of the chi squared per degree of freedom and its extensions are also provided without proof (see~\cite{BastosDiagnosticsGaussianProcess2009}).

Assume, as we do throughout this paper, that the EFT LECs have already been fit to data or are otherwise fixed quantities.
Moreover, assume that we have computed order-by-order predictions of some observable $\genobs_n$ at two sets of points, the $N$ training points $\kinparvecset_{\text{train}}$ and the $M$ validation points $\kinparvecset_{\text{val}}$.
The quality of the generated statistical diagnostics depends not only on the choices made in Sec.~\ref{sec:the_model} (see Table~\ref{tab:observable_assumptions}) but also upon 
this so-called ``train-test split.''
This partitioning can be revisited every time a workflow cycle completes.
One sign that the user may wish to revisit the train-test split is that the coefficient curves by eye look well-suited to a GP,
but the diagnostics say otherwise.
In that case, the user may change the partitioning and see whether the diagnostics change (e.g., perform cross-validation).
A model that proves robust under reasonable variation of the train-test split is one whose results can be taken as more 
credible than those of a model whose results are highly variable.

Using too few or too many training and testing points
may lead to unwarranted conclusions about robustness.
First, too few training or testing points can lead to a model that is undertrained and overtested or overtrained and undertested. 
Second, and somewhat counter-intuitively, the user cannot guard against these pitfalls by using many training and testing points.
Too many of either within a correlation length will lead to matrix ill-conditioning, which renders matrix inversion problematic.
The diagnostics then become very sensitive to the ``nugget'' used to regularize the matrix inversion.
Results that are sensitive to the choice of regularization of the (nearly) degenerate covariance matrix do not represent true tests of the statistical model that the $c_{n}$s are all drawn from a stationary GP.
To strike a safe path between these two perils
we used one or two training points and three or four testing points within a correlation length.

At the training points we have tuned our GP parameters $\param$ to its convergence pattern.
At the validation points we wish to assess whether we have accurately characterized our understanding of two cases, to which we apply the workflow and diagnostics in Table~\ref{tab:workflow}.
These are:
\begin{enumerate}

    \item The known low-order convergence pattern, by contrasting the $c_n$s with their tuned distribution.
    In this case, the validation data is $\fvalid \equiv \mathbf{c}_n$ for some $n \leq k$ and it is to be compared to the multivariate Gaussian with mean $\mathbf{m} = \mathbf{0}$ and covariance $K = \sdth^2 R$.
    This analysis helps answer the question of whether the EFT converges as we expect.
    The ``visualize the function'' diagnostic asks: ``Do coefficient curves look like they are all drawn from the same GP?''
    (In)consistency can be tricky to cleanly assess at this stage; for example,
    all curves may look to have similar length scale and variance (apparently consistent), or one (or more) curves can be seen to have a distinct variance or length scale (apparently inconsistent). But beware that fluctuations happen, so what looks dissimilar to humans may be the randomness.

    \item The unknown truncation error, by contrasting $\genobs_k$ with experimental data $\genobsexp$, including all sources of uncertainty. In this case, the validation data is $\fvalid \equiv \genobsexpset$ and it is to be compared to the multivariate Gaussian with mean $\mathbf{m} = \genobsset_k$ and covariance $K = \Sigma_{th} + \Sigma_{exp}$.
    This analysis helps answer the question of whether the EFT predictions are consistent with experiment when all errors are accounted for.
\end{enumerate}
With the notation for these cases defined, we can now step through the other diagnostics in Table~\ref{tab:workflow}.

\subsection{Mahalanobis distance}
\label{subsec:mahalanobis_distance}

The chi-squared statistic is an intuitive diagnostic to consider first, and can help introduce the idea of a reference distribution.
It is defined by
\begin{align} \label{eq:chisq_statistic}
    \chi^2 = \sum_{i=1}^M \frac{(f_{\text{val},i} - m_i)^2}{\sigma_i^2} .
\end{align}
If
the $\fvalid$ were really drawn from a Gaussian with mean $\mathbf{m}$ and uncorrelated noise $\sigma_i$, then this aptly named quantity would follow a $\chi_\nu^2$ distribution with $\nu \equiv M$ degrees of freedom.
It is important to stress that Eq.~\eqref{eq:chisq_statistic} has a \emph{distribution};
that is, even if we had correctly estimated $\fvalid$, it could be above or below the ``gold standard'' value, but it should not be too far away.
The mean of the $\chi_\nu^2$ distribution is $\nu$.
To na{\"i}ve $\chi^2_{\text{red}}$ apologists it then follows, as night follows day, that $\chi^2_{\text{red}} = \chi^2/\nu$ \emph{should} be 1.
This claim should be regarded with skepticism unless error bands are provided (and with extreme skepticism if one is using the $\chi^2_{\text{red}}$ as a goodness-of-fit criterion for a non-linear model, see~\cite{Andrae:2010gh}).
That is, the reference distribution $\chi_\nu^2$, both its mean and its uncertainties, can inform us what reasonable $\chi^2$ diagnostic sizes look like.

For the truncation error model described in Sec.~\ref{sec:the_model}, it is clear that the theoretical errors become correlated.
Thus, the vanilla chi-squared statistic no longer applies.
However, one can compute the (squared) Mahalanobis distance
\begin{align}
    \DMD^2(\fvalid) = \transpose{(\fvalid - \inputvec{m})} K^{-1} (\fvalid - \inputvec{m}) ,
\end{align}
of which the chi-squared is merely a special case when the errors are uncorrelated.
In calculating this distance, the covariance matrix of the GP at the validation points, $K$, plays the role of a metric tensor.
Interestingly, if $\fvalid$ were really drawn from a multivariate Gaussian with mean $\mathbf{m}$ and covariance $K$, then $\DMD^2$ still follows a $\chi_\nu^2$ distribution with $\nu=M$ degrees of freedom.

The $\DMD^2$ combined with its reference distribution can quantitatively tell us whether, for example, our observable coefficients $c_n$ follow a GP as we hypothesize.
We evaluate $\DMD^2(\fvalid)$ separately at each order (i.e., for each $c_n$ curve) for which we have a \chiEFT\ calculation of NN scattering; that is, we use the \chiEFT\ coefficients at the kinematics defined by the validation points to compute the Mahalanobis distance of those coefficients from the mean curve—--which is taken to be zero in our model of EFT coefficients. 
This allows us to see if all $n_c$ of the $\DMD^2(\mathbf{c}_n)$ values lie within a reasonable range (68\% or 95\%) of the reference distribution (which shows that the curves are consistent with our model) or whether there are outliers (which shows the opposite).
If one or more of the EFT coefficients correspond to a GP fit that is statistically too good then $\DMD^2$ for those $c_n$s will be markedly less than the number of degrees of freedom because errors in the GP are overestimated. In contrast, a $\DMD^2$ that is large compared to the number of degrees of freedom means the errors in the GP do not encompass the validation points in a statistically correct way.

\subsection{Pivoted Cholesky decomposition}
\label{subsec:pc_decomp}

Instead of relying on a one-number summary, such as the chi squared statistic, one could instead consider the weighted residual $(f_{\text{val},i} - m_i)^2/\sigma_i^2$ at each point.
Such a residual vector contains much more information than its sum, and permits one to inquire where exactly the theory is failing.
If $\mathbf{m}$ and $\sigma$ are correct, then there should be no pattern across the indices of the residual vector, and the reference distribution for each index is itself a standard Gaussian $\normal(0, 1)$.

The correlated analog requires a standard deviation matrix $K = GG^\trans$ from which one can compute
\begin{align}
    \DVARIT{G} = G^{-1}(\fvalid - \mathbf{m}).
\end{align}
The $G$ matrix is not unique, but we choose it to be the pivoted Cholesky decomposition~\cite{BastosDiagnosticsGaussianProcess2009}, and call $\DVAR{PC}$ the pivoted Cholesky (PC) diagnostic.
Each index of this vector still corresponds to one particular validation point, but the indices have been pivoted such that the first index has the largest variance, the second has the largest variance after one has conditioned on the first validation point, and so on~\cite{BastosDiagnosticsGaussianProcess2009}.
Again, the reference distribution at each index is a standard Gaussian, but this diagnostic can fail in illuminating ways.
In essence, misestimates of the variance $\sdth^2$ appear at all indices, and misestimates of the correlation structure show up at large index.

Table~\ref{tab:workflow} lists five possible patterns for this diagnostic, which are illustrated in Fig.~\ref{fig:diagnostic_cheatsheet}.
Interpretations of the corresponding plots of the $c_n$ and PC diagnostics include (with length scale $\ell$, standard deviation $\sigma$, GP-estimated quantities ``est,'' and actual underlying quantities ``true''):
\begin{enumerate}
    \item Correct: At a given index, coefficient values at different orders are Gaussian-distributed, with the same variance exhibited at all indices (see subplot (a) in Fig.~\ref{fig:diagnostic_cheatsheet}).
    \item $\ell_{\mathrm{est}} > \ell_{\mathrm{true}}$: EFT coefficients associated with different orders are correctly distributed at small index but their variance gets noticeably larger as the index increases (see subplot (b) in Fig.~\ref{fig:diagnostic_cheatsheet}), a phenomenon known as ``trumpeting.''
    \item $\ell_{\mathrm{est}} < \ell_{\mathrm{true}}$: EFT coefficients associated with different orders are correctly distributed at small index but their variance gets noticeably smaller as the index increases (see subplot (c) in Fig.~\ref{fig:diagnostic_cheatsheet}), a phenomenon known as ``funneling.''
    \item $\sigma_{\mathrm{est}} > \sigma_{\mathrm{true}}$: EFT coefficients at different orders exhibit scatter that is smaller than the estimated variance, and this happens across all validation points (see subplot (d) in Fig.~\ref{fig:diagnostic_cheatsheet}).
    \item $\sigma_{\mathrm{est}} < \sigma_{\mathrm{true}}$: EFT coefficients at different orders exhibit scatter that is larger than the estimated variance, and this happens for all validation points (see subplot (e) in Fig.~\ref{fig:diagnostic_cheatsheet}).
\end{enumerate}
In our EFT application we examine $\DVAR{PC}$ for each order in the EFT for which we have coefficient data---as we did with $\DMD^2$---since it may be that the variance or length-scale problems we are trying to diagnose show up at some orders but not others. 

\subsection{Credible interval diagnostic}
\label{subsec:credible_intervals}

\begin{figure*}[!t]
    \centering
    \includegraphics{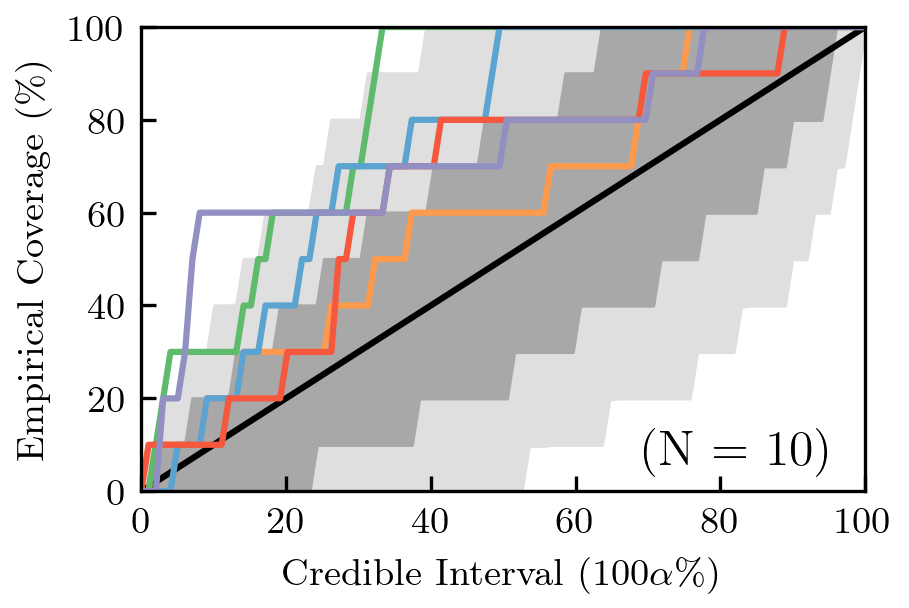}
    \includegraphics{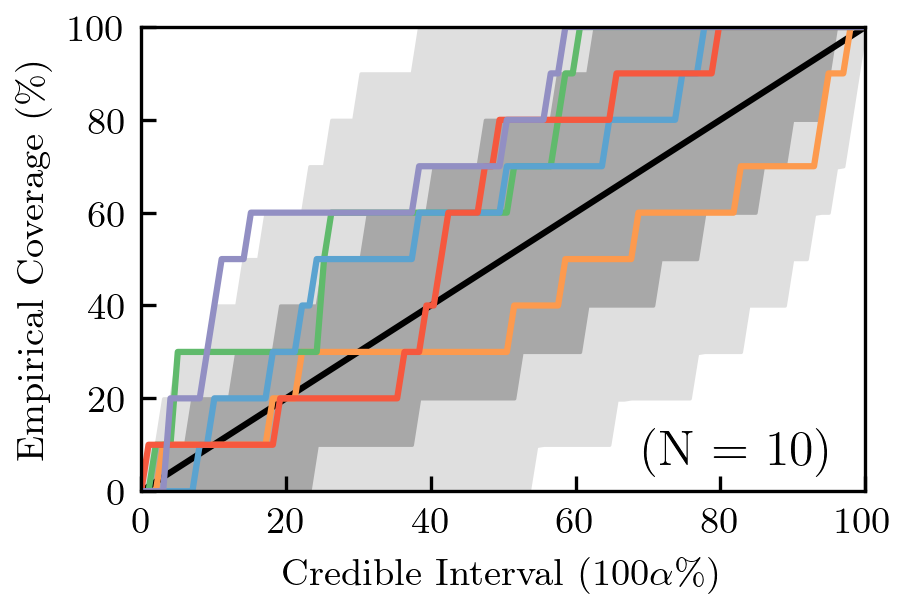}
    \phantomsublabel{-5.4}{1.6}{fig:sgt_dci_mpi138}
    \phantomsublabel{-2.3}{1.6}{fig:sgt_dci_mpi200}
     \caption{
     Plots of the credible intervals (``weather plots'') corresponding to the coefficients of the total cross section in Fig.~\ref{fig:sgt_coeff_Qsum_prel_mpi138} (a) and Fig.~\ref{fig:sgt_coeff_Qsum_prel_mpi200} (b).
     The concentration of curves above and to the left of the black midline in the lefthand figure implies that the GP variance is being overestimated so the credible intervals are too large.
     The curves on the righthand side track the midline better: The empirical coverage falls within the shaded region, so the error bands better capture the shift at the next EFT order.
    }
    \label{fig:credible_intervals}
\end{figure*}

We seek credible intervals that are accurate representations of our uncertainty.
For a given set of $c_{n}$s, the GP is trained on all the coefficients at their training points, the underlying distribution is calculated from the fitted GP. We assess the accuracy of our credible intervals by comparing the coefficients’ distributions at their validation (testing) points to this underlying distribution.
If our model for the uncertainty is accurate and we construct a $100\dcip\%$ credible interval for $\dcip\in[0, 1]$, it should approximately contain $100\dcip\%$ of the validation data.
Our final diagnostic implements this idea, and is known as the credible interval diagnostic $\DCI$, the empirical coverage probability plot, or, more colloquially, the weather plot~\footnote{The term ``weather plot'' is inspired by Ref.~\cite{silver2012signal}, which explains in more detail the origin of this nickname.}.
One can check this at a single value of $\dcip$ or all values between $[0,1]$.
We plot $\DCI$ for all $\dcip$, denoted $\DCI(\dcip)$, and create its reference distribution via sampling: This determines how far away $\DCI(\dcip)$ can be from $\dcip$ before the diagnostic signals statistical inconsistency~\cite{Melendez:2019izc}.

Examination of a plot of $\DCI(\dcip)$ vs.\ $\dcip$ will therefore quickly reveal whether the estimated GP variance is too large (entailing too-large error bands), in which case $\DCI(\dcip)$ grows faster than $\dcip$, or too small, leading to $\DCI(\dcip)$ consistently smaller than $\dcip$.
An example of the former case is shown in Fig.~\ref{fig:sgt_dci_mpi138}, in which too small a value of $\mpieff$ leads to too large a GP variance, a situation that is visibly rectified in Fig.~\ref{fig:sgt_dci_mpi200} when the value of $\mpieff$ is increased.

The $\DCI$ diagnostic is an internal check on the self-consistency of our model; it does not depend on experimental data.
We can, however, also compare EFT predictions with full uncertainty quantification to data~\cite{Melendez:2017phj, Epelbaum:2019zqc, Kejzlar:2023tlm}.
Such a comparison is made in the next section.

\subsection{Output: Statistically rigorous EFT error bands for observables}
\label{subsec:error_bands}

\begin{figure*}[!t]
    \centering
    \includegraphics{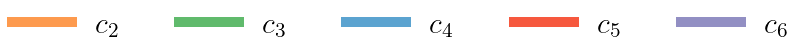}
    \includegraphics{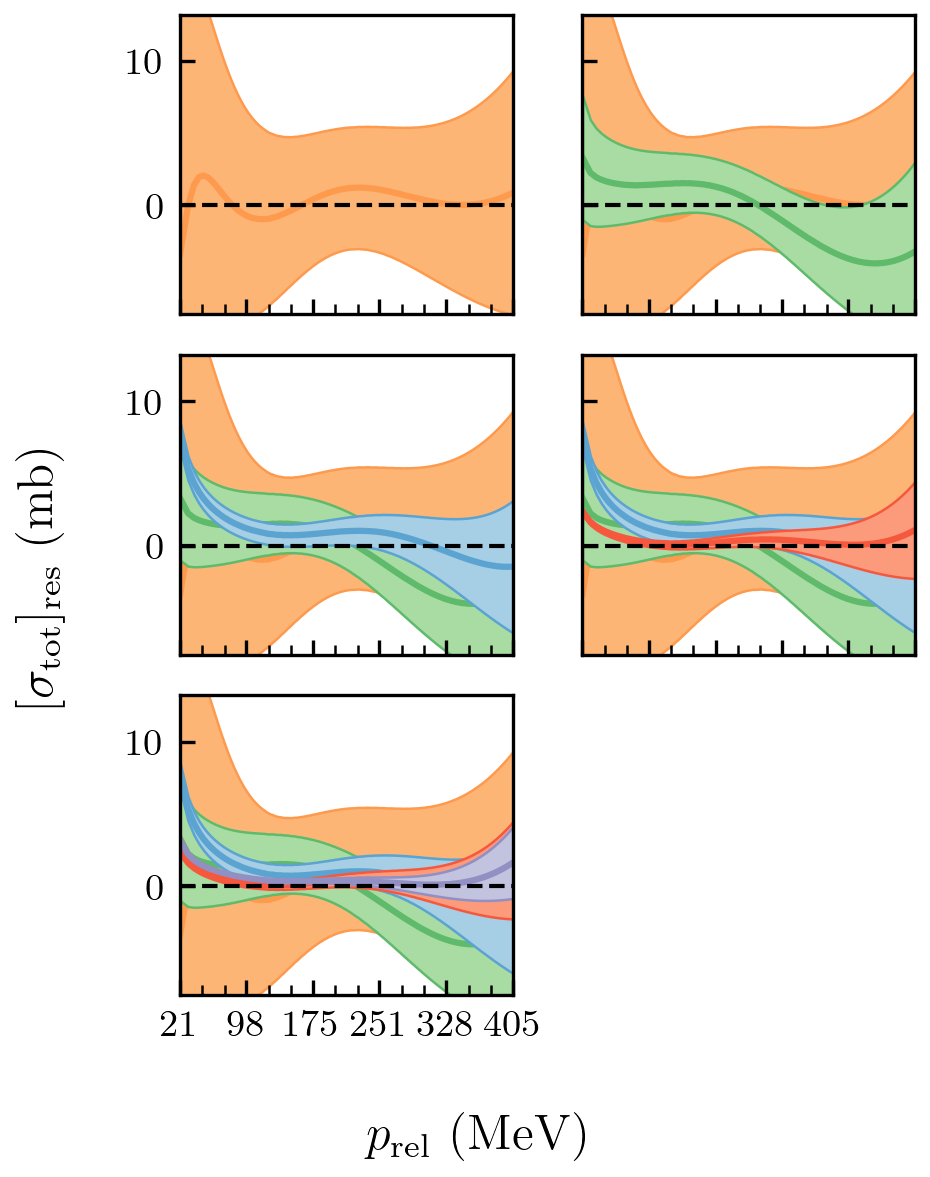}
    \includegraphics{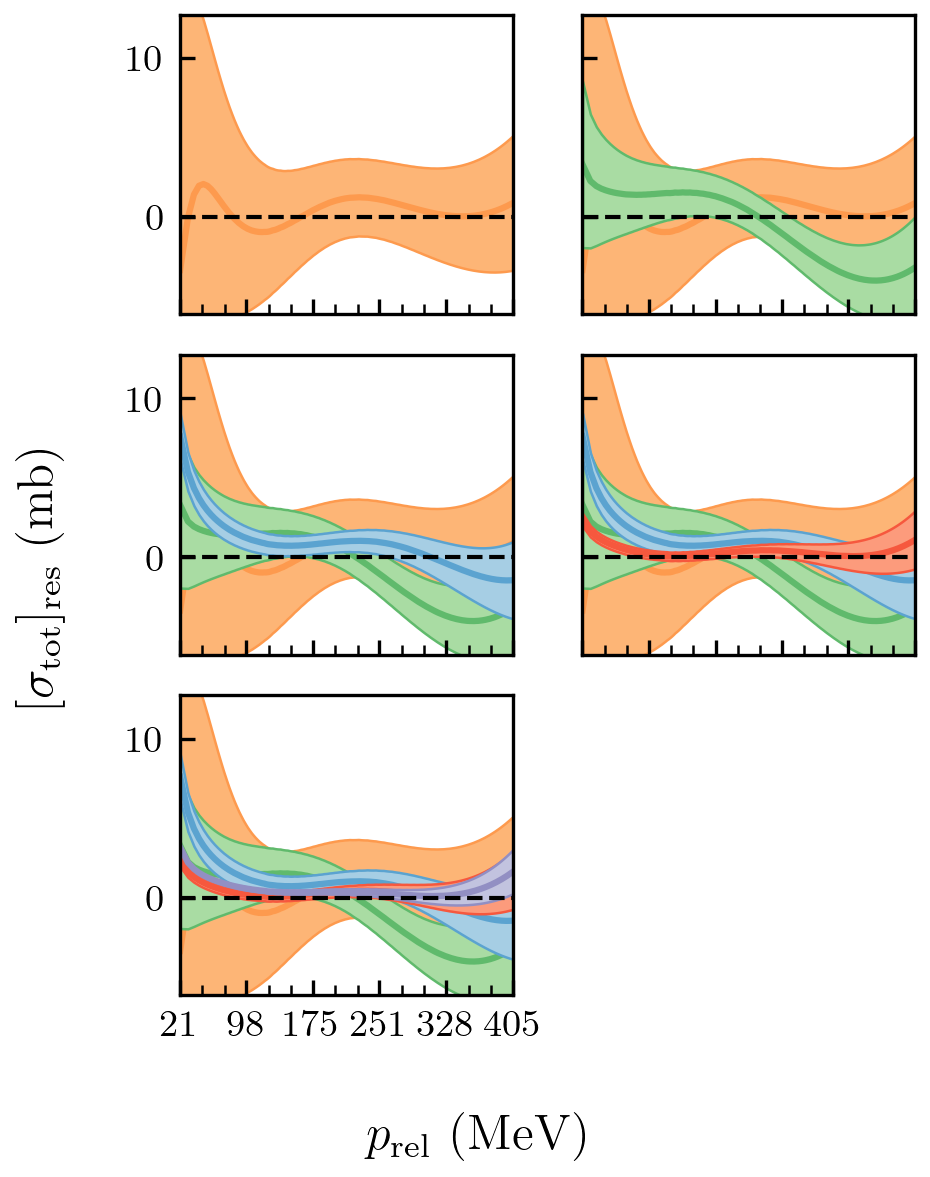}
    \phantomsublabel{-3.8}{1}{fig:sgt_exp_mpi138}
    \phantomsublabel{-0.65}{1}{fig:sgt_exp_mpi200}
     \caption{
     Plots of the residuals and truncation error corresponding to the coefficients of the total cross section in Fig.~\ref{fig:sgt_coeff_Qsum_prel_mpi138} (a) and Fig.~\ref{fig:sgt_coeff_Qsum_prel_mpi200} (b).
     Here, the differences between the experimental and predicted value of the total cross section at each order (the solid colored lines) with their associated same-color 68\% error bands, are plotted against zero (the dashed black line).
     This is useful for assessing whether the error model is converging properly (e.g., falling within the predicted 68\% error bands 68\% of the time) and where in the input space the predictions agree most and least with experiment.
     Here, one can observe improvement in agreement between the experimental and theoretical values for the observable as the value of $\mpieff$ is changed from left to right, but at higher orders one sees degraded performance at low momentum, which accords with our visual assessment of the corresponding coefficient plots.
    }
    \label{fig:truncation_error_residuals}
\end{figure*}

If the GP model passes all the other diagnostic tests, then we can use Eq.~\eqref{eq:discr_k_prior} to form statistically consistent EFT truncation uncertainty bands for the observable(s) from which the $c_n$s were extracted. 
We typically do this using the MAP values of $\Lambda_b$ and $\mpieff$. 
The resulting error bands then include not only the EFT truncation uncertainty, but also the GP uncertainties from interpolating between the training points. 
They do not include the uncertainties in the GP hyperparameters since we use point estimates for the length scales and $\cbar^2$.

For an example of truncation-error plots that builds on the 
last section's comparison of Figs.~\ref{fig:sgt_coeff_Qsum_prel_mpi138} and~\ref{fig:sgt_coeff_Qsum_prel_mpi200}, see Fig.~\ref{fig:truncation_error_residuals}.
This shows the residuals between the true and predicted values for $\sigmatot$ at each of the orders under test.
The difference between the two sets of plots arises from choosing $\mpieff = 138\,\mathrm{MeV}$ (\ref{fig:sgt_exp_mpi138}) versus $\mpieff = 200\,\mathrm{MeV}$ (\ref{fig:sgt_exp_mpi200}), which results in different error bands.
The improvement occasioned by the change in $\mpieff$'s value is visible in the superior agreement between the true and predicted values in the right-hand plot. 
That the low-momentum region is where the greatest discrepancy can be found flags that regime as an area where our statistical model may be less well-suited.

\subsection{Output: Posterior probability distributions of  \texorpdfstring{$\Lambda_b$ and $\mpieff$}{Lambdab-mpi}}
\label{subsec:corner_plots}

One circumstance in which the statistical diagnostics immediately reveal a problem is if the EFT expansion parameter has been misestimated. 
In that case the variance of the coefficients $c_n$ will either grow or shrink with EFT order $n$: grow if the expansion parameter has been chosen too small, and shrink if it has been chosen too large.
In either case the $\DCI$ plot shows a failure to correctly estimate Bayesian credible intervals, with that failure increasing in severity at higher EFT orders.
Of course, this kind of systematic trend in the $c_n$s can sometimes be seen when they are plotted together on the same scale.
This indication of trouble from the $c_n$ curves can be quantified, in this case via a formula that estimates the best value of $\Lambda_b$~\cite{Melendez:2017phj,Melendez:2019izc}.
The method employed in Refs.~\cite{Melendez:2017phj} was pointwise --- that is, it treated the estimate of the uncertainty at each testing point as independent of that at other testing points --- whereas the newer methods showcased in this paper are curvewise --- that is, they account for correlations among testing points at a given order by means of a length scale that characterizes a GP.

The best values of $\Lambda_b$ and $\mpieff$ are the ones that ``right-size'' the coefficients, i.e., ensure that the coefficients appear to be drawn from the same distribution.
The requirement that they do not show a systematic size trend thus translates into a probability distribution function (pdf) for the EFT breakdown momentum and effective soft scale (see the Appendices of Ref.~\cite{Melendez:2019izc}).

To calculate this pdf, we first set meshes for each of the $n_{r}$ hyperparameters.
In our case, $n_{r} = 4$: $\Lambdab$, $\mpieff$, the length scale in the lab-energy input space $\ell_{E}$, and the length scale in the scattering-angle input space $\ell_{\theta}$.
Then, we form a  $n_{r}$-dimensional mesh from the Cartesian product of these one-dimensional meshes. 
We have not listed the marginal variance $\cbar^2$ as one of the random variables that forms the mesh, because Gaussian processes can be marginalized over their variance analytically.
This yields a statistical object formed out of the Student $t$-distribution instead of one formed out of a Gaussian (normal) distribution---a TP rather than a GP. 
So, at each point in this $n_r$-dimensional mesh we calculate the log-likelihood that the \chiEFT\ coefficients $c_n$ for a given observable are described by a TP corresponding to the hyperparameters associated with that location in the mesh. 
We add the log-priors to the log-likelihood to find the log-posterior, and exponentiate \& normalize it to find the posterior pdf on the mesh. 
(When we learn from more than one observable at once we assume they are independent, by summing the log-likelihoods before combining the overall log-likelihood with the log-priors.)
Lastly, we marginalize over the length scales, since $\Lambdab$ and $\mpieff$ are physical parameters across observables and length scales are observable-specific. 
From the joint $\Lambdab$-$\mpieff$ posterior, we can extract MAP values for each of these two random variables, calculate their correlation coefficient, and also marginalize to find their one-dimensional posterior pdfs, including the respective means and variances of $\Lambdab$ and $\mpieff$.

A key question is then whether the $\Lambdab$ and $\mpieff$ values initially used to form the $c_n$s from the observable coefficients are consistent with this posterior.
If not, the analysis has to be repeated with values that are --- this would typically be the MAP value of the two-dimensional pdf ${\rm pr}(\Lambdab, \mpieff|\genobsset_{k},I)$.

The results of these calculations are discussed in detail in the next section, Sec.~\ref{sec:gp_stationarity}.
Note that we only use TPs for computing the posterior pdfs of $\Lambdab$ and $\mpieff$, since for those physical parameters we want to marginalize over $\cbar^2$.
The coefficients $c_n$ of the \chiEFT\ expansion are still fit with a GP whenever we generate the diagnostics that tell us whether the EFT coefficients conform to the BUQEYE statistical model or not. 

\begin{figure}
\centering
\includegraphics{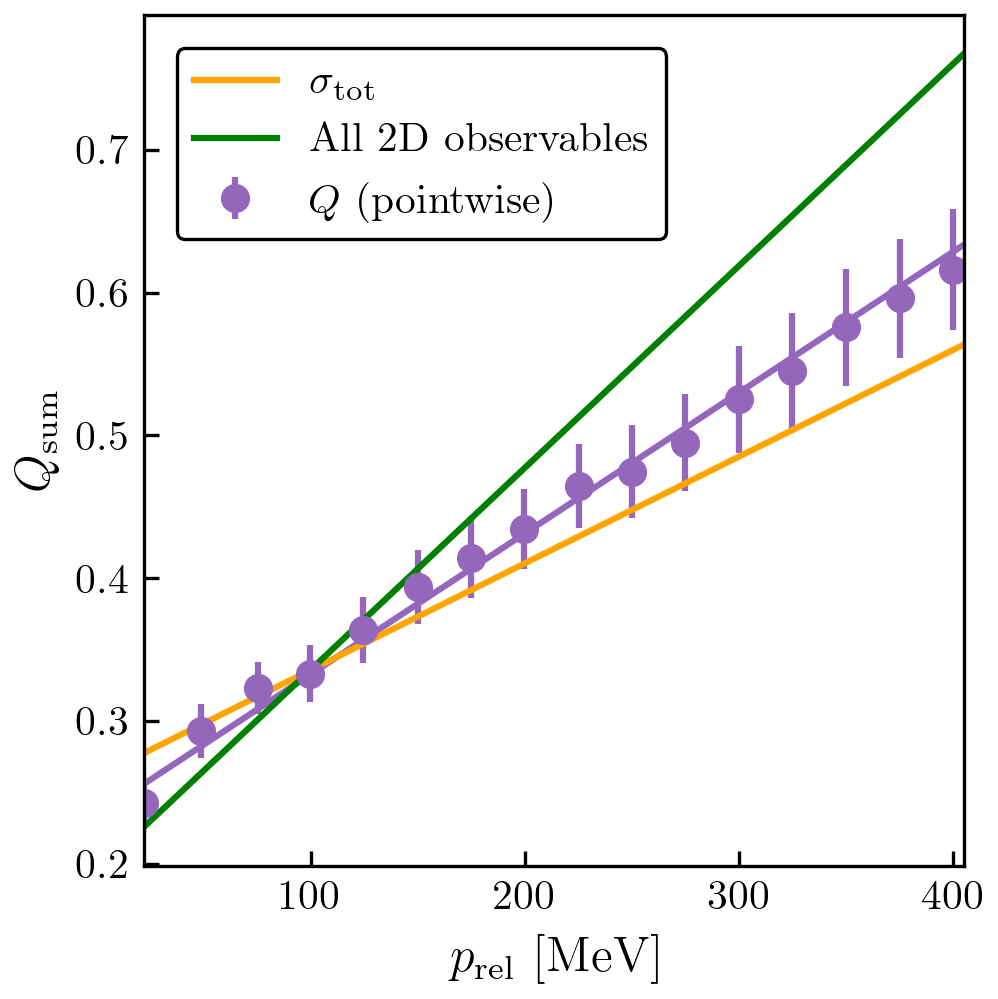}
    \caption{The MAP values of the dimensionless expansion parameter $Q$ extracted using the procedure outlined in Sec.~\ref{subsec:corner_plots} and corresponding $2\sigma$ error bars are shown in purple as a function of the relative momentum $\prel$. 
    The corresponding linear fit, which  is used to extract $\Lambdab^{\star}$ and 
    $\mpieff^{\star}$ in Eq.~\eqref{eq:opaat_fit_values}, is also shown in purple.
    Also shown are: the orange line $Q = \Qsum(p = \prel, \Lambdab, \mpieff)$, with $\Lambdab$ and $\mpieff$ MAP values extracted from posterior pdfs generated from $\sigmatot$ (see the ninth row of Table~\ref{tab:scale_values}), and the green line $Q = \Qsum(p = \prel, \Lambdab^{\prime}, \mpieff^{\prime})$, with $\Lambdab$ and $\mpieff$ values obtained as per the first row of Table~\ref{tab:scale_values} [see Eq.~\eqref{eq:2dallobs_fit_values}].
    All information shown in this figure comes from observable predictions up to and including the highest order under consideration, \NNNNLOp.
    }
    \label{fig:qvsp_linear}
\end{figure}

\begin{figure}[tb]
    \centering
    \includegraphics{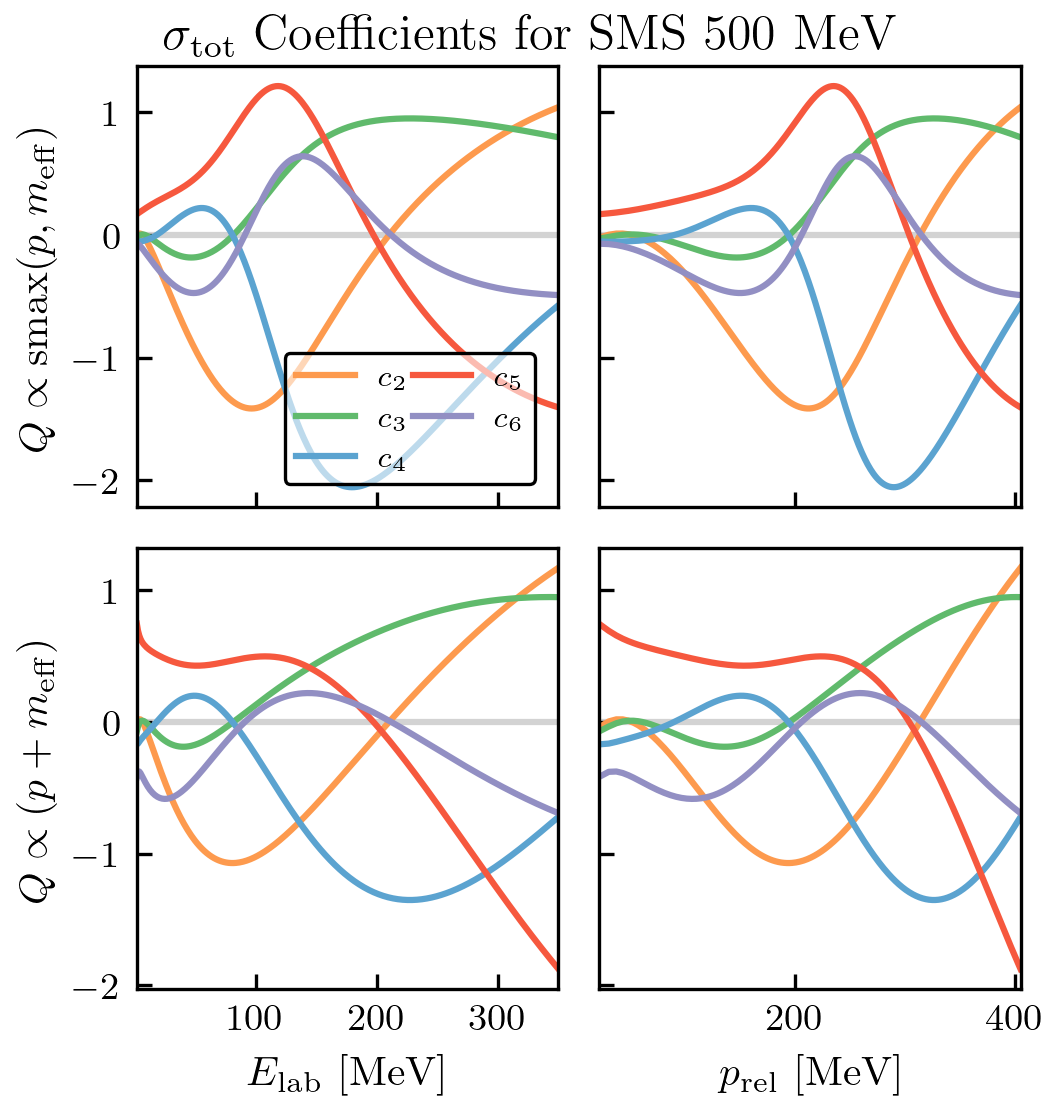}
    \phantomsublabel{-1.85}{3.00}{fig:sgt_coeff_Qmax_Elab_opt}
    \phantomsublabel{-0.34}{3.00}{fig:sgt_coeff_Qmax_prel_opt}
    \phantomsublabel{-1.85}{1.43}{fig:sgt_coeff_Qsum_Elab_opt}
    \phantomsublabel{-0.34}{1.43}{fig:sgt_coeff_Qsum_prel_opt}
    \caption{Observable coefficients for the total neutron-proton cross section ($\sigmatot$), under various assumptions for $Q(p)$ and the input space $\kinparvec$.
    The coefficients presented here are extracted the same way as those in Fig.~\ref{fig:sgt_coeff_assumptions_mpi200} except that the values of $\Lambdab$ and $\mpieff$ are the MAP values extracted from the corresponding joint posterior probability distributions (see Table~\ref{tab:scale_values}).
    Specifically, subplots (a) and (b) are plotted using $\Lambdab$ and $\mpieff$ values from the tenth row of Table~\ref{tab:scale_values} and subplots (c) and (d) using values from the ninth row.
    }\label{fig:sgt_coeff_assumptions_opt}
\end{figure}

\section{GP stationarity and \texorpdfstring{$\Lambdab$}{Lambdab} and \texorpdfstring{$\mpieff$}{mpieff}}
\label{sec:gp_stationarity}

Previous work placed strong (even delta-function) priors on the values of $\Lambdab$ and $\mpieff$, treating them as point values.
Continuing to treat them this way would be acceptable as long as we had ironclad intuition on a range in which those values were likely to fall and the results of our analysis were not sensitive to them, but our intuition is not strong enough and our outcomes not insensitive enough to justify that approach.
Indeed, we have already seen in Fig.~\ref{fig:credible_intervals} the 
improvement made in the stationarity and naturalness of the $\sigmatot$ coefficients when the value of $\mpieff$ is changed from 138 to 200~MeV from Fig.~\ref{fig:sgt_coeff_assumptions_mpi138} to Fig.~\ref{fig:sgt_coeff_assumptions_mpi200}.
Thus, more care is needed in specifying the values of those parameters for the purposes of generating graphical and statistical diagnostics.

We begin with an extraction of the expansion parameter $Q$ that uses separate and independent one-dimensional TPs to model the EFT coefficients at several different energies. 
In the terminology deployed above, this approach is pointwise in $\kinparvec_{E}$ and curvewise in $\kinparvec_{\theta}$.
We considered the EFT coefficients of the differential cross section and the five spin observables $D$, $A_{xx}$, $A_{yy}$, $A$, and $A_{y}$ at different fixed values of $\prel$ from 25 to 400 MeV in increments of 25 MeV and employed the procedure discussed in Sec.~\ref{subsec:corner_plots} ({\it mutatis mutandis} since we only have the angular length scale) to extract a posterior pdf for $Q$ at each $\prel$.
Figure~\ref{fig:qvsp_linear} shows the $2\sigma$ confidence intervals for $Q(\prel)$ that result from these 16 one-dimensional analyses.
There is a clear linear dependence on $\prel$ that strongly supports the $\Qsum$ parametrization of the EFT expansion parameter.
The straight-line fit to the $\Q$ data is shown in purple in the figure and yields:\footnote{Unless otherwise noted, these intervals and others discussed in this section are 68\% credibility intervals, which corresponds to $1\sigma$.}
\begin{align}
    \label{eq:opaat_fit_values}
    \Lambdab^{\star} = 780 \pm 20\,\mathrm{MeV} \quad \mathrm{and} \quad
    \mpieff^{\star} = 240 \pm 10\,\mathrm{MeV}.
\end{align}

\begin{table*}[tb]
    \centering
    \renewcommand{\arraystretch}{1.7}
    \begin{ruledtabular}
    \begin{tabular}{ c c c c c c c c }
        \multicolumn{8}{c}{Extracted Values of $\Lambdab$ and $\mpieff$ in MeV for Different Parametrizations} \\
        $Q$ & $p$ & $x_{E}$ & $x_{\theta}$ & $\Lambdab$ & $\mpieff$ & Observable(s) & Comments \\
        \hline
        $\Qsum$ & $\prel$ & $\prel$ & $\negcos$ & $570 \pm 10$ & $138 \pm 3$ & All 2D obs. & \makecell{
        Figs.~\ref{fig:qvsp_linear}, \ref{fig:dsg_60degrees_SMS500MeV_success}--%
        \ref{fig:d_150MeV_SMS500MeV_Qsum_cos_Qofprel}, \ref{fig:ay_60degrees_SMS500MeV_success}, \\
         \ref{fig:axx_90degrees_SMS500MeV_failure}--\ref{fig:a_50MeV_SMS500MeV_cos_backward}, 
        \ref{fig:d_100MeV_SMS500MeV_failure}--%
        \ref{fig:ayy_50MeV_SMS500MeV_Qsum_prel_cos},
        \ref{fig:dsg_150MeV_dci_success},
        \\
        \ref{fig:ay_60deg_dci_success}, 
        \ref{fig:ay_200MeV_SMS500MeV_cos}--\ref{fig:ay_200MeV_SMS500MeV_cos_middle},  \ref{fig:d_50MeV_SMS500MeV_success}--\ref{fig:axx_150MeV_SMS500MeV_success}.} \\
        $\Qpoly$ & $\prel$ & $\prel$ & $\negcos$ & $378 \pm 5$ & $106 \pm 0$ & All 2D obs. & \makecell{
        Fig.~\ref{fig:dsg_60degrees_SMS500MeV_failure}.} \\
        $\Qsum$ & $\prel$ & $\Elab$ & $\negcos$ & $610 \pm 10$ & $186 \pm 4$ & All 2D obs. & \makecell{
        Fig.~\ref{fig:ay_60degrees_SMS500MeV_failure}, \ref{fig:ay_60deg_dci_failure}.} \\
        $\Qpoly$ & $\prel$ & $\Elab$ & $\negcos$ & $459 \pm 6$ & $155 \pm 1$ & All 2D obs. &  \\
        $\Qsum$ & $\psmax(\prel, \qcm)$ & $\prel$ & $\negcos$ & $660 \pm 10$ & $172 \pm 4$ & All 2D obs. & \makecell{
        Figs.~\ref{fig:dsg_150MeV_dci_failure}, \ref{fig:dsg_150MeV_SMS500MeV_Qsum_cos_Qofpq}.} \\
        $\Qsum$ & $\prel$ & $\prel$ & $\qcm$ & $650 \pm 10$ & $184 \pm 5$ & All 2D obs. & \makecell{
        Fig.~\ref{fig:d_150MeV_SMS500MeV_Qsum_qcm_Qofprel}.} \\
        $\Qsum$ & $\prel$ & $\prel$ & $\theta$ & $590 \pm 10$ & $144 \pm 2$ & All 2D obs. & \makecell{
        Figs.~\ref{fig:a_50MeV_SMS500MeV_deg}, \ref{fig:ay_200MeV_SMS500MeV_deg}.} \\
        $\Qsum$ & $\prel$ & $\prel$ & $\negcos$ & $530 \pm 10$ & $120. \pm 3$ & All 2D obs. & \makecell{$c_{6}$ omitted. \\ Fig.~\ref{fig:ayy_50MeV_SMS500MeV_Qsum_prel_cos_noc6}.} \\
        \hline
        $\Qsum$ & $\prel$ & $\prel$ & & $990 \pm 90$ & $350 \pm 40$ & $\sigmatot$ & \makecell{        Figs.~\ref{fig:qvsp_linear}, \ref{fig:sgt_coeff_Qsum_Elab_opt}--\ref{fig:sgt_coeff_Qsum_prel_opt}.} \\
        $\Qpoly$ & $\prel$ & $\prel$ & & $670 \pm 70$ & $250 \pm 40$ & $\sigmatot$ & \makecell{
        Figs.~\ref{fig:sgt_coeff_Qmax_Elab_opt}--\ref{fig:sgt_coeff_Qmax_prel_opt}.} \\
    \end{tabular}
    \end{ruledtabular}
    \caption{
    This table includes information on the posteriors (as calculated with a TP per Sec.~\ref{subsec:corner_plots}) of the breakdown scale $\Lambdab$ and soft scale $\mpieff$ under different analysis choices.
    The first through fourth columns list the choices of parametrization for $Q$, $p$, the lab-energy input space $x_{E}$, and the scattering-angle input space $x_{\theta}$; the fifth and sixth columns list the mean values and standard deviations for the fully marginalized $\Lambdab$ and $\mpieff$ posterior pdfs; the seventh column shows the observable(s) from which the values are derived; and the eighth column lists the figures in this paper in which these MAP values are used.
    For all calculations involving the 2D observables, the $\Lambdab$ mesh ranges from 200 to 900 MeV, the $\mpieff$ mesh from 1 to 350 MeV, the $\ell_{E}$ mesh from 1 to 150 MeV, and the $\ell_{\theta}$ mesh from 0.01 to 2 times the total length of the scattering-angle input space.
    For all calculations involving $\sigmatot$, the length scale meshes are the same as for calculations involving the 2D observables but the $\Lambdab$ mesh is evenly spaced from 450 to 1150 MeV and the $\mpieff$ mesh evenly spaced from 100 to 450 MeV for the calculations done in the ninth row.
    Training points are located at \{1, 12, 33, 65, 108, 161, 225, 300\} MeV lab energy and \{41, 60, 76, 90, 104, 120, 139\}$^{\circ}$ scattering angle.
    Uniform log-priors are placed on the length scales over all positive values, a uniform log-prior is placed on $\Lambdab$ from 200 to 900 MeV, and a uniform log-prior is placed on $\mpieff$ from 1 to 350 MeV.
    }
    \label{tab:scale_values}
\end{table*}

How do these values relate to those we get from the total cross section, which was not part of the set of observables used to extract $Q(\prel)$? 
The intervals for $\Lambdab$ and $\mpieff$ obtained when a TP is used to analyze the EFT convergence of only $\sigmatot$ are given in the ninth and tenth rows of Table~\ref{tab:scale_values}, where the ninth (tenth) row uses the $\Qsum$ ($\Qmax$) 
parametrization of the expansion parameter. 
Other rows show those intervals (also extracted via TP) for different analysis choices and are discussed below.
The intervals from the $\sigmatot$ analysis are not consistent with the numbers in Eq.~\eqref{eq:opaat_fit_values}, but the resulting plot of $Q$ that the $\sigmatot$ values yield (plotted as an orange line in Fig.~\ref{fig:qvsp_linear}) is somewhat similar (within $2\sigma$ error bands) to those yielded by the numbers in Eq.~\eqref{eq:opaat_fit_values}.

Furthermore, these extracted values for the breakdown and soft scales correspond to curves that meet our criteria for stationarity and naturalness. 
In Fig.~\ref{fig:sgt_coeff_assumptions_opt} we replot the $\sigmatot$ coefficients from Figs.~\ref{fig:sgt_coeff_assumptions_mpi138} and~\ref{fig:sgt_coeff_assumptions_mpi200} using the MAP values for $\Lambdab$ and $\mpieff$ for both $Q$ parametrizations. 
Figures~\ref{fig:sgt_coeff_Qmax_Elab_opt} and~\ref{fig:sgt_coeff_Qmax_prel_opt} show the coefficients using the values in the $\Qsum$ $\sigmatot$ (ninth) row of Table~\ref{tab:scale_values} and Figs.~\ref{fig:sgt_coeff_Qsum_Elab_opt} and~\ref{fig:sgt_coeff_Qsum_prel_opt} those obtained with the values in the $\Qsmax$ $\sigmatot$ (tenth) row.
The result is coefficient functions that are stationary and natural to the unaided eye across momenta. 

However, the values extracted from $\sigmatot$ data alone are not consistent with Epelbaum's suggestion of $\Lambdab \approx 650\mbox{--}700\, \mathrm{MeV}$ and $\mpieff \approx 200\mbox{--}225\, \mathrm{MeV}$. 
These breakdown and soft scale ranges were 
obtained from 
empirical coverage plots
that optimize $\mpieff$ for the highest-order predictions of $\sigmatot$ alone~\cite{Epelbaum:2019wvf}.

What does a full two-dimensional analysis of the EFT coefficient curves reveal? 
The posterior pdfs for the breakdown scale $\Lambdab$ and soft scale $\mpieff$  obtained from data across the full two-dimensional input spaces $(x_{E}, x_{\theta}) = (\prel, \negcos)$ (the total cross section $\sigmatot$, for which we have only one-dimensional data, is omitted from the set of considered observables) are summarized in the first row of Table~\ref{tab:scale_values}. 
The intervals are:
\begin{align}
    \label{eq:2dallobs_fit_values}
    \Lambdab^{\prime} = 570 \pm 10\,\mathrm{MeV} \quad \mathrm{and} \quad    \mpieff^{\prime} = 138 \pm 3\,\mathrm{MeV}.
\end{align}
In this first row we have used $\Qsum$, as well as making several other analysis choices that the next section will show improve the statistical consistency of our GP description. 
Other analysis choices are represented in the second to eighth rows of the table. All but one have in common that, when we learn from all the 2D observables at once, $\Lambdab$'s 68\% interval lies between 450 and 670 MeV and $\mpieff$'s between 115 and 190 MeV. 

The stark differences between $(\Lambdab^{\star}, \mpieff^{\star})$ and $(\Lambdab^{\prime}, \mpieff^{\prime})$ immediately prompt the question: ``Why is there so great an inconsistency?''
The difference arises from the imposition of a stationary TP on what turns out to be a nonstationary set of coefficient curves.
In Fig.~\ref{fig:qvsp_linear}, the $Q$ posterior pdfs (from which $\Lambdab^{\star}$ and $\mpieff^{\star}$ are extracted) are calculated using data at different fixed momenta, with the resulting coefficients from which the $Q$ distributions are extracted depending on $x_{\theta}$ alone.
Thus, while the TP used to extract the $Q$ posterior distribution at each fixed $\prel$ ``knows'' 
about correlations in length scale and variance across $x_{\theta}$, it doesn't ``know'' about correlations across $x_{E}$ (i.e., between fixed $\prel$).
Only the TP used to calculate $\Lambdab$ and $\mpieff$ posteriors 
from the two-dimensional data on EFT coefficients takes into account their correlations across $x_{E}$ {\it and} $x_{\theta}$.

In principle, such a two-dimensional analysis is superior, since it leverages information across energies in a way that accounts for correlations. 
But it is only superior if those correlations are well modeled. 
Recall that we assumed a stationary GP, in which the correlations between coefficients across angles persist to the same extent irrespective of the value of the momentum.
However, 
in the lower-energy regions of $x_{E} = \prel$, coefficients in $x_{\theta}$ are best fit by GPs with longer length scales, while in higher-energy regions coefficients in $x_{\theta}$ are best fit by GPs with shorter length scales.
An explanation for $\ell_{\theta}$'s dependence on $x_{E}$ is
that, semiclassically, the highest partial-wave contribution $L_{\mathrm{max}}$ accessible by a state depends upon the energy of that state; thus, at low energy, $L_{\mathrm{max}}$ too is low.
The lower the quantum number $L$ of the partial wave, the less structure it has and the more slowly it changes, which renders it better characterized by a GP with a long length scale.
Thus, the length scale in the angle-dependent input space $x_{\theta}$ should depend on the energy at which the observable is fixed and specifically be long when that energy is low. These observations and this physics argument imply that correlations in angle are not, in fact, independent of energy and a nonstationary model for GP coefficients would be a better choice for our 2D analysis. 

\begin{figure}
\centering
\includegraphics{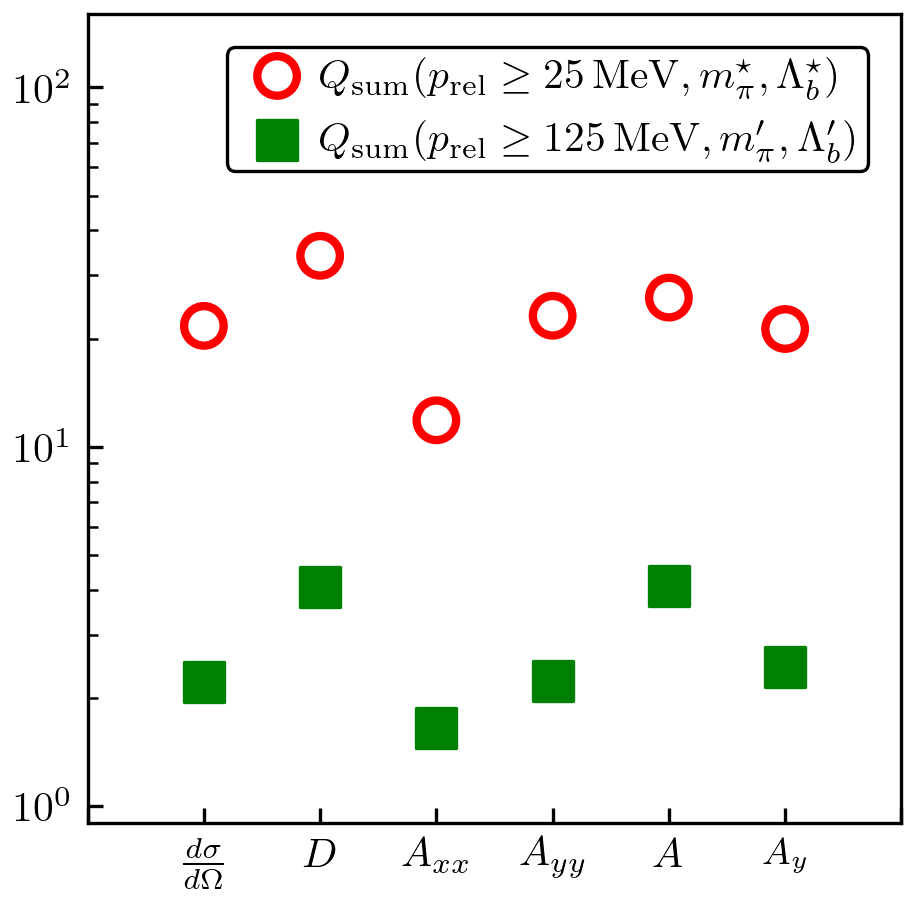}
    \caption{
    Plot of the ratio of the greatest to lowest value of $\cbar^{2}$ of the coefficients at each of the fixed $\prel$ from Fig.~\ref{fig:qvsp_linear}.
    Red markers correspond to those generated with the values in Eq.~\eqref{eq:opaat_fit_values} and green to those generated with the values from Eq.~\eqref{eq:2dallobs_fit_values}, while circle markers correspond to those generated taking into account all fixed $\prel$ and square to those generated taking into account only fixed $\prel \geq 125\,\mathrm{MeV}$.
    They are shown for the combination of all 2D observables, namely the differential cross section and the spin observables $D$, $A_{xx}$, $A_{yy}$, $A$, and $A_{y}$.
    The clear conclusion is that the results in green are more conducive to an assumption of naturalness than those in red, 
    since the former are generally closer to 1 than the latter.
    }
    \label{fig:cbar2_ratios}
\end{figure}

A gauge of rough stationarity in $\cbar^{2}$, which is the GP variance and thus the characteristic size of the coefficients, can be found by taking the ratio between the greatest and least $\cbar^{2}$ values of a fitted GP at different training points across $x = (x_{E}, x_{\theta})$.
A sign that the coefficients are right-sized across $x$---i.e., stationary in $\cbar^{2}$---is that the values of that ratio hover around 1, with values much greater signaling nonstationarity in coefficient size.
To assess nonstationarity in $\kinparvec_{E}$, we compare $\cbar^{2}$ at each of the fixed $\prel$ from the analysis that produced Fig.~\ref{fig:qvsp_linear}.
Specifically, we run this analysis for the differential cross section and the five spin observables.
For each, we take the ratio, across the GPs fitted at each fixed momentum, of the greatest variance to the least.

We plot these values in Fig.~\ref{fig:cbar2_ratios}, where the red circles correspond to values of this ratio extracted using $\Lambdab^{\star}$ and $\mpieff^{\star}$ from Eq.~\eqref{eq:opaat_fit_values}.
These exhibit a high degree of nonstationarity.
One might think that, since the nonstationarity in $\ell_{\theta}$ is most notable for relatively low $\prel$, we can reevaluate the ratio but with the results from $\prel \leq 100\,\mathrm{MeV}$ omitted from the assessment (this is also the momentum region where pionless EFT applies to NN scattering).
But even then, 
the ratio between the maximum and minimum is still too high.
However, if we perform the reevaluation using $\Lambdab^{\prime}$ and $\mpieff^{\prime}$ from Eq.~\eqref{eq:2dallobs_fit_values}, which were determined in an analysis that took into account correlations in $\cbar^{2}$ across $x$ \emph{and} we omit $\prel \leq 100\,\mathrm{MeV}$ when computing the ratio, then 
the results correspond to the green square markers in Fig.~\ref{fig:cbar2_ratios}.
These coefficients 
are reasonably stationary.

This test is important for developing a preference of values for $\Lambdab$ and $\mpieff$ and determining the physical regime in which our model is most reliable, but the true test is the one undertaken in the next section, where coefficients are extracted using different values of the breakdown and soft scales as well as different choices of parametrization and their consistency with the BUQEYE model assessed with the aid of graphical and statistical diagnostics.
In that section, we will show that the choices of parametrization from the first row of Table~\ref{tab:scale_values} [namely, $Q(p) = \Qsum(\prel)$ and $(x_{E}, x_{\theta}) = (\prel, \negcos)$] that give rise to $\Lambdab^{\prime}$ and $\mpieff^{\prime}$ of Eq.~\eqref{eq:2dallobs_fit_values} are generally superior. 
When comparing this with other analysis choices in each case we adopt the corresponding MAP values of 
$\Lambdab$ and $\mpieff$, since those values  right-size coefficients across momenta and angle. 
We also decline to look at coefficients at fixed $\prel \leq 100\,\mathrm{MeV}$ due to the strength of nonstationarity's effects there.
We will investigate this nonstationarity further in future work that will also address other potentials.

\section{Assessing the BUQEYE model for chiral EFT}
\label{sec:results_model}

\begin{figure*}[!p]
    \centering
    \includegraphics{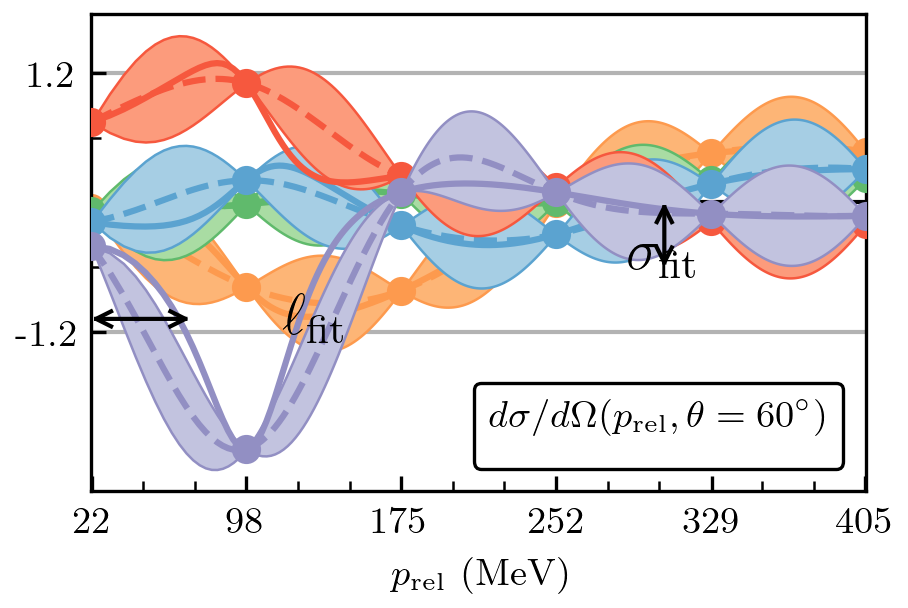}
    \includegraphics{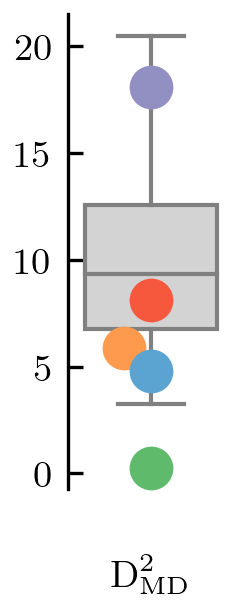}
    \includegraphics{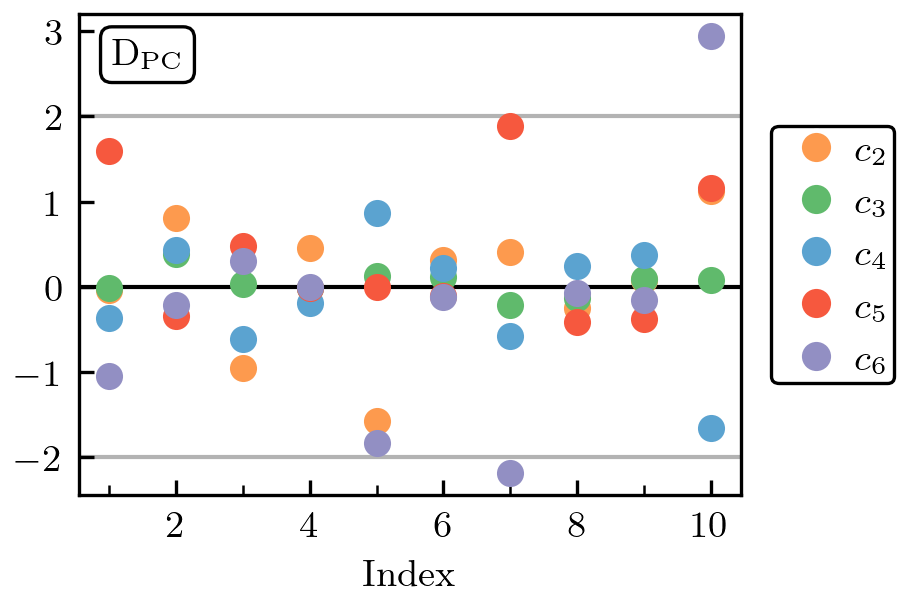}
    
     \caption{Diagnostics for the differential cross section at $\theta=60^{\circ}$.
    Here, the coefficients are plotted with $\kinparvec_{E} = \prel$ and $Q = \Qsmax(p = \prel, \mpieff = 106\, \mathrm{MeV}, \Lambdab = 378\, \mathrm{MeV})$ (optimal values of $\mpieff$ and $\Lambdab$ from Table~\ref{tab:scale_values}).
    The statistical diagnostics are calculated with 6 training points and 10 testing points.
    Note the coefficients' generally higher variance and shorter length scale in the left half of the input space by comparison to the right half (especially for $c_{6}$), and $c_{3}$'s low $\DMD^2$ value and tendency to have $\DVAR{PC}$ values close to 0 in the diagnostics.
    }
    \label{fig:dsg_60degrees_SMS500MeV_failure}

\vspace*{0.2in}

    \centering
\includegraphics{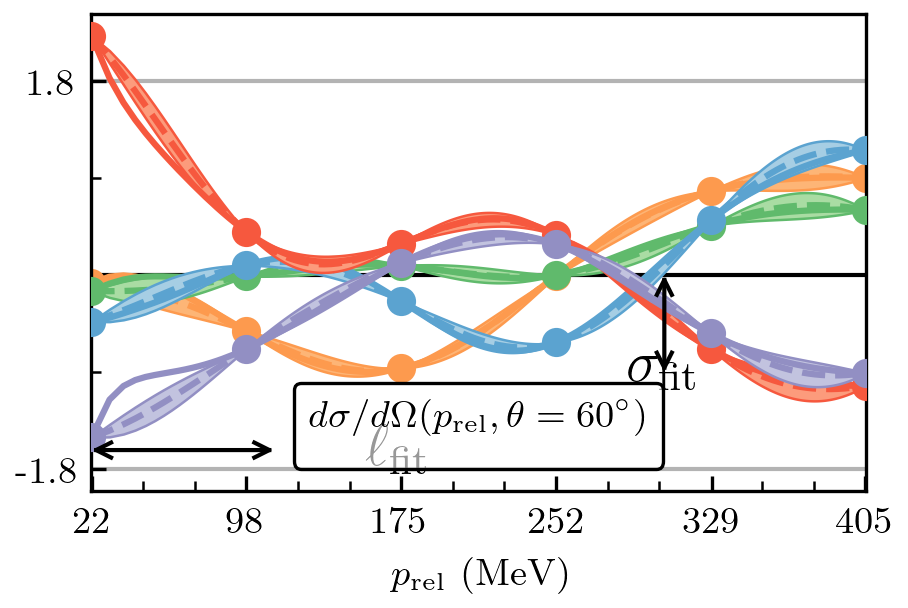}
    \includegraphics{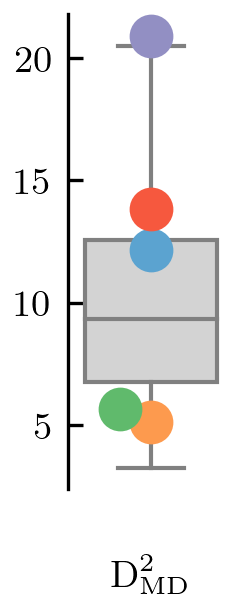}
    \includegraphics{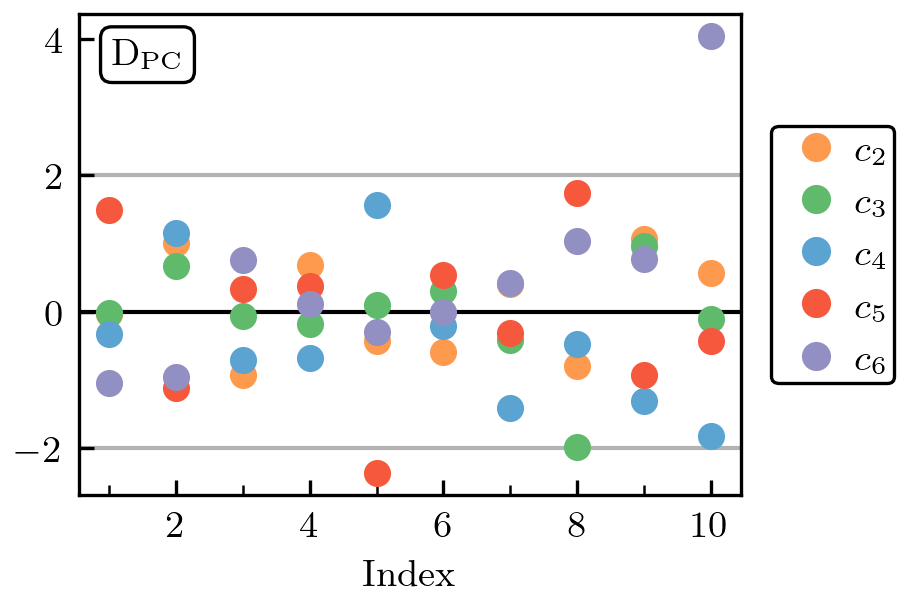}
    \caption{
    Figures here are generated with the same choices as those in Fig.~\ref{fig:dsg_60degrees_SMS500MeV_failure}, but with $Q = \Qsum(p = \prel, \mpieff = 138\, \mathrm{MeV}, \Lambdab = 570\, \mathrm{MeV})$ (optimal values of $\mpieff$ and $\Lambdab$ from Table~\ref{tab:scale_values}).
    The change in $Q$ parametrization from 
    Fig.~\ref{fig:dsg_60degrees_SMS500MeV_failure} 
    remediates nonstationarity (except at the lowest momenta) in the coefficients, as seen in the improved diagnostics.
    }
    \label{fig:dsg_60degrees_SMS500MeV_success}

\vspace{0.2in}

    \centering

\includegraphics{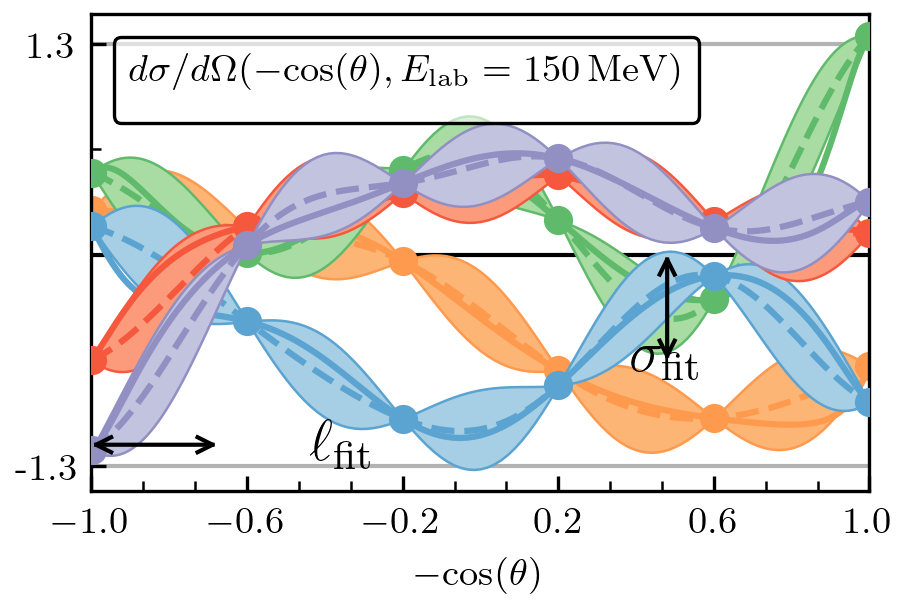}
        \includegraphics{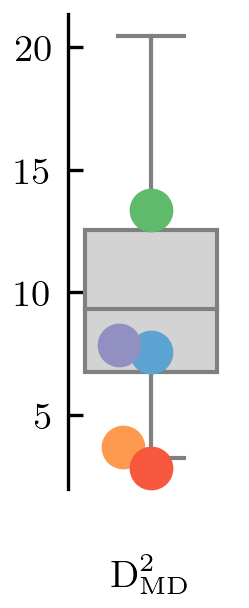}
    \includegraphics{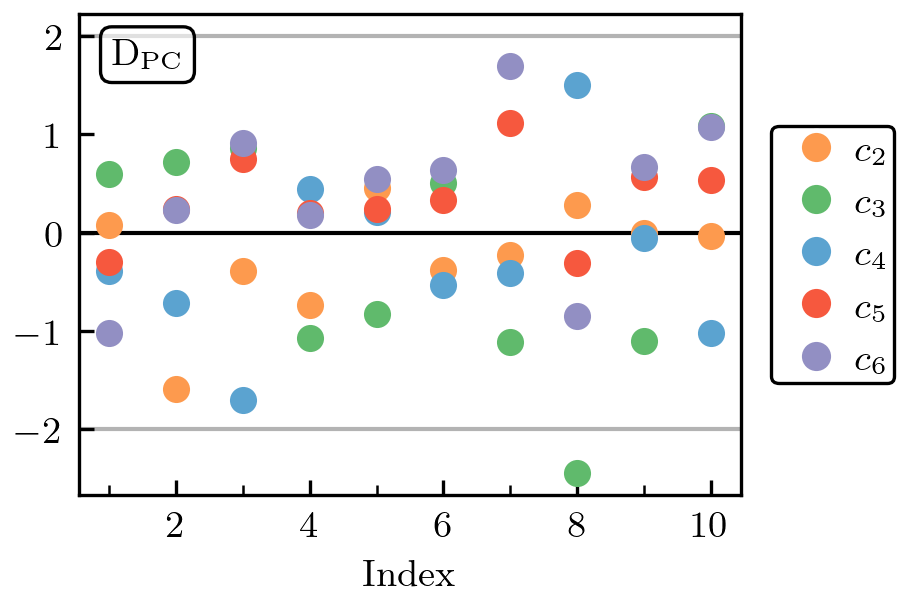}
    \caption{
    Diagnostics for the differential cross section at $\Elab = 150\,\mathrm{MeV}$.
    Here, the coefficients are plotted with $\kinparvec_{\theta} = \negcos$ and $Q = \Qsum(p = \prel, \mpieff = 138\, \mathrm{MeV}, \Lambdab = 570\, \mathrm{MeV})$ (optimal values of $\mpieff$ and $\Lambdab$ from Table~\ref{tab:scale_values}).
    The statistical diagnostics are calculated with 6 training points and 10 testing points.
    }
    \label{fig:dsg_150MeV_SMS500MeV_Qsum_cos_Qofprel}
\end{figure*}

Now we can use the diagnostics from Sec.~\ref{sec:diagnosticsandoutputs}
to make a wide-ranging assessment of the choices of parametrization in Sec.~\ref{subsec:parametrizations}, which lead to the favored choices summarized in Table~\ref{tab:observable_assumptions}.
Our goal here is to answer the question, ``Are the dimensionless coefficients extracted according to Eq.~\eqref{eq:obs_k_truncation} consistent with the hypothesis that they are random draws from the same GP?''
If they are, that implies a pattern for the coefficients that can be inductively generalized to obtain a (correlated) truncation error for the EFT series.
As we have stressed, we advocate a combination of graphical and statistical  diagnostics that quantify and standardize the assessment.

In Secs.~\ref{subsec:Q_param}-\ref{subsec:yref}, 
we present a series of case studies applying the workflow of Table~\ref{tab:workflow}.
Starting from uninformed choices for input parametrizations, we iterate 
different choices of $y_{\rm ref}$, $Q(p)$, and $x$,
looking for general consistency with our model from the
coefficients and diagnostics (especially the $\DMD^2$ and $\DVAR{PC}$ plots) across the kinematic range and observables of interest.
To show the BUQEYE model's broader applicability, additional examples of consistency are given 
in Appendix~\ref{app:examples}.
We stress that although only one potential with one regulator scheme and scale is tested here, any \chiEFT\ potential is amenable to such an analysis,
which is facilitated by an accompanying Jupyter notebook.

\begin{figure*}[!p]
\centering

\includegraphics{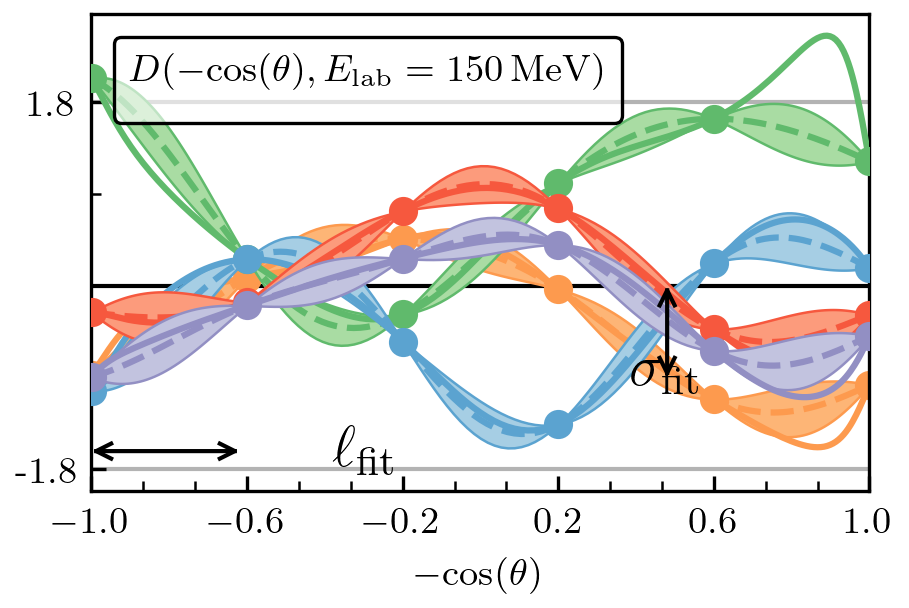}
        \includegraphics{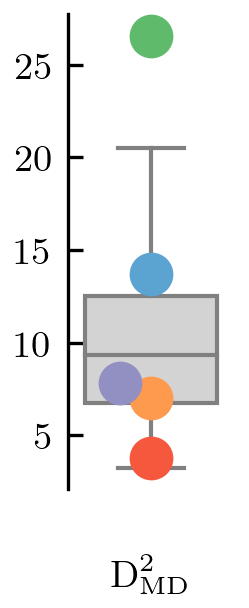}
    \includegraphics{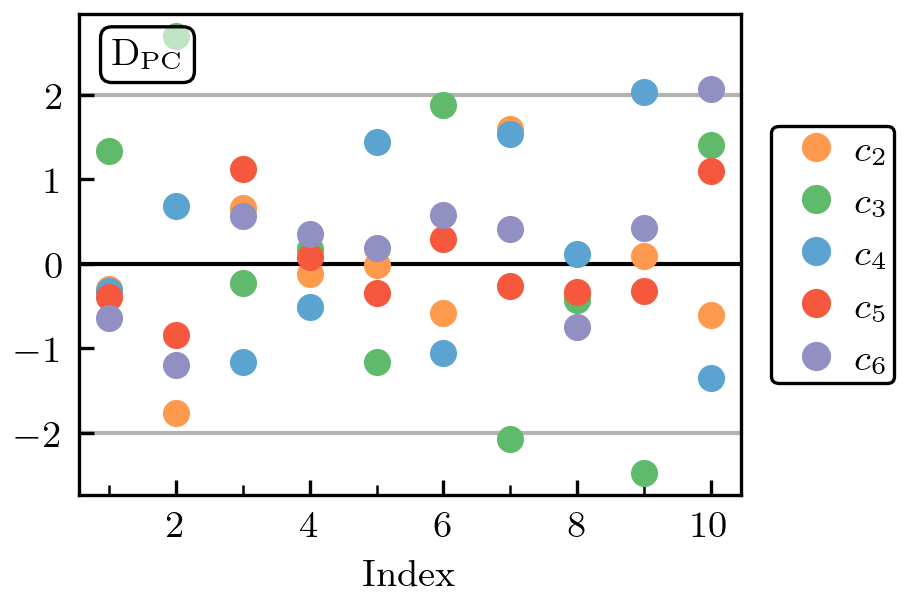}
    \caption{Diagnostics for the spin observable $D$ at $\Elab = 150\,\mathrm{MeV}$.
    Here, the coefficients are plotted with $\kinparvec_{\theta} = \negcos$ and $Q = \Qsum(p = \prel, \mpieff = 138\, \mathrm{MeV}, \Lambdab = 570\, \mathrm{MeV})$ (optimal values of $\mpieff$ and $\Lambdab$ from Table~\ref{tab:scale_values}).
    The statistical diagnostics are calculated with 6 training points and 10 testing points.
    }
    \label{fig:d_150MeV_SMS500MeV_Qsum_cos_Qofprel}

\vspace{0.2in}

    \includegraphics{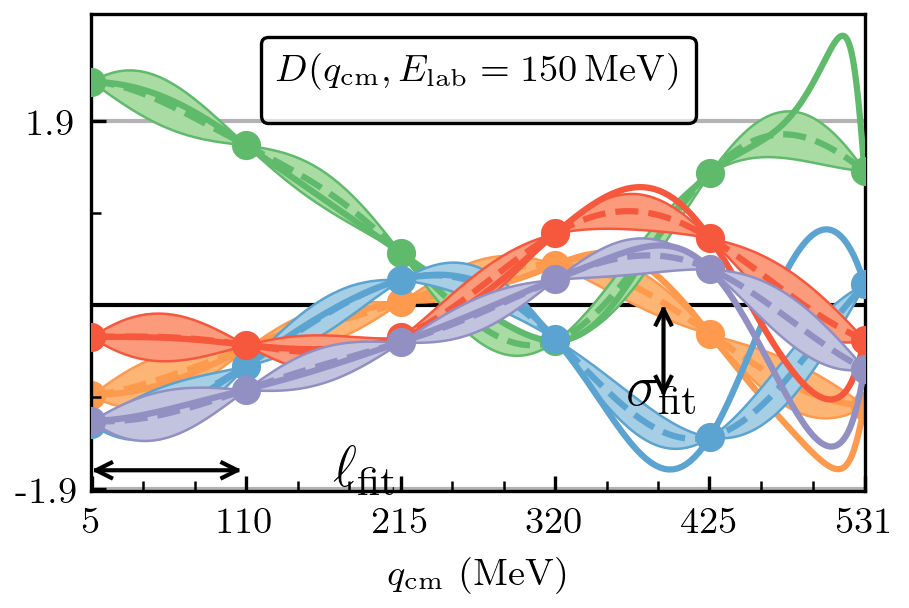}
        \includegraphics{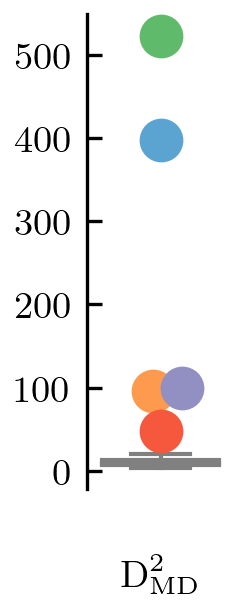}
    \includegraphics{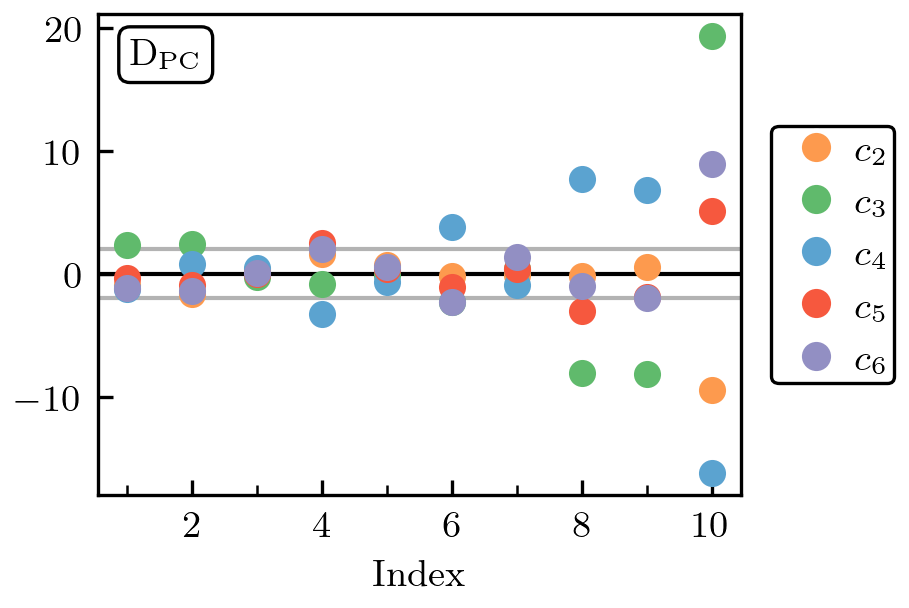}
    \caption{
    Figures here are generated with the same choices as those in Fig.~\ref{fig:d_150MeV_SMS500MeV_Qsum_cos_Qofprel}, but with $\kinparvec_{\theta} = \qcm$ and $Q = \Qsum(p = \prel, \mpieff = 184\, \mathrm{MeV}, \Lambdab = 650\, \mathrm{MeV})$ (optimal values of $\mpieff$ and $\Lambdab$ from Table~\ref{tab:scale_values}).
    Note the nonstationarity in the coefficient plots and the trumpeting in the $\DVAR{PC}$ plot (especially when compared to Fig.~\ref{fig:d_150MeV_SMS500MeV_Qsum_cos_Qofprel}), which are signs that something is amiss with the length scale.
    }
    \label{fig:d_150MeV_SMS500MeV_Qsum_qcm_Qofprel}

\vspace{0.2in}
    \centering\includegraphics{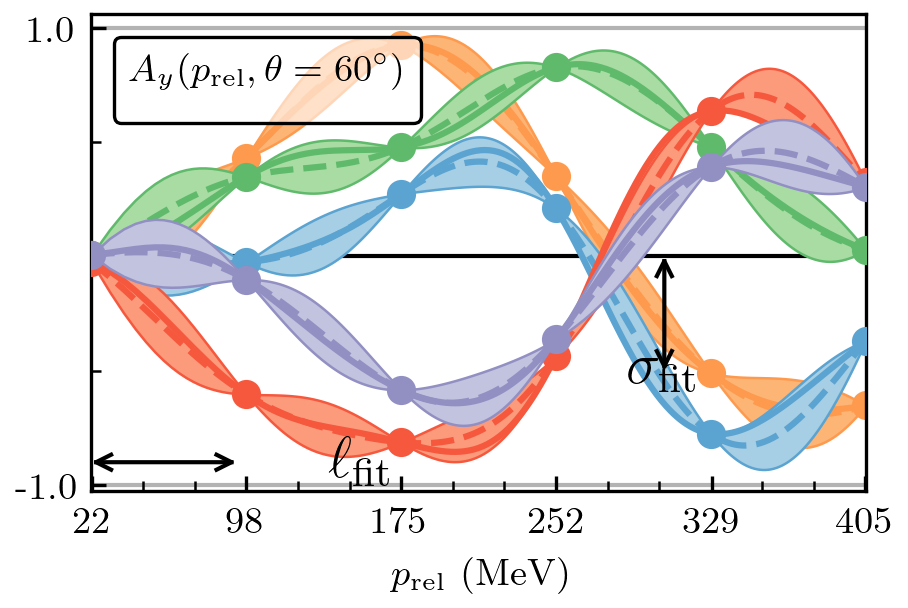}
    \includegraphics{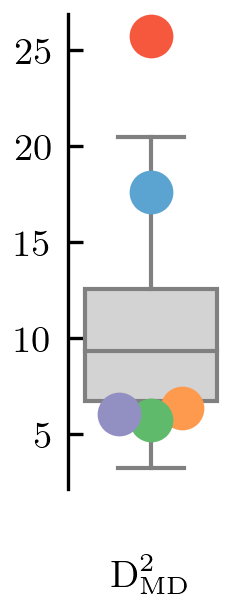}
    \includegraphics{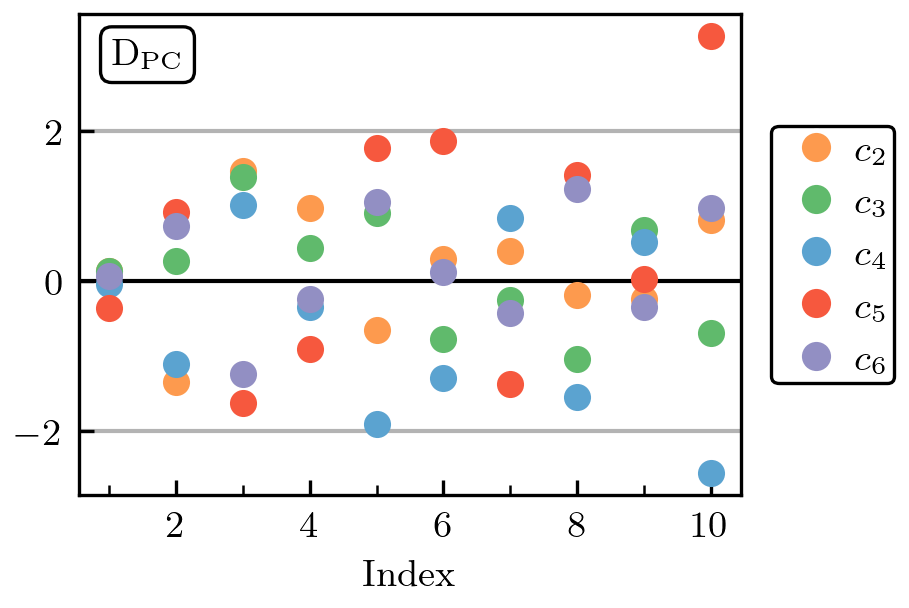}
    \caption{Diagnostics for the spin observable $A_{y}$ at $\theta = 60^{\circ}$.
    Here, the coefficients are plotted with $\kinparvec_{E} = \prel$ and $Q = \Qsum(p = \prel, \mpieff = 138\, \mathrm{MeV}, \Lambdab = 570\, \mathrm{MeV})$ (optimal values of $\mpieff$ and $\Lambdab$ from Table~\ref{tab:scale_values}).
    The statistical diagnostics are calculated with 6 training points and 10 testing points.
    }
    \label{fig:ay_60degrees_SMS500MeV_success}
\end{figure*}

\subsection{Parametrizing the expansion parameter}
\label{subsec:Q_param}

Our first case study examines how the choice of dimensionless expansion parameter $Q$ impacts the coefficients; we examine the differential cross section at a fixed  $\theta=60^{\circ}$ in Figs.~\ref{fig:dsg_60degrees_SMS500MeV_failure} and \ref{fig:dsg_60degrees_SMS500MeV_success} as a representative example. 
Here and throughout this section we show the coefficients, extracted as described in the captions, plotted as solid lines in the left panel. 
A GP is fit at training points located at the major $x$-axis ticks (denoted as circles on each curve), with testing points located at the minor ticks. 
The GP mean for each curve is shown as a dashed line and the $2 \sigma$ (95\% confidence interval) GP bands are shown as colored regions. 
The width of the ``bubbles'' will vary depending on how the GP length scale compares to the test point spacing.
In the middle and right panels of each figure are the $\DMD^2$ and $\DVAR{PC}$ diagnostics for this GP fit.

At first sight, coefficients in both Fig.~\ref{fig:dsg_60degrees_SMS500MeV_failure} and Fig.~\ref{fig:dsg_60degrees_SMS500MeV_success} appear generally natural and the diagnostics show none of the extreme pathologies in Fig.~\ref{fig:diagnostic_cheatsheet}, which implies that the BUQEYE model is applicable under different choices of $Q$ parametrization.
However, inspection of the coefficients in Fig.~\ref{fig:dsg_60degrees_SMS500MeV_failure}, which uses $Q = \Qsmax$ [see Eq.~\eqref{eq:Q_poly}], indicates nonstationarity, namely that the variance is larger and the length scale shorter at low momentum than at high momentum.
This is confirmed by the $\DMD^2$ and $\DVAR{PC}$ plots, which show evidence of nonstationarity: The values for $c_{3}$ are very low ($\DMD^2$ plot) and cluster very close to zero ($\DVAR{PC}$ plot).
Figure~\ref{fig:dsg_60degrees_SMS500MeV_success} shows improvement upon switching from parametrizing $Q$ with $\Qpoly$ [Eq.~\eqref{eq:Q_poly}] to $\Qsum$ [Eq.~\eqref{eq:Q_sum}]. 
This is correlated with the increase in the fit variance and length scale.

As detailed in Sec.~\ref{subsec:parametrizations},
we find that, once optimal values for $\Lambdab$ and $\mpieff$ are determined, $\Qpoly$ and $\Qsum$ are generally on a par in generating 
sets of coefficients consistent with the BUQEYE model across many different choices of observable 
and other parametrizations of $p$ and $x$.
However, given the 
compelling
evidence from Sec.~\ref{sec:gp_stationarity} for an underlying $\Qsum$-like structure in the dependence of $Q$ on $\prel$, we will use $\Qsum$ in subsequent figures.

Next we examined three parametrizations of the characteristic momentum $p$ in $Q(p, \mpieff)$: $p = \prel$, $p = \qcm$, and $p = \psmax(\prel, \qcm)$.
Overall, $p=\prel$ performs the best; an exemplary case is shown in Fig.~\ref{fig:dsg_150MeV_SMS500MeV_Qsum_cos_Qofprel}, where the apparent stationarity of the coefficients is backed up by the statistical diagnostics.
In contrast, $p = \qcm = \prel \sqrt{1 - \cos(\theta)}$ is not a good choice: It has a value of roughly zero at forward angles and roughly $\prel$ at backward angles, which heavily exaggerates the size of coefficients at forward angles compared to at backward angles, with predictably deleterious effects on the statistical diagnostics.
Finally, $p = \psmax(\prel, \qcm)$ does not offer improved behavior and can even degrade the performance (see Fig.~\ref{fig:dsg_150MeV_SMS500MeV_Qsum_cos_Qofpq} in Appendix A).

\subsection{Parametrizing the input spaces}
\label{subsec:input_space}

\begin{figure*}[!p]
\centering

\includegraphics{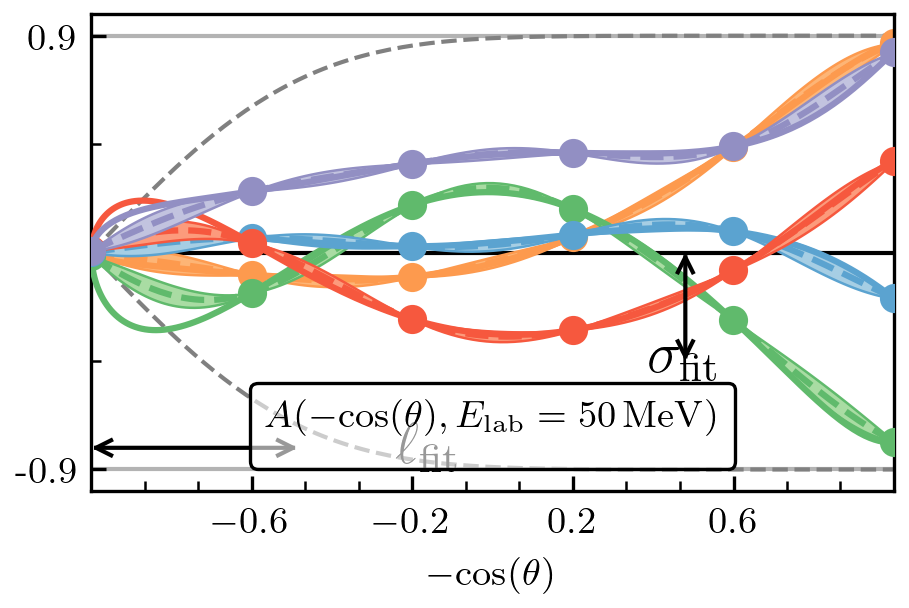}
    \includegraphics{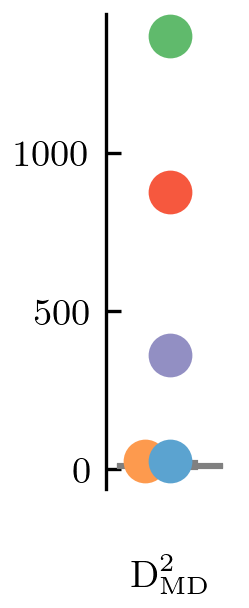}
    \includegraphics{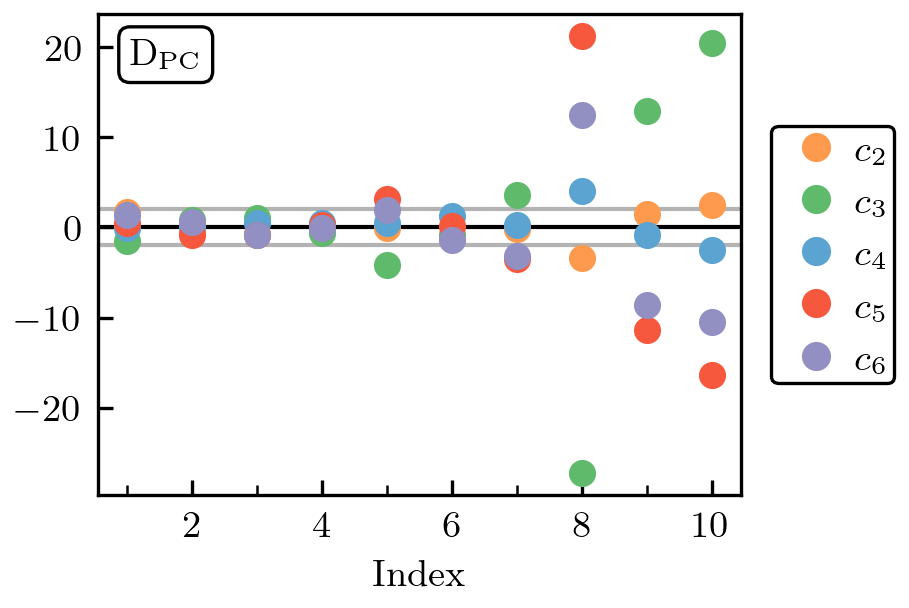}
    \caption{Diagnostics for the spin observable $A$ at $\Elab = 50\,\mathrm{MeV}$.
    Here, the coefficients are plotted with $\kinparvec_{\theta} = \negcos$ and $Q = \Qsum(p = \prel, \mpieff = 138\, \mathrm{MeV}, \Lambdab = 570\, \mathrm{MeV})$ (optimal values of $\mpieff$ and $\Lambdab$ from Table~\ref{tab:scale_values}).
    The statistical diagnostics are calculated with 6 training points and 10 testing points.
    Constraints lead to bunching of coefficients at forward angle that resolves as angle increases, leading to a 
     nonstationary length scale and diagnostics that announce nonstationarity.
    }
    \label{fig:a_50MeV_SMS500MeV_cos}

\vspace*{.2in}

\includegraphics{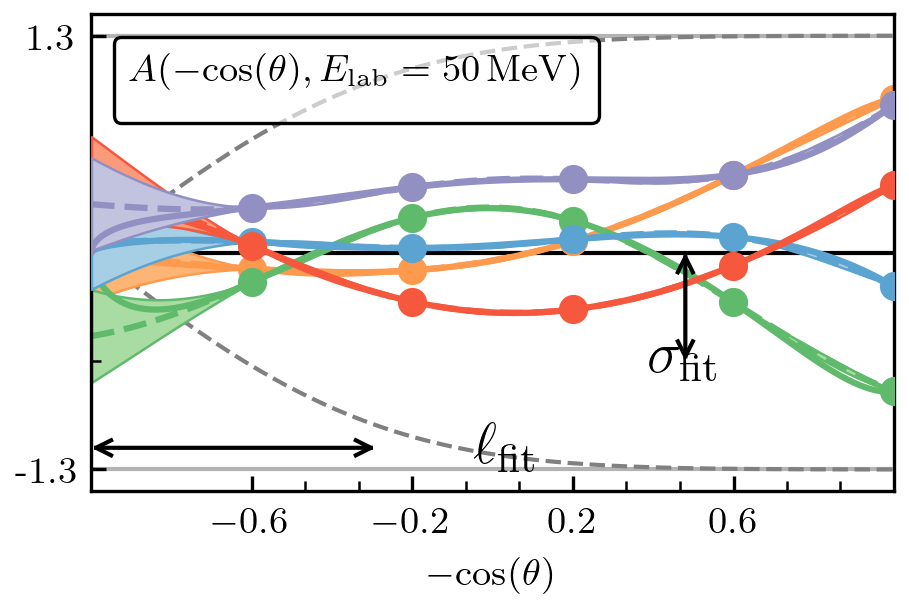}
    \includegraphics{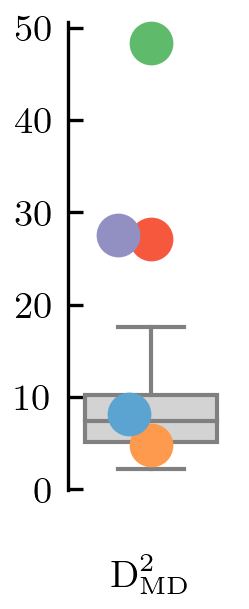}
    \includegraphics{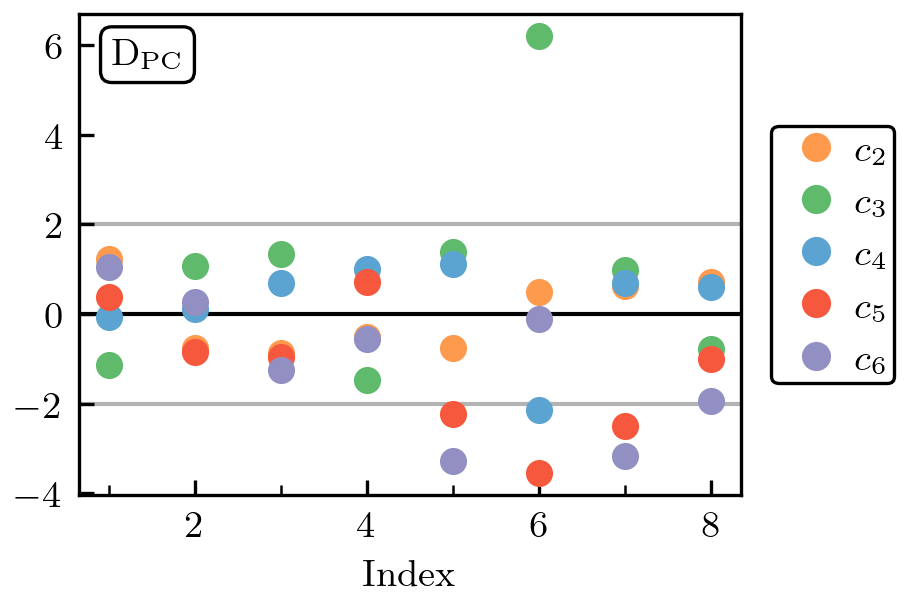}
    \caption{
    Figures here are generated with the same choices as those in Fig.~\ref{fig:a_50MeV_SMS500MeV_cos}, but with 5 training points and 8 testing points.
    With the lack of training and testing points at forward angles, the situation of stationarity, as shown by the diagnostics, improves compared to Fig.~\ref{fig:a_50MeV_SMS500MeV_cos} but is not ideal.
    }
    \label{fig:a_50MeV_SMS500MeV_cos_backward}

\vspace*{.2in}

\includegraphics{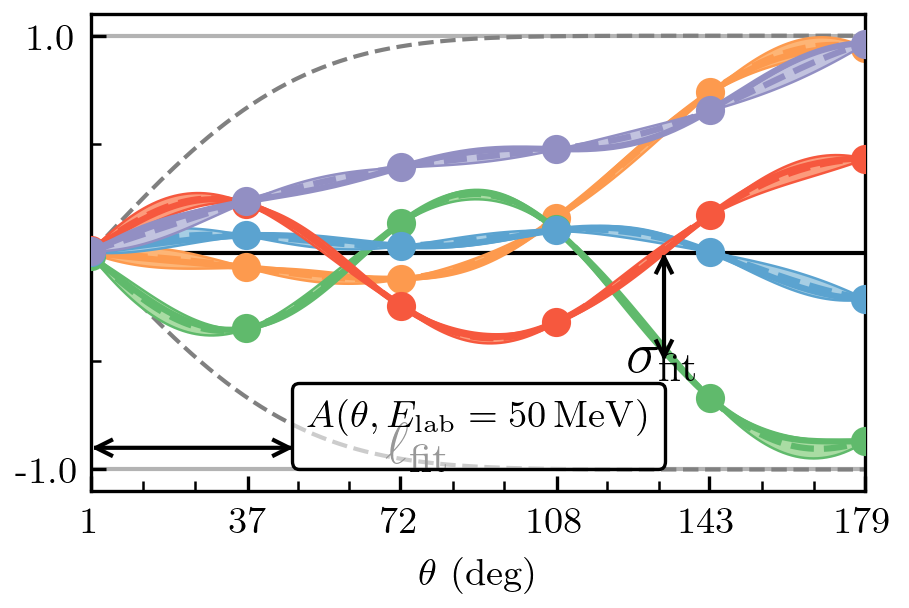}
    \includegraphics{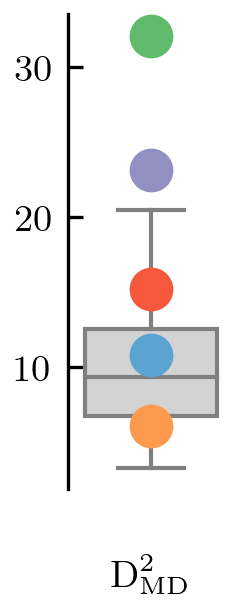}
    \includegraphics{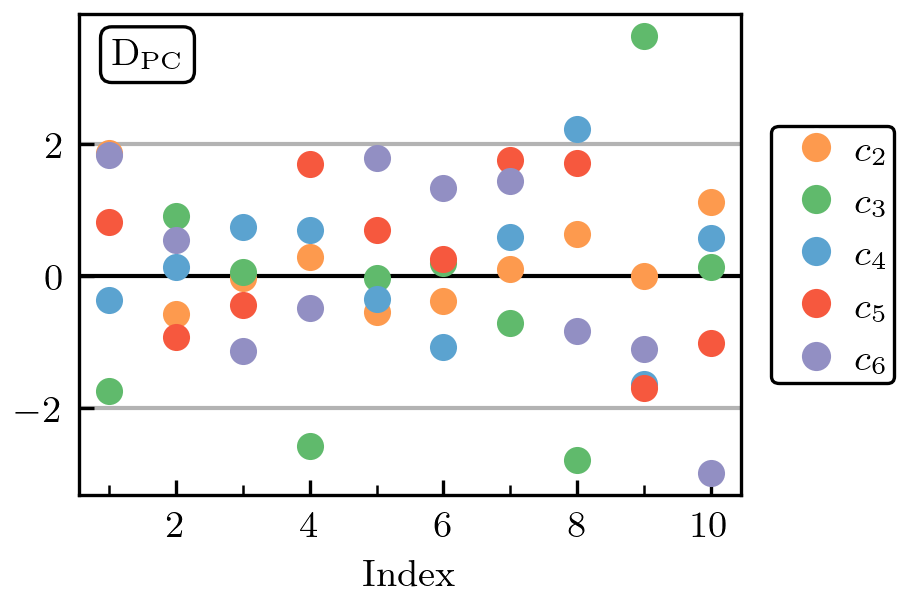}
    \caption{
    Figures here are generated with the same choices as those in Fig.~\ref{fig:a_50MeV_SMS500MeV_cos}, but with $\kinparvec_{\theta} = \theta$ and $Q = \Qsum(p = \prel, \mpieff = 144\, \mathrm{MeV}, \Lambdab = 590\, \mathrm{MeV})$ (optimal values of $\mpieff$ and $\Lambdab$ from Table~\ref{tab:scale_values}).
    A simple change of input space from Fig.~\ref{fig:a_50MeV_SMS500MeV_cos} yields much improvement in the convergence pattern.
    }
    \label{fig:a_50MeV_SMS500MeV_deg}

\end{figure*}

We also tested four different input spaces for the scattering-angle dimension $\kinparvec_{\theta}$ of the differential cross section and spin observables.
Coefficients generated with $x = \negcos$ showed consistency with our model broadly across angles and energies.
One such example is shown in Fig.~\ref{fig:d_150MeV_SMS500MeV_Qsum_cos_Qofprel}.
Because of their similar functional form, $x = \negcos$ and $x = \qcm^{2} = \prel^{2} (1 - \cos(\theta))$ gave nearly identical results in the statistical diagnostics.
We also find that $x = \theta$ provides a suitable input space in many cases, including where $x = \negcos$ may fail to give model-consistent coefficients (see Figs.~\ref{fig:a_50MeV_SMS500MeV_deg} and~\ref{fig:ay_200MeV_SMS500MeV_deg}).

However, the choice of input space $x = \qcm$ can lead to problems; e.g., see Fig.~\ref{fig:d_150MeV_SMS500MeV_Qsum_qcm_Qofprel}.
In this case, there is obvious nonstationarity in the length scale, which manifests as rapidly varying coefficients at high momentum and visible trumpeting in the $\DVAR{PC}$ plot.

\begin{figure*}[!p]
    \includegraphics{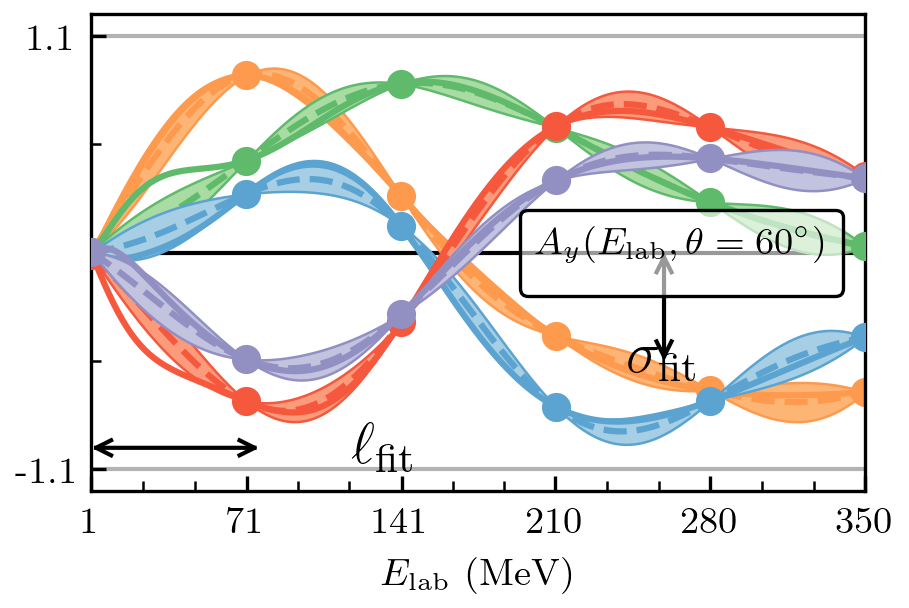}
    \includegraphics{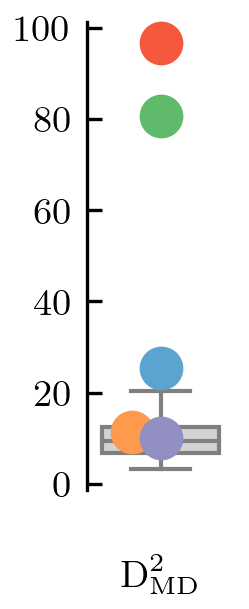}
    \includegraphics{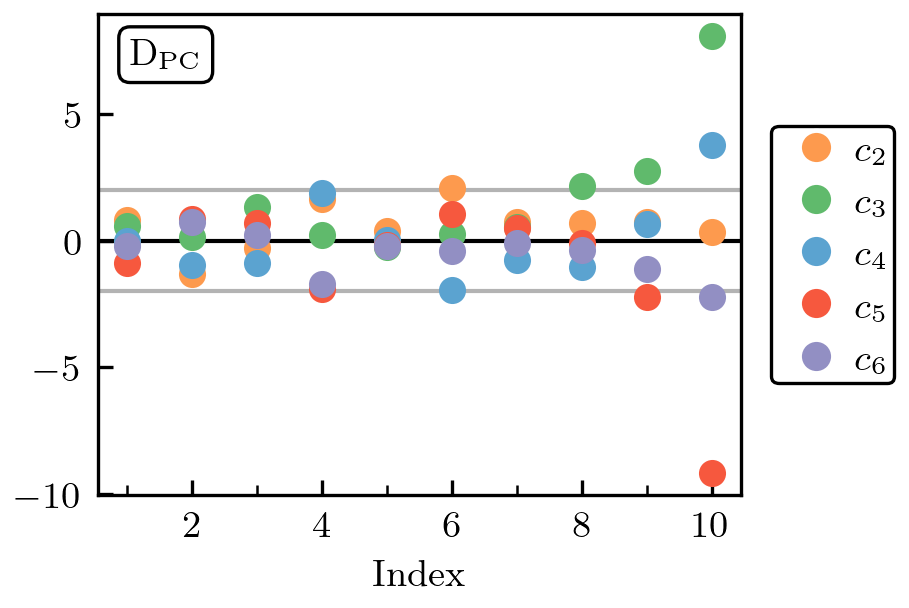}
    
     \caption{
    Figures here are generated with the same choices as those in Fig.~\ref{fig:ay_60degrees_SMS500MeV_success}, but with $\kinparvec_{\theta} = \Elab$ and $Q = \Qsum(p = \prel, \mpieff = 186\, \mathrm{MeV}, \Lambdab = 610\, \mathrm{MeV})$ (optimal values of $\mpieff$ and $\Lambdab$ from Table~\ref{tab:scale_values}).
    Note that $c_{3}$ and $c_{5}$ seem to present particular issues in their long length scale, which the $\DMD^2$ plot puts in stark relief.
    }
    \label{fig:ay_60degrees_SMS500MeV_failure}

\end{figure*}

Additionally, we tested two input spaces for the total cross section and for the lab-energy dimension of the differential cross section and spin observables: $x = \Elab$ and $x = \prel$.
We saw that the total cross section analysis in Sec.~\ref{subsec:parametrizations} favored $x=\prel$ as the input space. 
We also find cases of other observables where the stretching effect of $x=\prel$ is preferred; 
for example, $A_y$ in Fig.~\ref{fig:ay_60degrees_SMS500MeV_success} with $x=\prel$ is more stationary than Fig.~\ref{fig:ay_60degrees_SMS500MeV_failure}
with $x=\Elab$.
While we can also find examples where $x=\Elab$ is preferred, our general choice is $x=\prel$ because, across a wide range of hyperparameter choices, it tends to perform better. 

\begin{figure*}[!p]

\includegraphics{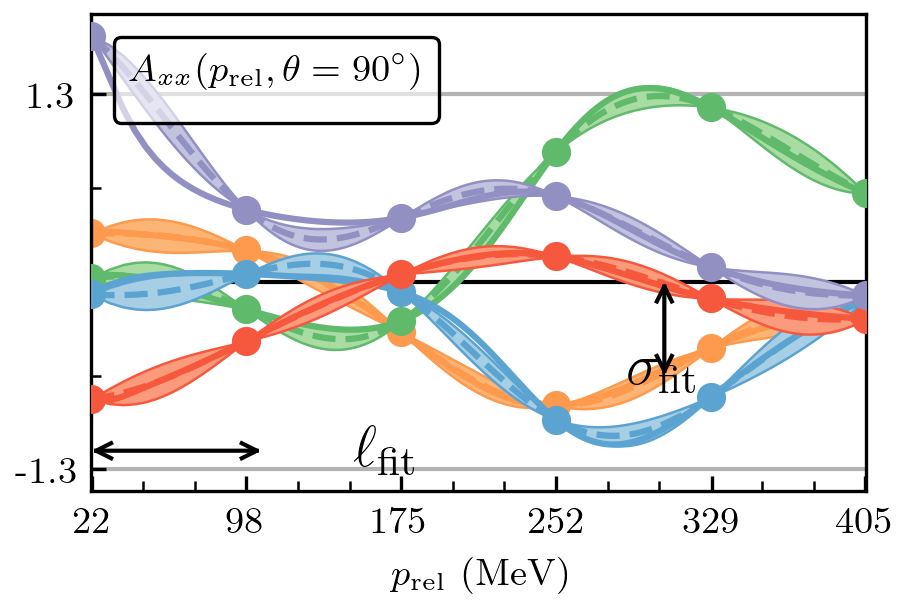}
    \includegraphics{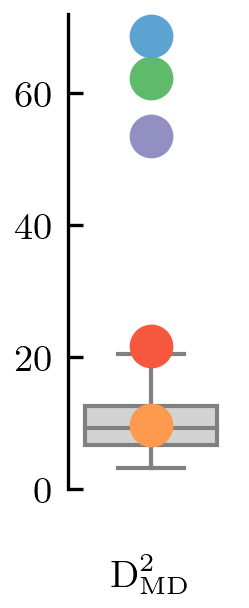}
    \includegraphics{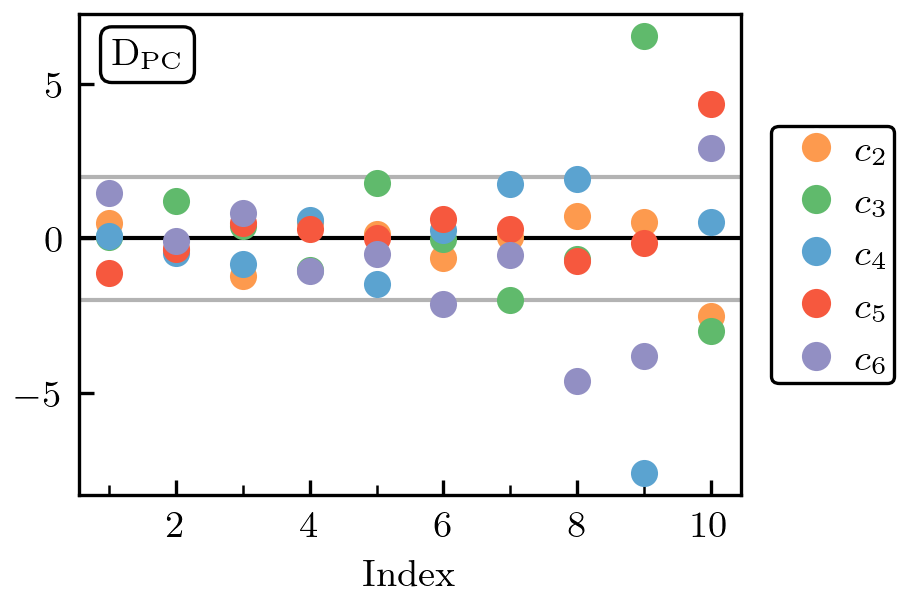}
    \caption{Diagnostics for the spin observable $A_{xx}$ at $\theta = 90^{\circ}$.
    Here, the coefficients are plotted with $\kinparvec_{\theta} = \prel$ and $Q = \Qsum(p = \prel, \mpieff = 138\, \mathrm{MeV}, \Lambdab = 570\, \mathrm{MeV})$ (optimal values of $\mpieff$ and $\Lambdab$ from Table~\ref{tab:scale_values}).
    The statistical diagnostics are calculated with 6 training points (starting near 22\,MeV) and 10 testing points.
    }
    \label{fig:axx_90degrees_SMS500MeV_failure}

\vspace*{.5in}

\includegraphics{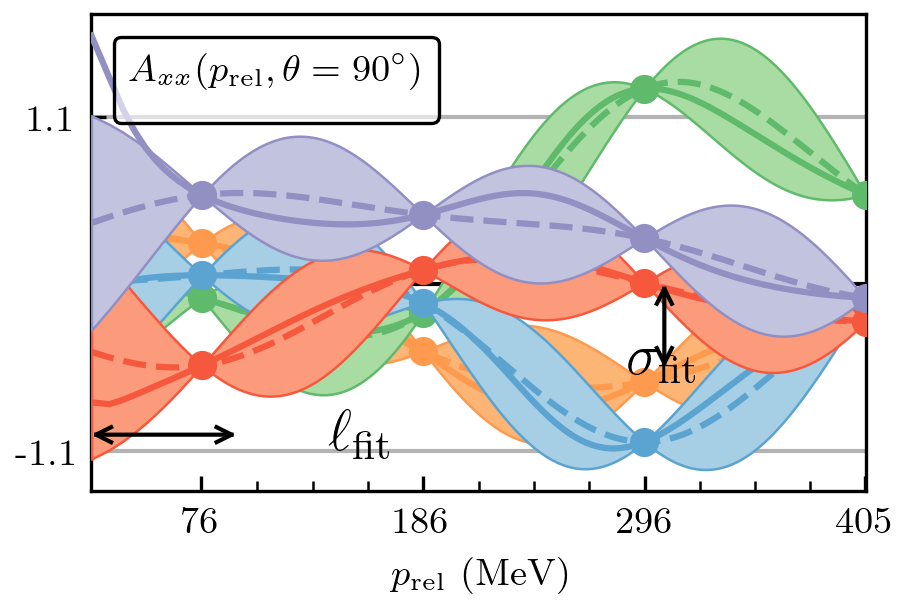}
    \includegraphics{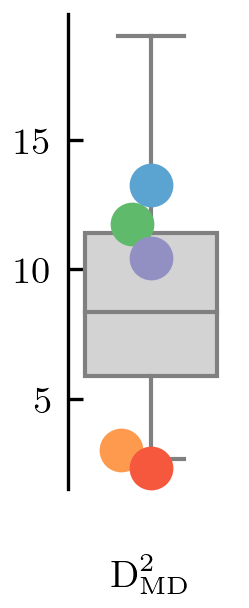}
    \includegraphics{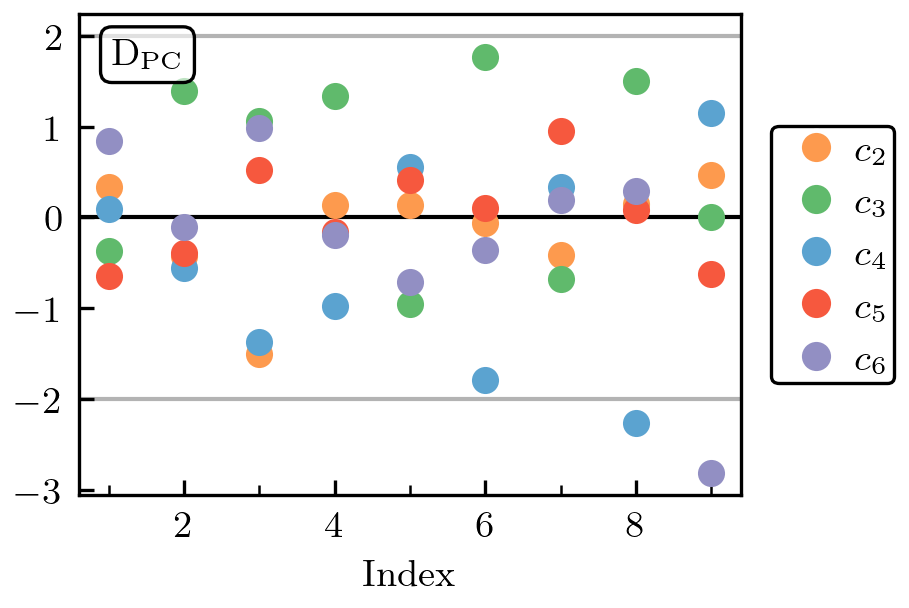}
    \caption{
    Figures here are generated with the same choices as those in Fig.~\ref{fig:axx_90degrees_SMS500MeV_failure}, but with 4 training points and 9 testing points.
    The omission of training and testing points for momenta below 75 MeV removes the trumpeting  of $c_3$, $c_4$, and $c_6$ seen in Fig.~\ref{fig:axx_90degrees_SMS500MeV_failure} and so yields a consistent $\DMD^2$ distribution.
    }
    \label{fig:axx_90degrees_SMS500MeV_success}
\end{figure*}

\begin{figure*}[!p]
    \centering
    
    \includegraphics{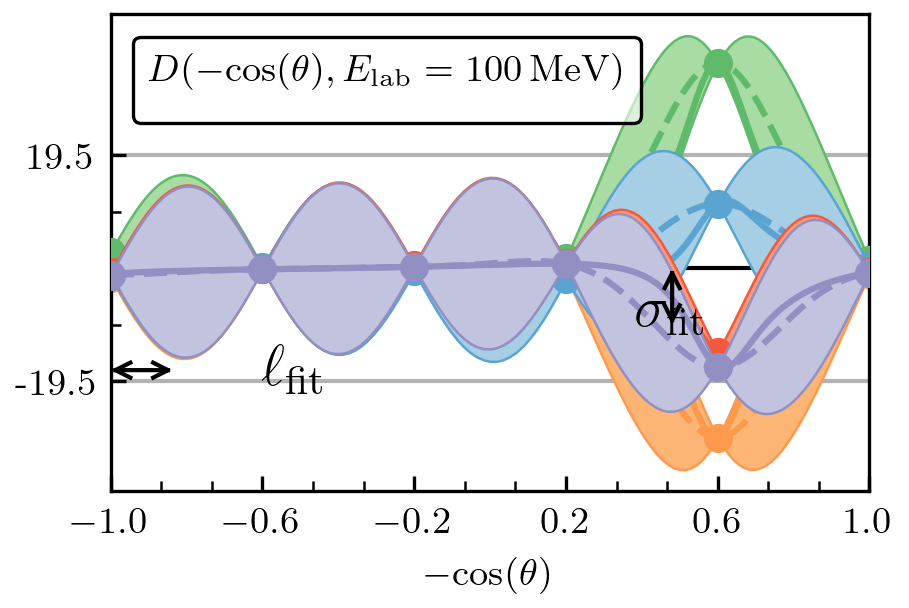}
    \includegraphics{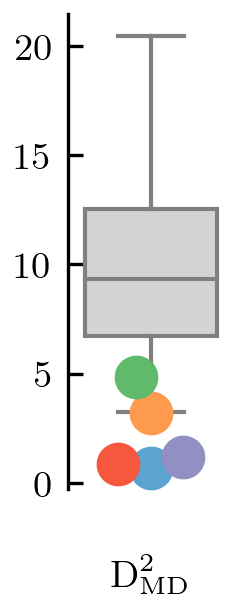}
    \includegraphics{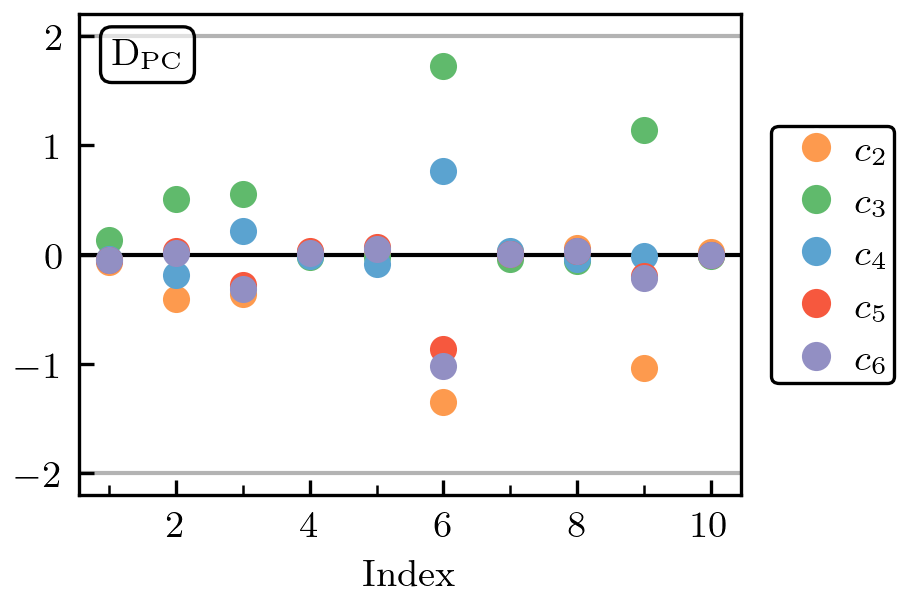}
    \caption{
    Diagnostics for the spin observable $D$ at $\Elab = 100\,\mathrm{MeV}$.
    Here, the coefficients are plotted with $\kinparvec_{\theta} = \negcos$ and $Q = \Qsum(p = \prel, \mpieff = 138\, \mathrm{MeV}, \Lambdab = 570\, \mathrm{MeV})$ (optimal values of $\mpieff$ and $\Lambdab$ from Table~\ref{tab:scale_values}).
    The statistical diagnostics are calculated with 6 training points and 10 testing points.
    The failure arises from the fact that $y_{\rm ref}$ is set to the highest order of data that exists, which crosses zero somewhere in the input space, instead of 1.
    This mistake leads to extremely nonstationary and unnatural coefficients.
    }
    \label{fig:d_100MeV_SMS500MeV_failure}

\vspace*{.2in}

    \includegraphics{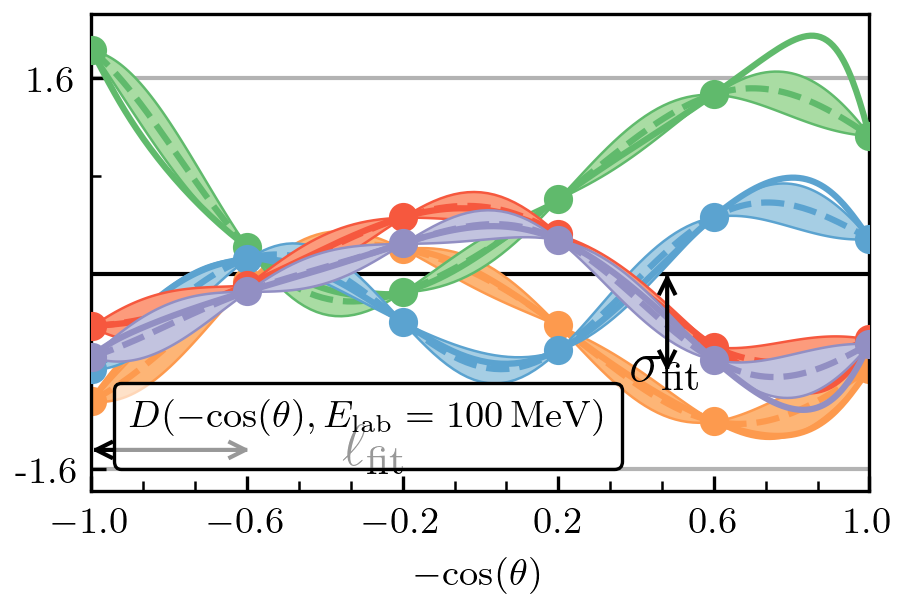}
    \includegraphics{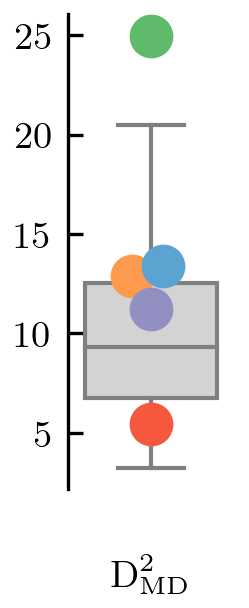}
    \includegraphics{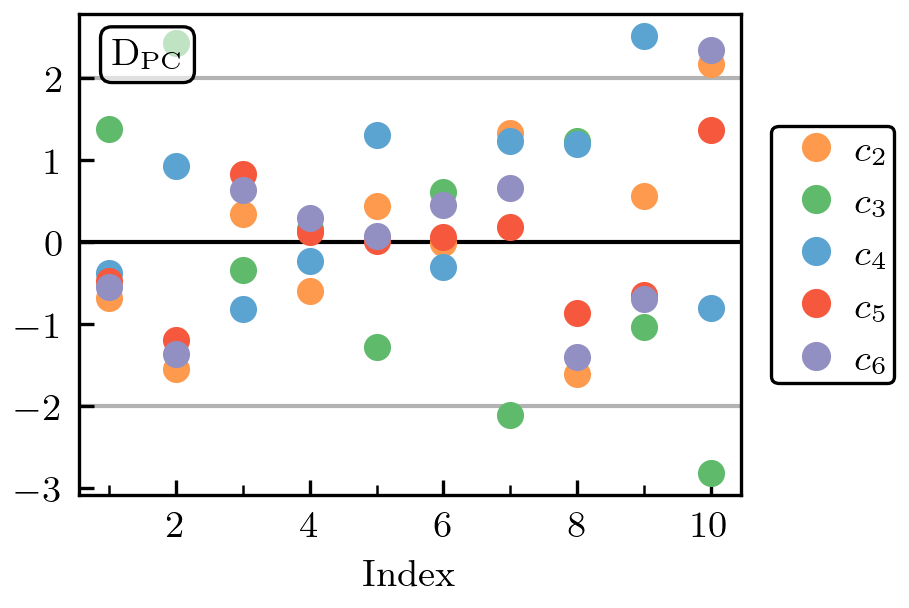}
    \caption{
    Figures here are generated with the same choices as those in Fig.~\ref{fig:d_100MeV_SMS500MeV_failure}, but with $y_{\rm ref} = 1$.}
    
    \label{fig:d_100MeV_SMS500MeV_success}

\end{figure*}

\subsection{Train-test split and constraints}
\label{subsec:traintestsplit}

Though not an explicit feature of Eq.~\eqref{eq:obs_k_expansion}, the train-test split features importantly in the workflow in Table~\ref{tab:workflow}.
In some cases, the location of training and testing points may be informed by theoretical or empirical knowledge about the physical regimes where an EFT expansion fares poorly.
For example, we find that the coefficients are often at least somewhat nonstationary at low energy/momentum (see Figs.~\ref{fig:sgt_coeff_assumptions_mpi138} and~\ref{fig:sgt_coeff_assumptions_mpi200}).
Even for a widely reliable choice of parametrizations and optimized values for $\Lambdab$ and $\mpieff$, when we train and test over all momenta in Fig.~\ref{fig:axx_90degrees_SMS500MeV_failure}, we find the nonstationarity of the coefficients (mainly $c_{5}$) reflected starkly in the diagnostics.
But when the region below 75 MeV relative momentum is omitted from both the training and testing sets in Fig.~\ref{fig:axx_90degrees_SMS500MeV_success}, a much stronger pattern of stationarity emerges.

Aside from the exclusion of regimes where the convergence or suitability of an EFT may be in doubt, more idiosyncratic considerations may bear on the choice of train-test split.
As mentioned in Sec.~\ref{subsec:buqeye_recap}, two spin observables have constraints in their values 
due to time-reversal symmetry: $A(\Elab, \theta = 0^{\circ}) = 0$, and $A_{y}(\Elab, \theta = \left\{0^{\circ}, 180^{\circ}\right\}) = 0$.
This does not affect the visual evidence of (non)stationarity 
in the coefficients, but it can present problems for the statistical diagnostics.
Because the constraints eliminate the  uncertainty that the GP can assign at those points,
the diagnostics for these observables paint a very poor picture (Fig.~\ref{fig:a_50MeV_SMS500MeV_cos}) regardless of how the coefficients look.
There are at least two options for remediating the situation.
One might simply omit the training and testing points in the region of the constraints, as shown in Fig.~\ref{fig:a_50MeV_SMS500MeV_cos_backward}, without changing any choices of parametrization; this approach improves consistency with our model.
Alternatively, a change of input space, such as the switch from $x = \cos(\theta)$ to $x = \theta$ in Fig.~\ref{fig:a_50MeV_SMS500MeV_deg}, may also improve consistency without having to omit a portion of the input space in the train-test split.
Figures~\ref{fig:ay_200MeV_SMS500MeV_cos}--\ref{fig:ay_200MeV_SMS500MeV_deg} (see Appendix~\ref{app:examples}) tell a similar story.

\subsection{Reference scale}
\label{subsec:yref}

Some care is needed in the choice of $y_{\rm ref}$ in Eq.~\eqref{eq:obs_k_expansion}.
The total cross section and differential cross section are always positive and the total cross section spans many orders of magnitude in value, so it makes sense to set $y_{\rm ref}$ equal to the values of the highest calculated order, as we do in all cases in this paper.
However, spin observables ($D$, $A_{xx}$, $A_{yy}$, $A$, and $A_{y}$) can take positive and negative values over a given input space, which makes this approach to $y_{\rm ref}$ problematic, as dividing by the set of data for the highest calculated order when it has a zero-crossing can lead to divergences.
An example of the deleterious consequences is seen in Fig.~\ref{fig:d_100MeV_SMS500MeV_failure} for the spin observable $D$, where $\genobsref$ is set to the highest order.
We strongly recommend instead setting $\genobsref = 1$ for spin observables; see Fig.~\ref{fig:d_100MeV_SMS500MeV_success} for the corresponding plot to Fig.~\ref{fig:d_100MeV_SMS500MeV_failure} but with $\genobsref = 1$.

\begin{figure*}[!p]
\centering

\includegraphics{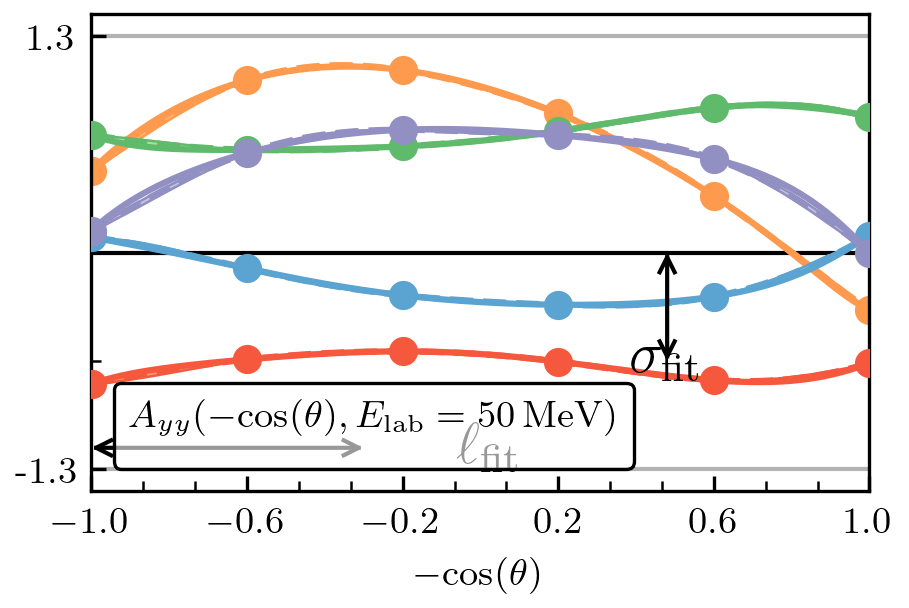}
    \includegraphics[scale=0.75]{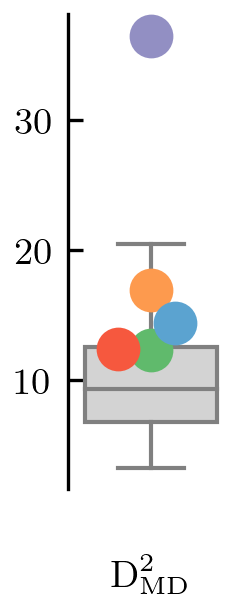}
    \includegraphics{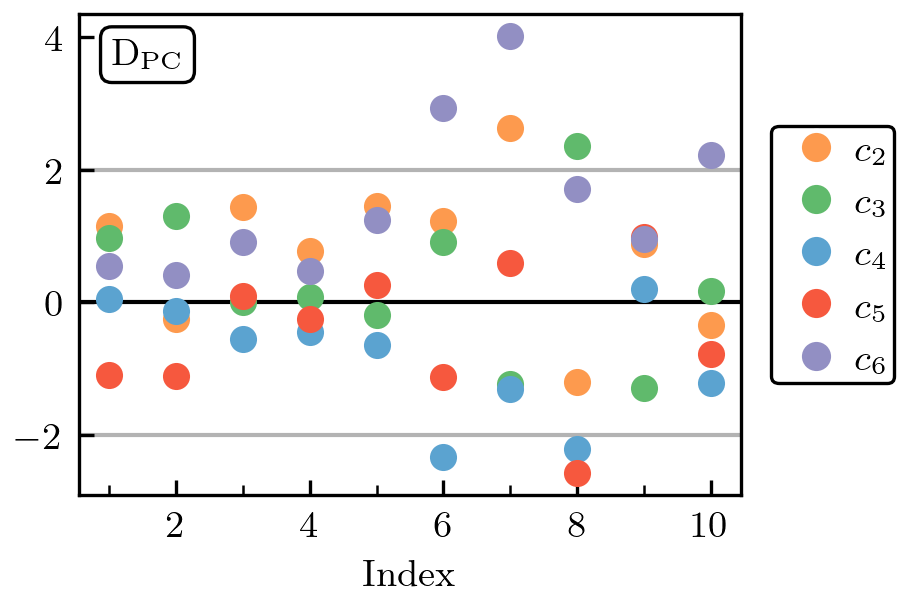}
    \caption{
    Diagnostics for the spin observable $A_{yy}$ at $\Elab = 50\,\mathrm{MeV}$.
    Here, the coefficients are plotted with $\kinparvec_{\theta} = \negcos$ and $Q = \Qsum(p = \prel, \mpieff = 138\, \mathrm{MeV}, \Lambdab = 570\, \mathrm{MeV})$ (optimal values of $\mpieff$ and $\Lambdab$ from Table~\ref{tab:scale_values}).
    The statistical diagnostics are calculated with 6 training points and 10 testing points.
    Note that $c_{6}$'s $\DMD^2$ value shows it to be an outlier.
    }
    \label{fig:ayy_50MeV_SMS500MeV_Qsum_prel_cos}

\vspace*{.2in}

\includegraphics{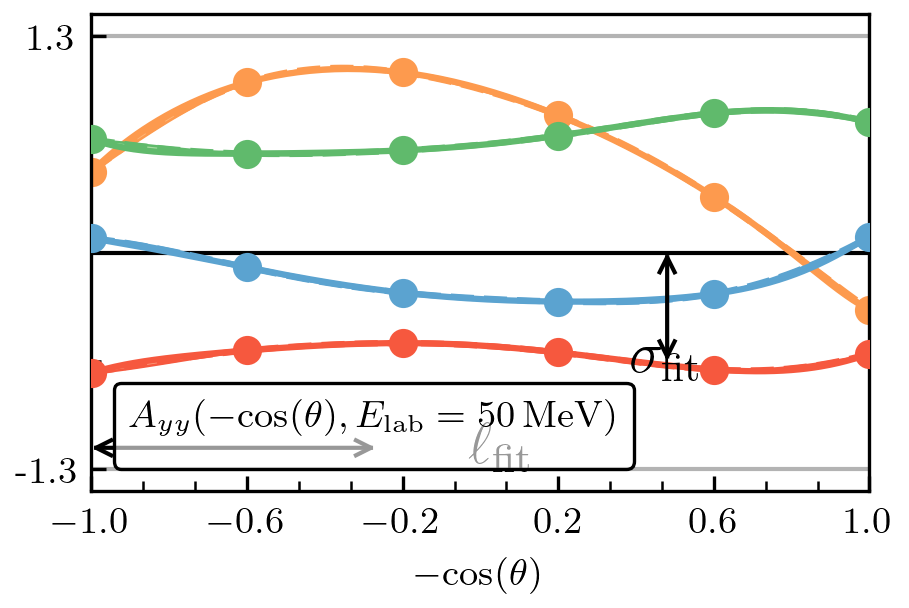}
        \includegraphics[scale=0.75]{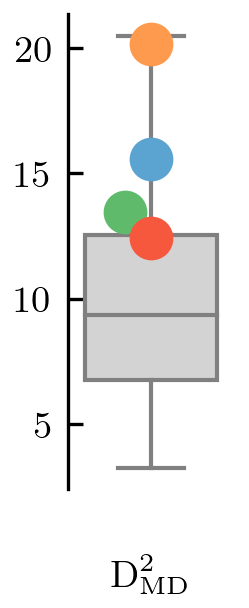}
    \includegraphics{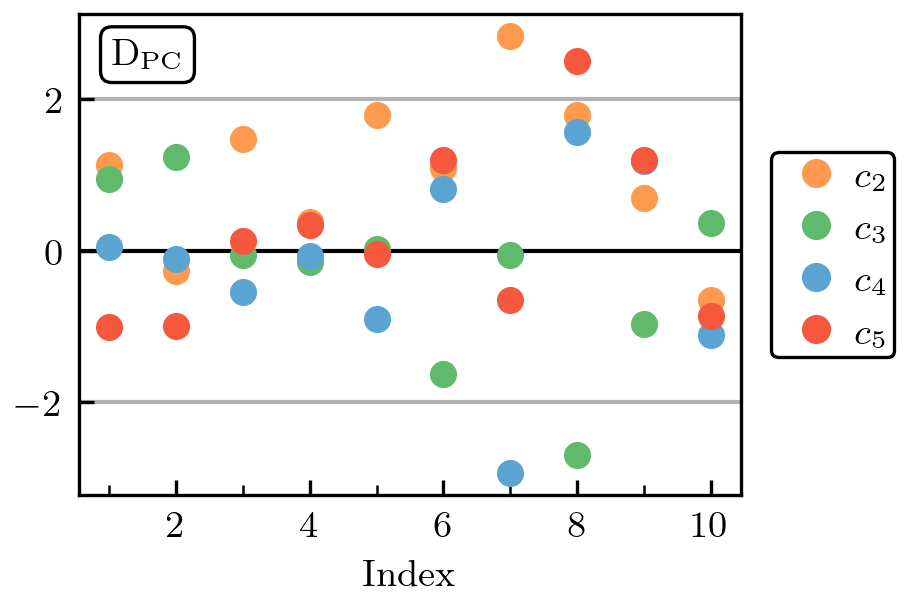}
    \caption{
    Diagnostics for the spin observable $A_{yy}$ at $\Elab = 50\,\mathrm{MeV}$.
    Here, the coefficients are plotted with $\kinparvec_{\theta} = \negcos$ and $Q = \Qsum(p = \prel, \mpieff = 120\, \mathrm{MeV}, \Lambdab = 528\, \mathrm{MeV})$ (optimal values of $\mpieff$ and $\Lambdab$ from Table~\ref{tab:scale_values}).
    The statistical diagnostics are calculated with 6 training points and 10 testing points.
    The coefficient function $c_{6}$ is omitted from all calculations.
    Note that this omission results in improvements in consistency shown in the diagnostics.
    }
    \label{fig:ayy_50MeV_SMS500MeV_Qsum_prel_cos_noc6}
\end{figure*}

\section{Practical applications} \label{sec:applications}

Here we consider two applications of our analysis workflow.

\subsection{Treatment of \NNNNLOp}
\label{subsec:n4lo+}

The SMS potential includes five complete orders (LO through \NNNNLO) as well as an additional incomplete order known as \NNNNLOp, which contains numerically important corrections to $D$- and $F$-wave scattering processes.
But is \NNNNLOp subject to the same power-counting scheme as the lower, complete orders?
We can essay an answer to that question by consulting coefficient plots and their associated statistical diagnostics.
Frequent, pronounced inconsistency with our model across different sets of input parameters would be taken as a sign that this order is not compliant with our power-counting paradigm (however much the inclusion of \NNNNLOp physics makes observable predictions more accurate), but overall compliance with our model would indicate that \NNNNLOp should be treated in the same fashion as a complete order.
We examine what happens when the sixth-order (i.e., \NNNNLOp) coefficient $c_{6}$ is omitted from the analysis in cases of inconsistency with our model.

The vast majority of parametrization choices lead to model-consistency when $c_{6}$ is included. 
In those cases the omission of $c_{6}$ leaves that consistency intact and makes little difference.
There are cases where model-inconsistency is observed and $c_{6}$ is the culpable coefficient (e.g., see Fig.~\ref{fig:ayy_50MeV_SMS500MeV_Qsum_prel_cos}); there, omitting it from the GP model of the coefficients can remediate the situation (e.g., see Fig.~\ref{fig:ayy_50MeV_SMS500MeV_Qsum_prel_cos_noc6}).
However, care is required in this exercise: As the highest-order coefficient under consideration, $c_{6}$ is most susceptible to wrong-sizing by choices of inappropriate values for $Q$.
This sensitivity should caution against definitive conclusions that $c_{6}$ is a pathological order \emph{per se}.
Since \NNNNLOp does not seem significantly more likely than the other orders to be pathological and its omission does not decisively affect consistency, we conclude that it ought to be treated, at least provisionally, as a full order subject to the usual power-counting scheme.

\subsection{Do \texorpdfstring{$\DCI$}{DCI} plots work?}

\begin{figure*}[!t]
    \centering
    \includegraphics{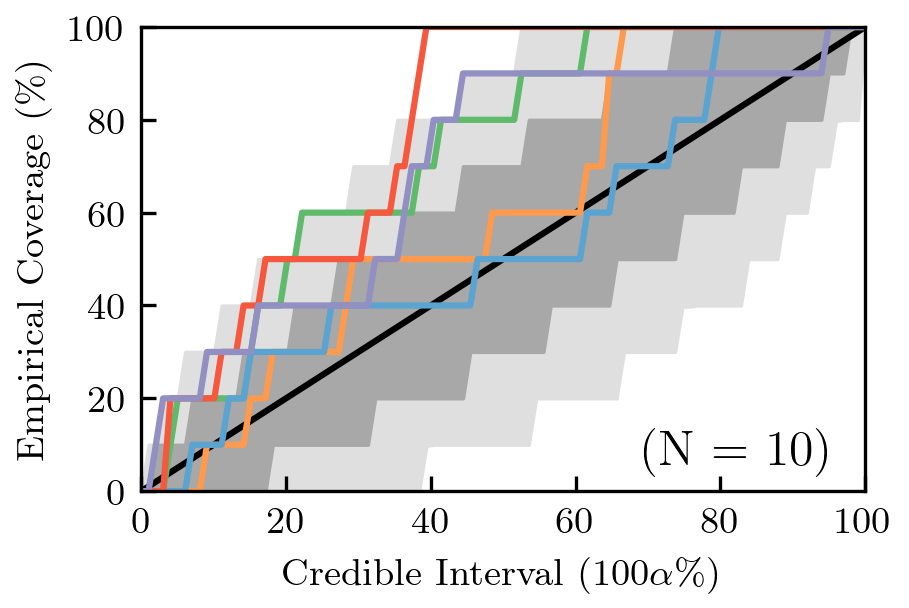}
    \includegraphics{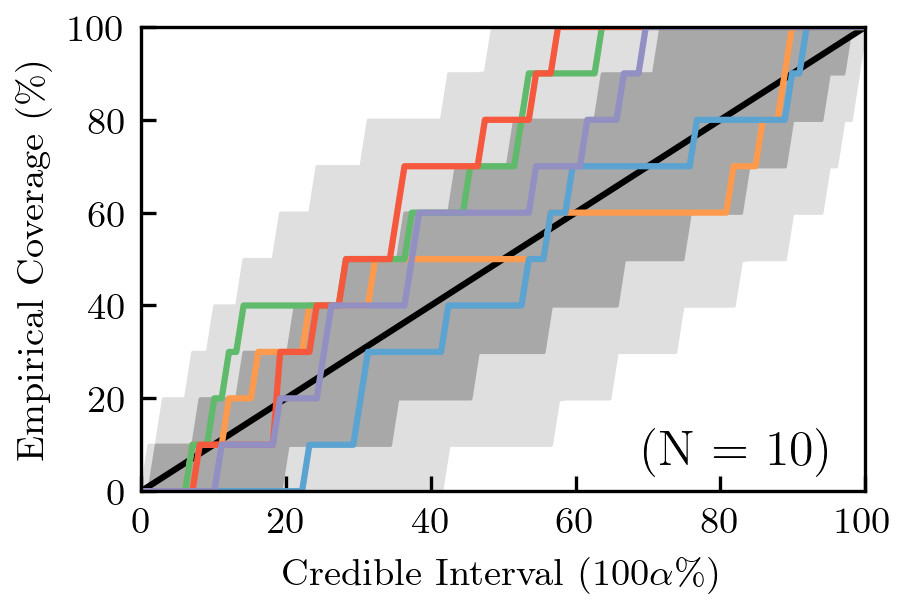}
    \phantomsublabel{-5.4}{1.6}{fig:dsg_150MeV_dci_failure}
    \phantomsublabel{-2.3}{1.6}{fig:dsg_150MeV_dci_success}
     \caption{
     Plots of the credible intervals (``weather plots'') corresponding to the coefficients of the differential cross section at $\Elab = 150\,\mathrm{MeV}$ in Fig.~\ref{fig:dsg_150MeV_SMS500MeV_Qsum_cos_Qofpq} (a) and Fig.~\ref{fig:dsg_150MeV_SMS500MeV_Qsum_cos_Qofprel} (b).
     The concentration of curves above and to the left of the black midline in the lefthand figure is an indication that the truncation error is being overestimated and the error model is too conservative.
     Those on the righthand side track the midline better, within the shaded confidence bounds, and thus show a better agreement between estimated and actual error bars.
    }
    \label{fig:credible_intervals_a}
\end{figure*}

\begin{figure*}[!t]
    \centering
    \includegraphics{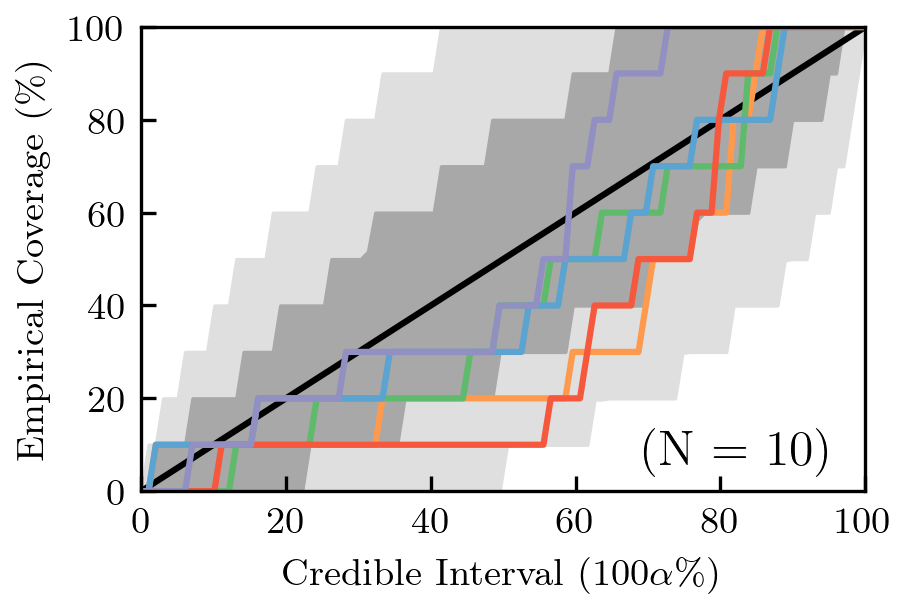}
    \includegraphics{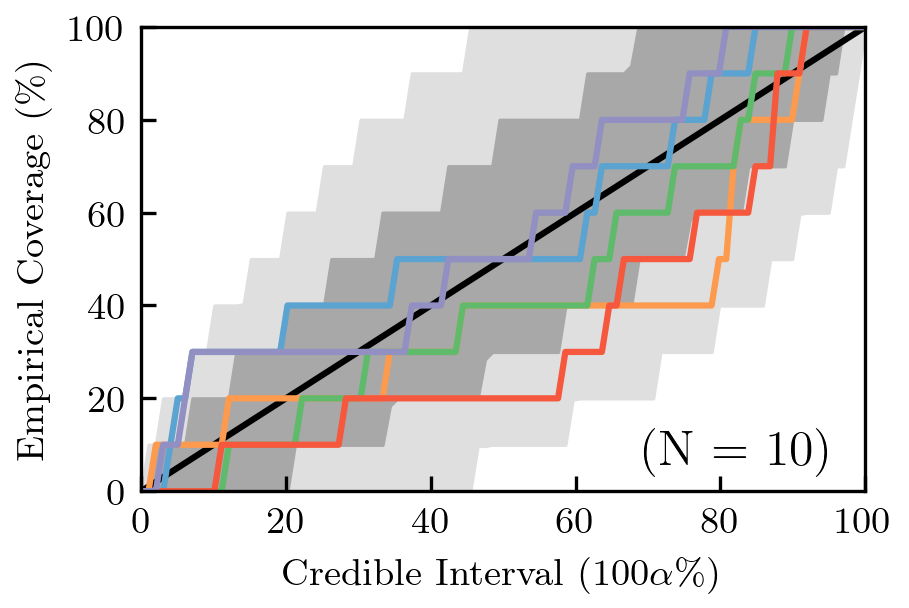}
    \phantomsublabel{-5.4}{1.6}{fig:ay_60deg_dci_failure}
    \phantomsublabel{-2.3}{1.6}{fig:ay_60deg_dci_success}
     \caption{
     Plots of the credible intervals (``weather plots'') corresponding to the coefficients of the spin observable $A_{y}$ at $\theta = 60^{\circ}\,\mathrm{MeV}$ in Fig.~\ref{fig:ay_60degrees_SMS500MeV_failure} (a) and Fig.~\ref{fig:ay_60degrees_SMS500MeV_success} (b).
     The concentration of curves below and to the right of the black midline in the lefthand figure is an indication that the truncation error is being underestimated and the error model is not conservative enough.
     Those on the righthand side track the midline better, within the shaded confidence bounds, and thus show a better agreement between estimated and actual error bars.
    }
    \label{fig:credible_intervals_b}
\end{figure*}

In Sec.~\ref{subsec:credible_intervals} we introduced the $\DCI$ (weather) plot to assess how well the variance of the GP that we fit to our training data captures the validation data.
This purpose is fundamental to the BUQEYE model's goal: to obtain statistically rigorous error bands for \chiEFT\ predictions of physical observables.
To do this, we need to check whether the error band from the fitted GP's variance truly encompasses the assumed percentage of the validation data; 
i.e., that our model is working as expected.
Furthermore, it is important to verify that the $\DCI$ plots actually concur with the other diagnostics. 

Here we offer two examples from this paper and highlight the insight that $\DCI$ plots can share.
Credible interval plots can point out cases in which the EFT error bands are assessed too conservatively or not conservatively enough.
An example of error bands that are too conservative occurs in the comparison of Fig.~\ref{fig:dsg_150MeV_SMS500MeV_Qsum_cos_Qofpq}, in which the characteristic momentum is parametrized by $p = \psmax(\prel, \qcm)$, and Fig.~\ref{fig:dsg_150MeV_SMS500MeV_Qsum_cos_Qofprel}, where the more proper choice $p = \prel$ is made.
Figure~\ref{fig:credible_intervals_a} shows that when the change is made, the assessment of the error goes from too conservative (which corresponds to curves in the upper-left of the weather plot) to more evenhanded (which corresponds to curves along the midline).
For an example of error bands that are not conservative enough, see Fig.~\ref{fig:credible_intervals_b}, which is based on Figs.~\ref{fig:ay_60degrees_SMS500MeV_failure} and~\ref{fig:ay_60degrees_SMS500MeV_success}.
Here, the curves do the opposite: They move from the lower right to the midline, which signals that the uncertainty is more properly quantified with the choice of $x_{E} = \prel$ than $x_{E} = \Elab$.
The $\DCI$ plots corresponding to the other graphical and statistical figures in Sec.~\ref{sec:results_model} can be found in a Jupyter notebook with the other files associated with this paper~\cite{modern_nn_potentials_package}.

\section{Summary and outlook}
\label{sec:outlook}

A full Bayesian parameter estimation of the low-energy constants (LECs) that characterize interactions in chiral effective field theory (\chiEFT) requires rigorous assessment of theoretical errors, which in turn requires assessing
whether and when a correlated truncation-error model is suitable for the data at hand.
We have shown using graphical and statistical diagnostics that the semi-local momentum-space \chiEFT\ nucleon-nucleon (NN) potential of Reinert, Krebs, and Epelbaum with cutoff 500 MeV~\cite{Reinert:2017usi}, with appropriate choices of Gaussian process (GP) parametrizations,
shows order-by-order convergence consistent with the BUQEYE model.

This complements the recent results of Svennson {\it et al.}, who fitted two-dimensional GPs to the \NNLO coefficient $c_3$ in a \chiEFT\ treatment of NN scattering that included an explicit $\Delta(1232)$ isobar degree of freedom.
Svennson {\it et al.} did this for
15 different $np$ scattering observables in the energy range $30~{\rm MeV} \leq T_{\rm lab} \leq 290~{\rm MeV}$ and found that a stationary-GP description of $c_3$ passed statistical consistency checks in every case.

Both our findings and those of Ref.~\cite{Svensson:2023twt}
imply that robust statistical estimates of the truncation error can be made. 
Indeed Svennson {\it et al.} inserted the correlated \chiEFT\ truncation error derived from their GP fit to $c_3$ in the likelihood they used to estimate the EFT LECs. 
We illustrated some more basic applications of a correlated EFT error model in 
Sec.~\ref{sec:applications}.

In demonstrating the BUQEYE model's applicability, we gave guidance on where to start when choosing GP parametrizations, the signs of (in)consistency in a GP description that can be seen in statistical diagnostics, what those signs point to, and how to iterate through a workflow to assess the robustness of the parametrization choices and train-test split.
We examined coefficients corresponding to several different EFT orders, and so were able to employ
the requirement that they be described by a common GP to infer 
a \chiEFT\ breakdown scale $\Lambdab$ in the range 500--700 MeV and an EFT soft scale $\mpieff$ in the range 100--200 MeV. 
The posteriors for these scales have 68\% intervals narrower than this if specific GP parametrizations and forms of the EFT expansion parameter are invoked (see Table~\ref{tab:scale_values}).

These lower values of $\Lambdab$ and $\mpieff$ contrasted with those extracted from an analysis of the expansion parameter $Q$ that included correlations in scattering angle, but employed independent GPs at a number of $\prel$ values between 25 and 400 MeV (see Fig.~\ref{fig:qvsp_linear}). 
The range of GP variances seen across $\prel$ in this pointwise-in-$\prel$ analysis strongly suggests that modeling the correlations with stationary GPs is too restrictive.

We have studied only one exemplary potential scheme and scale in this paper, but similar analyses are 
in progress for other potentials.
An important part of this analysis will be to relax the assumption of stationary length scales and explore how nonstationary GPs might match the underlying structure in the NN observables.
Additionally, we have left unsettled the question of how best to treat behavior in regimes corresponding to regions of input space where one model seems to fail, as in Fig.~\ref{fig:axx_90degrees_SMS500MeV_failure} in  Sec.~\ref{subsec:traintestsplit} where our model fails with energy-dependent observables in the low-momentum regime.
In our case, we assessed with statistical diagnostics how often our model was consistent with the data including and excluding training and testing points from that region, presented both, and compared them, but that is not the only solution.

Future work will build on the insight gleaned from the present work and Ref.~\cite{Svensson:2023twt} to implement full Bayesian analyses of \chiEFT\ for nuclear observables.
This will be facilitated by 
recent advances in devising emulators that drastically reduce calculation times for 
repeated Monte Carlo sampling for Bayesian methods (e.g., see~\cite{Drischler:2022ipa,Giuliani:2022yna,Melendez:2022kid,Odell:2023cun,Duguet:2023wuh}).
The deficiencies we observed here at relative momenta well below $\mpieff$ motivates applying the BUQEYE model to pionless EFT~\cite{Bub:2024}, which is tailor-made to reproduce observables in this momentum regime.
It may be feasible to use Bayesian model mixing to statistically combine
chiral and pionless EFT predictions to better reproduce observables across a full range of momenta.



\begin{acknowledgments}
We gratefully acknowledge the contributions of S.~Wesolowski to an earlier version of this work.
We thank  Christian Forss{\'e}n, Mostofa Hisham, and Simon Sundberg for useful feedback on the manuscript.
The work of PJM and RJF was supported in part by the National Science Foundation Award Nos.~PHY-1913069 and PHY-2209442 and the NUCLEI SciDAC Collaboration under U.S. Department of Energy MSU subcontract no.~RC107839-OSU\@.
The work of DRP was supported by the US Department of Energy under
contract DE-FG02-93ER-40756 and by the Swedish Research Council via a Tage Erlander Professorship (Grant No 2022-00215).
The work of RJF, DRP, and MTP was supported in part by the National Science Foundation CSSI program under Award
No.~OAC-2004601 (BAND Collaboration~\cite{BAND_Framework}).
\end{acknowledgments}

\bibliography{EMN_Correlations_Refs,bayesian_refs}

\clearpage

\appendix
\section{Additional examples}
\label{app:examples}

\begin{figure*}[!p]
\centering
    \includegraphics{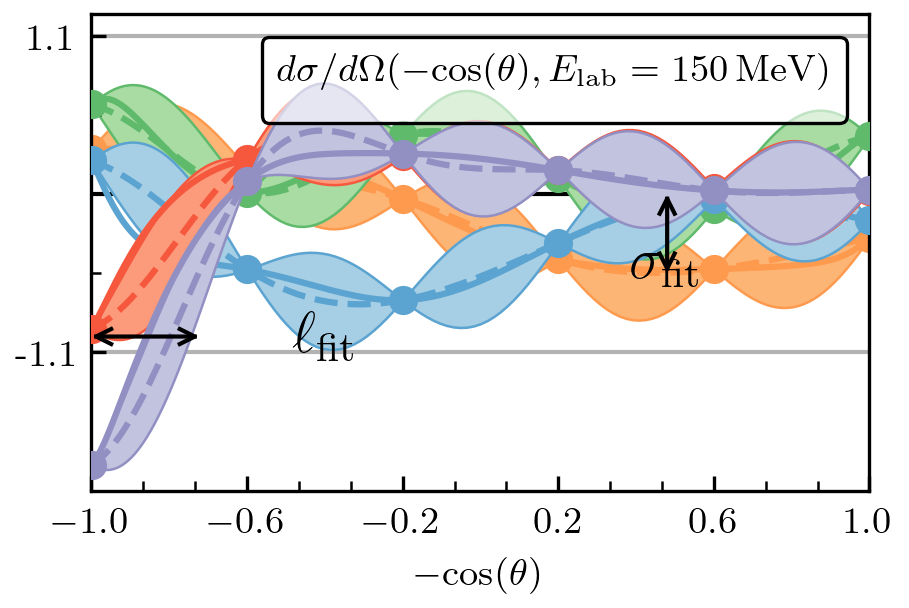}
        \includegraphics{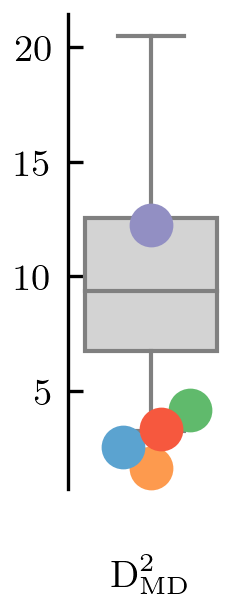}
    \includegraphics{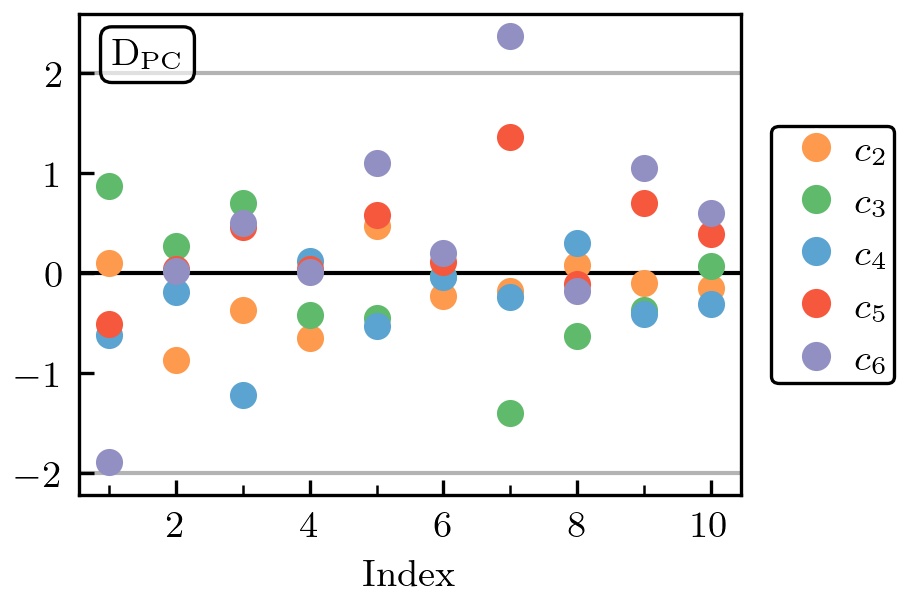}
    \caption{
    Diagnostics for the differential cross section at $\Elab = 150\,\mathrm{MeV}$.
    Here, the coefficients are plotted with $\kinparvec_{\theta} = \negcos$ and $Q = \Qsum(p = \psmax(\prel, \qcm), \mpieff = 172\, \mathrm{MeV}, \Lambdab = 660\, \mathrm{MeV})$ (optimal values of $\mpieff$ and $\Lambdab$ from Table~\ref{tab:scale_values}).
    The statistical diagnostics are calculated with 6 training points and 10 testing points.
    The choice to change $p = \prel$, as in Fig.~\ref{fig:dsg_150MeV_SMS500MeV_Qsum_cos_Qofpq}, to $p = \psmax(\prel, \qcm)$ fails and even backfires by flattening the coefficients past the second training point and causing the length scale to be underestimated (see Fig.~\ref{fig:diagnostic_cheatsheet}, ``Understimated~$\ell$''), with predictable results that are especially visible in the cluster of very low values for the $\DMD^2$ plot.
    }
    \label{fig:dsg_150MeV_SMS500MeV_Qsum_cos_Qofpq}

\vspace*{.3in}

\includegraphics{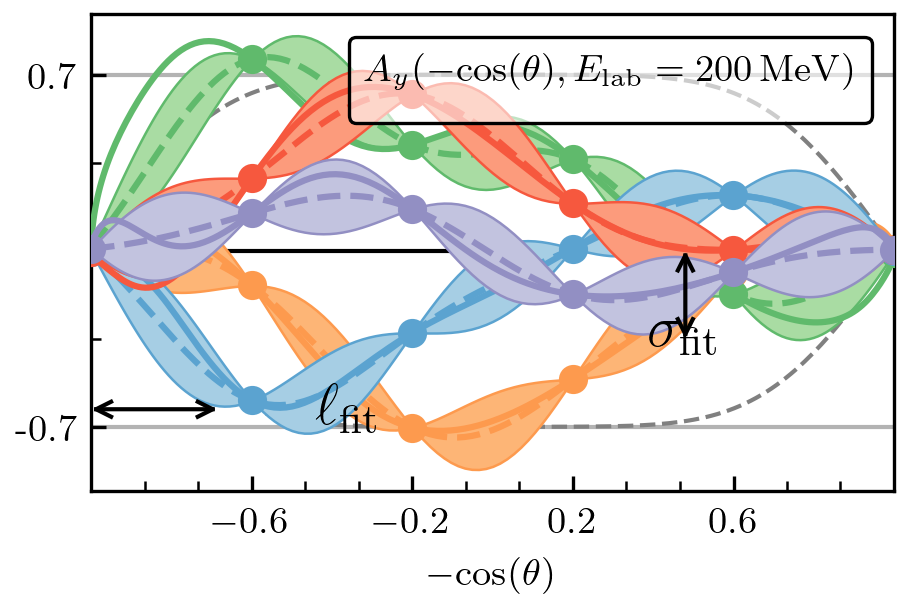}
    \includegraphics{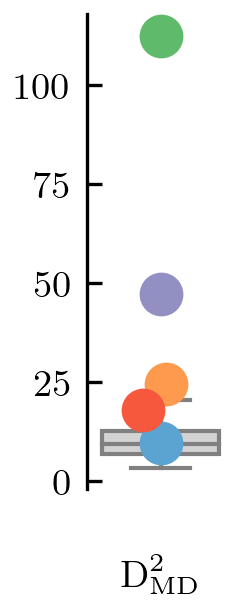}
    \includegraphics{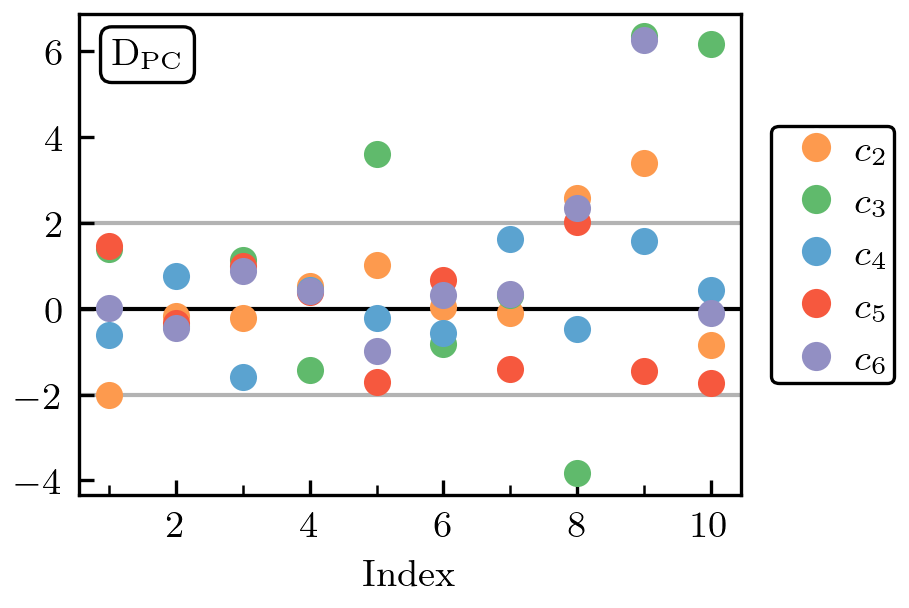}
    \caption{
    Diagnostics for the spin observable $A_{y}$ at $\Elab = 200\,\mathrm{MeV}$.
    Here, the coefficients are plotted with $\kinparvec_{\theta} = \negcos$ and $Q = \Qsum(p = \prel, \mpieff = 138\, \mathrm{MeV}, \Lambdab = 570\, \mathrm{MeV})$ (optimal values of $\mpieff$ and $\Lambdab$ from Table~\ref{tab:scale_values}).
    The statistical diagnostics are calculated with 6 training points and 10 testing points.
    Constraints lead to bunching of coefficients at forward and backward angles that resolves for angles between these extremes, leading to a 
     nonstationary length scale and diagnostics that announce nonstationarity.
    }
    \label{fig:ay_200MeV_SMS500MeV_cos}

\vspace*{.3in}

\includegraphics{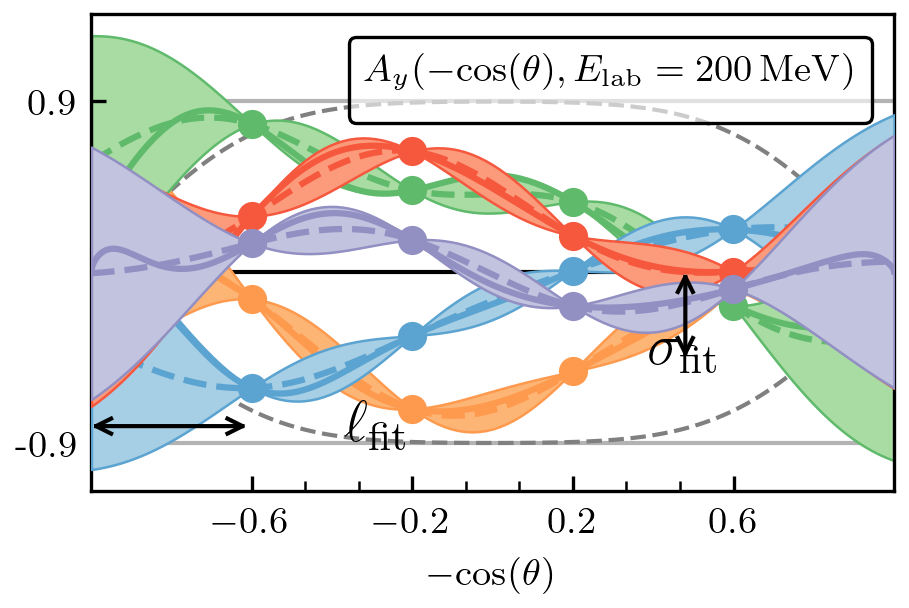}
    \includegraphics{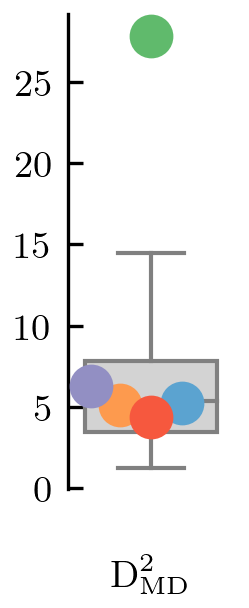}
    \includegraphics{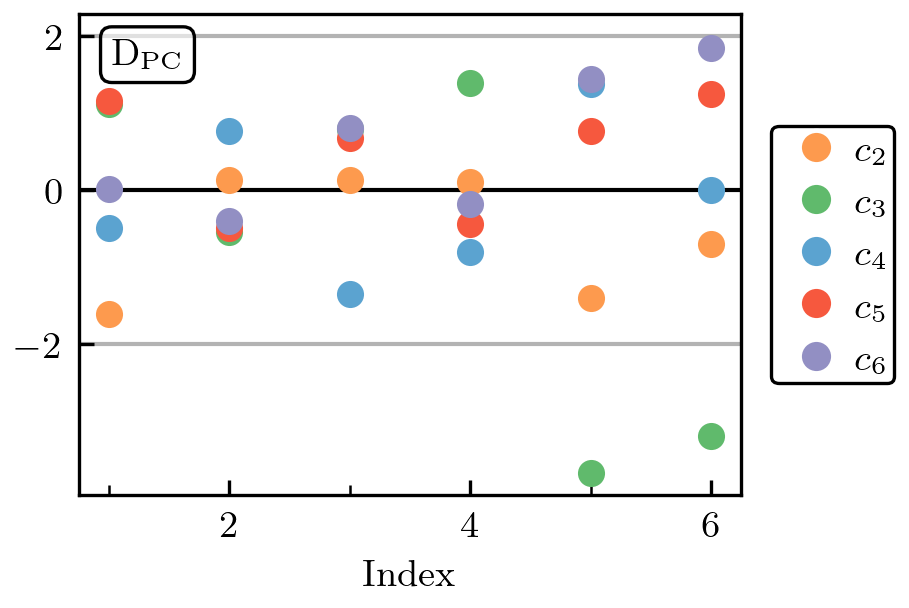}
    \caption{
    Figures here are generated with the same choices as those in Fig.~\ref{fig:ay_200MeV_SMS500MeV_cos}, but with 4 training points and 6 testing points.
    With the lack of training and testing points at forward and backward angles, the situation of stationarity, as shown by the diagnostics, improves but is not ideal.
    }
    \label{fig:ay_200MeV_SMS500MeV_cos_middle}
\end{figure*}


\begin{figure*}
\includegraphics{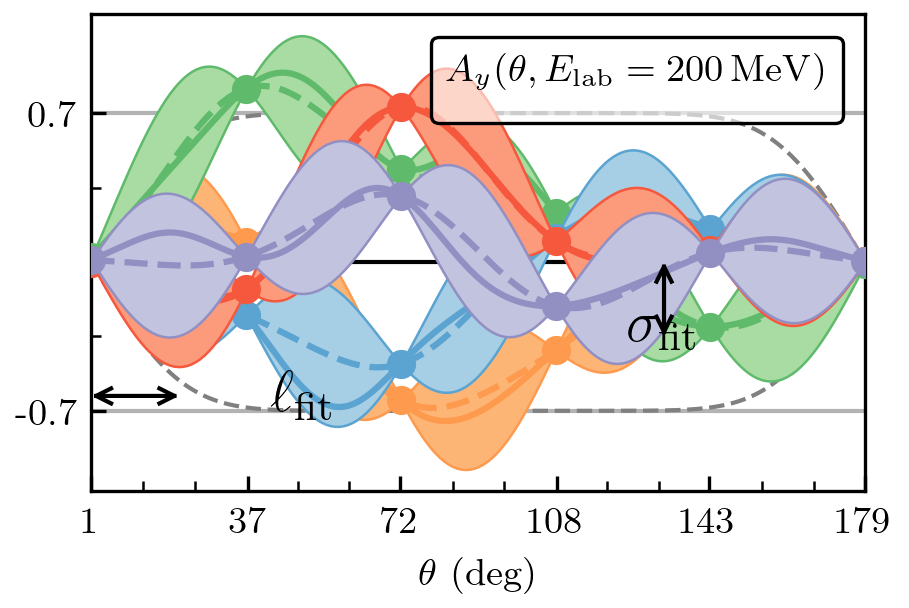}
    \includegraphics{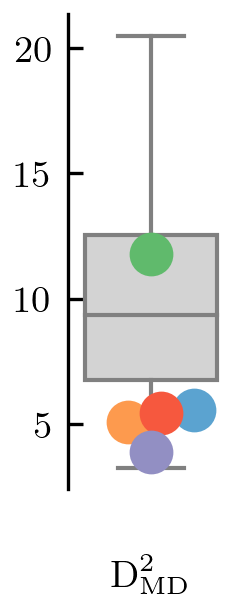}
    \includegraphics{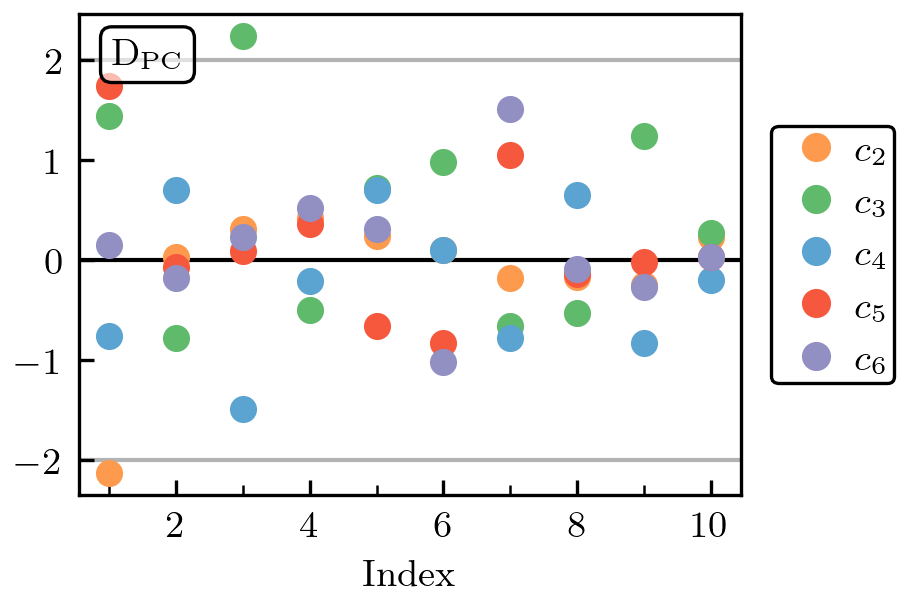}
    \caption{
    Figures here are generated with the same choices as those in Fig.~\ref{fig:ay_200MeV_SMS500MeV_cos}, but with $\kinparvec_{\theta} = \theta$ and $Q = \Qsum(p = \prel, \mpieff = 144\, \mathrm{MeV}, \Lambdab = 590\, \mathrm{MeV})$ (optimal values of $\mpieff$ and $\Lambdab$ from Table~\ref{tab:scale_values}).
    The omission of training and testing points at very forward and very backward angles is salutary for the convergence pattern, as seen in the statistical diagnostics.
    }
    \label{fig:ay_200MeV_SMS500MeV_deg}

\vspace*{.5in}

\includegraphics{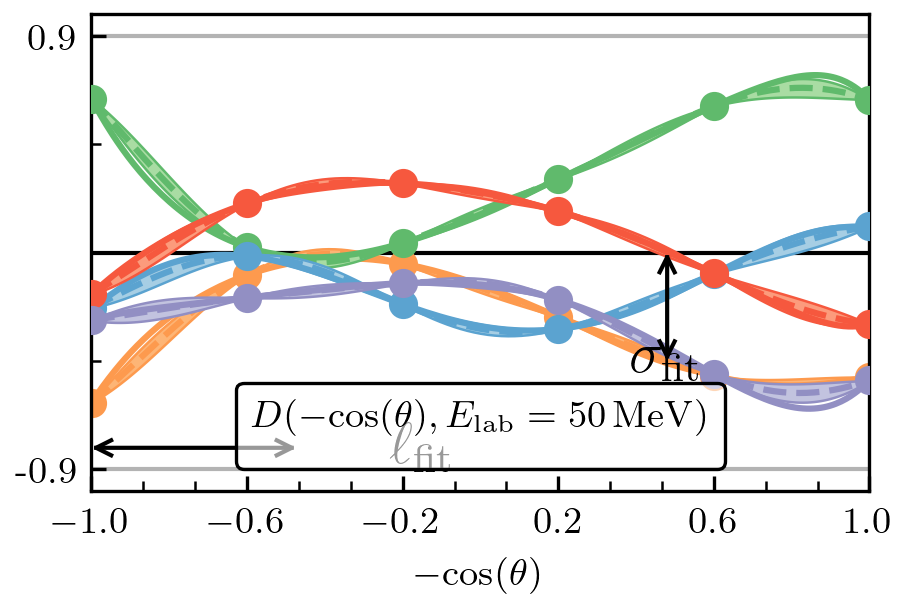}
    \includegraphics{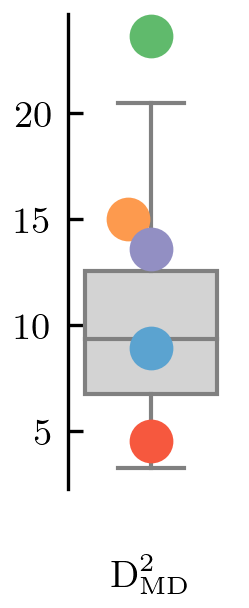}
    \includegraphics{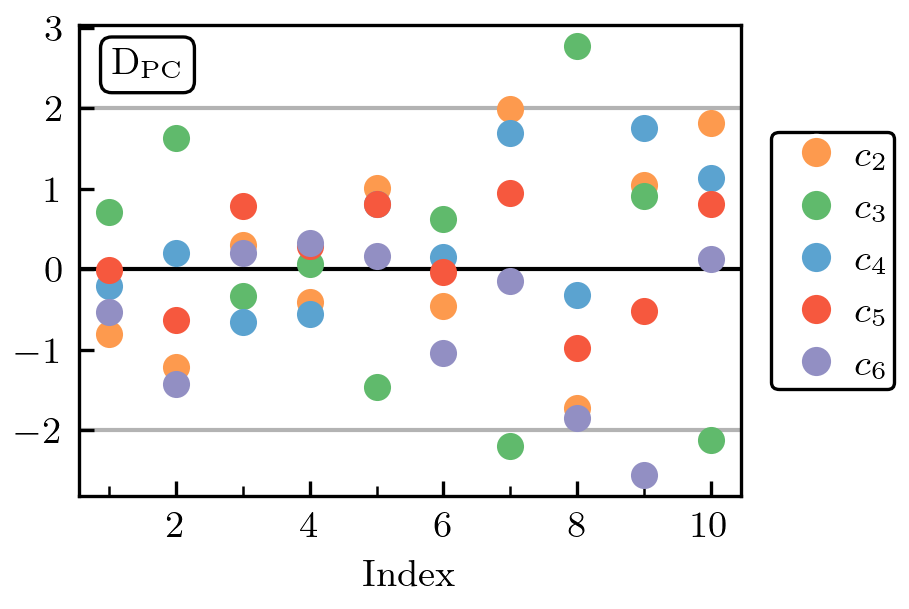}
    \caption{
    Diagnostics for the spin observable $D$ at $\Elab = 50\,\mathrm{MeV}$.
    Here, the coefficients are plotted with $\kinparvec_{\theta} = \negcos$ and $Q = \Qsum(p = \prel, \mpieff = 138\, \mathrm{MeV}, \Lambdab = 570\, \mathrm{MeV})$ (optimal values of $\mpieff$ and $\Lambdab$ from Table~\ref{tab:scale_values}).
    The statistical diagnostics are calculated with 6 training points and 10 testing points.
    }
    \label{fig:d_50MeV_SMS500MeV_success}

\vspace*{.5in}

\includegraphics{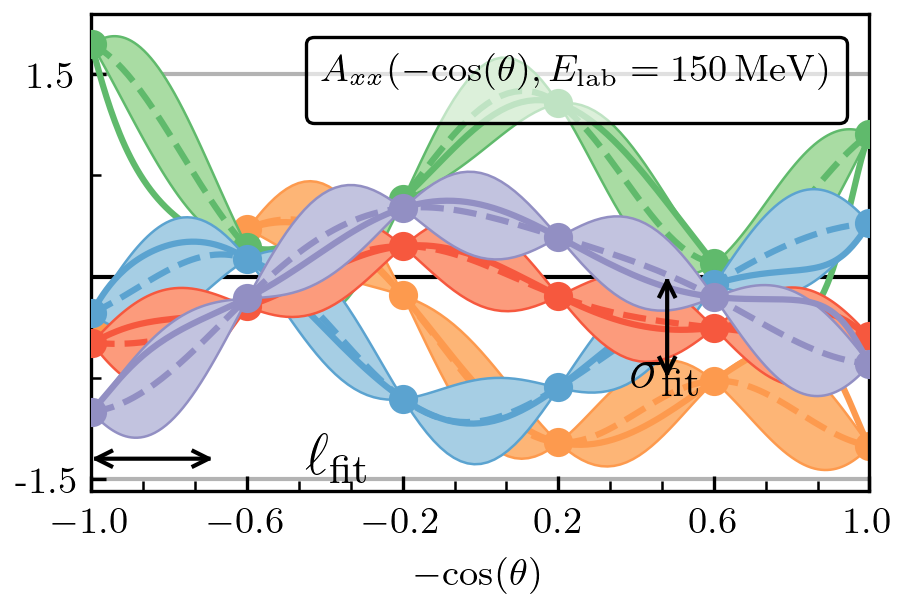}
    \includegraphics{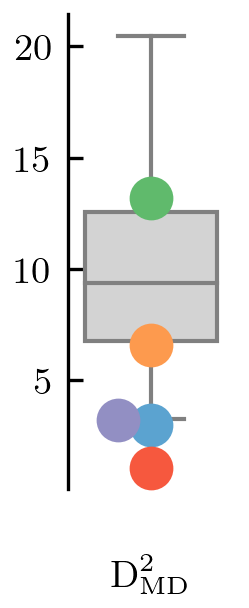}
    \includegraphics{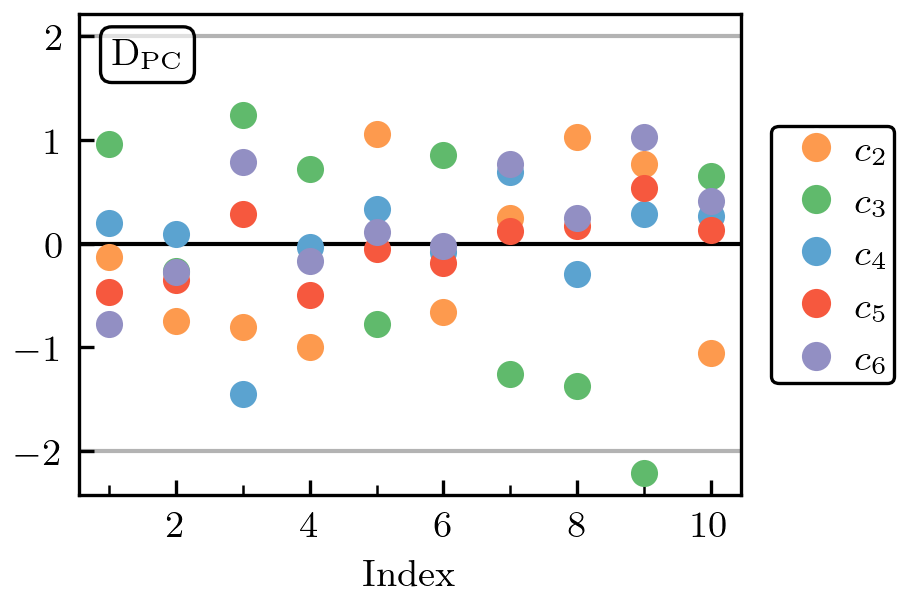}
    \caption{
    Diagnostics for the spin observable $A_{xx}$ at $\Elab = 150\,\mathrm{MeV}$.
    Here, the coefficients are plotted with $\kinparvec_{\theta} = \negcos$ and $Q = \Qsum(p = \prel, \mpieff = 138\, \mathrm{MeV}, \Lambdab = 570\, \mathrm{MeV})$ (optimal values of $\mpieff$ and $\Lambdab$ from Table~\ref{tab:scale_values}).
    The statistical diagnostics are calculated with 6 training points and 10 testing points.
    }
    \label{fig:axx_150MeV_SMS500MeV_success}
\end{figure*}

\

Following the discussion in Sec.~\ref{subsec:Q_param}, we include Fig.~\ref{fig:dsg_150MeV_SMS500MeV_Qsum_cos_Qofpq} to show the lack of improvement when the characteristic momentum in $Q(p,\mpieff)$ is parametrized by $p = \psmax(\prel, \qcm)$ instead of $p = \prel$.

Additionally, we provide in Figs.~\ref{fig:ay_200MeV_SMS500MeV_cos}--\ref{fig:ay_200MeV_SMS500MeV_deg} an exploration of handling constrained observables along the same lines as Figs.~\ref{fig:a_50MeV_SMS500MeV_cos}--\ref{fig:a_50MeV_SMS500MeV_deg} (see Sec.~\ref{subsec:traintestsplit}).

To demonstrate further the wide applicability of the BUQEYE model, we also have included here examples of when consistency can be observed with the BUQEYE model in the cases of 
the spin observable $D$ sliced in energy (Fig.~\ref{fig:d_50MeV_SMS500MeV_success}), and the spin observable $A_{xx}$ sliced in energy (Fig.~\ref{fig:axx_150MeV_SMS500MeV_success}).

\end{document}

%% file: buqeye_macros.tex
\makeatletter
\newcommand\newsubcommand[3]{\newcommand#1{#2\sc@sub{#3}}}
\def\sc@sub#1{\def\sc@thesub{#1}\@ifnextchar_{\sc@mergesubs}{_{\sc@thesub}}}
\def\sc@mergesubs_#1{_{\sc@thesub#1}}

\newcommand\newsupcommand[3]{\newcommand#1{#2\sc@sup{#3}}}
\def\sc@sup#1{\def\sc@thesup{#1}\@ifnextchar^{\sc@mergesups}{^{\sc@thesup}}}
\def\sc@mergesups^#1{^{\sc@thesup#1}}
\makeatother

\DeclareMathAlphabet{\mathbcal}{OMS}{cmsy}{b}{n}

\newcommand{\cbar}{\bar c}
\newcommand{\sdth}{\cbar}

\newcommand{\diagnostic}{\textup{D}}
\DeclareMathOperator{\CI}{CI}
\newcommand{\DCI}{\diagnostic_{\CI}}
\newcommand{\MD}{\textup{MD}}
\newcommand{\DMD}{\diagnostic_{\MD}}
\newcommand{\DVAR}[1]{\inputvec{D}_{\textup{#1}}}
\newcommand{\DVARIT}[1]{\inputvec{D}_{#1}}

\newcommand{\indicator}{\mathbf{1}}

\newcommand{\fvalid}{\inputvec{f}_{\textup{val}}}

\newcommand{\discrcorr}[1]{R_{\delta #1}}

\newcommand{\NNLO}{\ensuremath{{\rm N}{}^2{\rm LO}}\xspace}
\newcommand{\NNNLO}{\ensuremath{{\rm N}{}^3{\rm LO}}\xspace}
\newcommand{\NNNNLO}{\ensuremath{{\rm N}{}^4{\rm LO}}\xspace}
\newcommand{\NNNNLOp}{\ensuremath{{\rm N}{}^4{\rm LO}+}\xspace}

\DeclareMathOperator{\GP}{\mathcal{GP}}

\newcommand{\param}{\boldsymbol{\theta}}

\newcommand{\ordervec}{\vec}
\newcommand{\inputdimvec}{}
\newcommand{\inputvec}{\mathbf}

\newsubcommand{\ckvec}{\ordervec{c}}{k}

\newsubcommand{\bkvec}{\ordervec{b}}{k}
\newcommand{\kinparvec}{\inputdimvec{x}}
\newcommand{\kinparvecset}{\inputdimvec{\inputvec{x}}}

\newsubcommand{\ckvecset}{\ordervec{\inputvec{c}}}{k}

\newsubcommand{\ckvecapprox}{\mathbf{c}'}{k}
\newsubcommand{\ckvecapproxset}{\mathbf{C}'}{k}

\newsubcommand{\bkvecapprox}{\mathbf{b}'}{k}
\newsubcommand{\bkvecset}{\mathbf{B}}{k}
\newsubcommand{\bkvecapproxset}{\mathbf{B}'}{k}

\newcommand{\genobs}{y}

\newsubcommand{\genobsvec}{\ordervec{\genobs}}{k}
\newsubcommand{\genobsvecset}{\ordervec{\inputvec{\genobs}}}{k}
\newcommand{\genobsset}{\inputvec{\genobs}}

\newcommand{\genobsexp}{\genobs_{\textup{exp}}}        
\newcommand{\genobsexpset}{\inputvec{\genobs}_{\textup{exp}}}

\newsubcommand{\akvec}{\mathbf{a}}{k}

\newsubcommand{\akvecapprox}{\mathbf{a}'}{k}
\newsubcommand{\akvecset}{\mathbf{A}}{k}
\newsubcommand{\akvecapproxset}{\mathbf{A}'}{k}

{}  

\DeclareMathOperator{\pr}{pr} 
\newcommand{\given}{\,|\,}  

\newcommand{\Q}{\ensuremath{Q}}

\newcommand{\Qmax}{\Q_{\rm max}}

\newcommand{\Elab}{E_{\rm lab}}

\newcommand{\prel}{p_{\rm rel}}

\newcommand{\normal}{\mathcal{N}}

\newcommand{\transpose}[1]{{#1}^{\intercal}}
\newcommand{\trans}{\intercal}

\newcommand{\chiEFT}{$\chi$EFT}

\newcommand{\genobsref}{\ensuremath{y_{\mathrm{ref}}}}

\newcommand{\sigmatot}{\sigma_{\text{tot}}}

\def\diffd{\mathrm{d}}  

\DeclareDocumentCommand\differential{ o g d() }{ 
    \IfNoValueTF{#2}{
        \IfNoValueTF{#3}
            {\diffd\IfNoValueTF{#1}{}{^{#1}}}
            {\mathinner{\diffd\IfNoValueTF{#1}{}{^{#1}}\argopen(#3\argclose)}}
        }
        {\mathinner{\diffd\IfNoValueTF{#1}{}{^{#1}}#2} \IfNoValueTF{#3}{}{(#3)}}
    }

\newcommand{\pathd}{\mathcal{D}}  

\DeclareDocumentCommand\pathdifferential{ o g d() }{ 
    \IfNoValueTF{#2}{
        \IfNoValueTF{#3}
            {\pathd\IfNoValueTF{#1}{}{^{#1}}}
            {\mathinner{\pathd\IfNoValueTF{#1}{}{^{#1}}\argopen(#3\argclose)}}
        }
        {\mathinner{\pathd\IfNoValueTF{#1}{}{^{#1}}#2} \IfNoValueTF{#3}{}{(#3)}}
    }